\documentclass{article}

\usepackage{arxiv}
\usepackage[table,xcdraw]{xcolor}
\usepackage{amsmath,amsthm,amssymb,amscd}
\usepackage[all,cmtip]{xy}
\usepackage[utf8]{inputenc} 
\usepackage[T1]{fontenc}    
\usepackage{hyperref}       
\usepackage{url}            
\usepackage{booktabs}       
\usepackage{amsfonts}       
\usepackage{nicefrac}       
\usepackage{microtype}      
\usepackage{lipsum}
\usepackage{mathrsfs}
\usepackage{fancyhdr}
\usepackage{color}
\usepackage{subfig}
\usepackage{graphicx}
\usepackage{textcomp}
\usepackage{ytableau}
\usepackage{tikz}
\usetikzlibrary{calc,intersections,through,backgrounds,patterns}
\newtheorem{exam.}[subsection]{Example}
\newtheorem{def.}[subsection]{Definition}

\newtheorem{prop.}[subsection]{Proposition}
\newtheorem{coro}[subsection]{Corollary}
\newtheorem{theorem}[subsection]{Theorem}
\newtheorem*{remark}{Remark}

\usepackage{algorithm}
\usepackage{algpseudocode}

\newcommand{\id}{{\rm{id}}}

\newcommand{\calF}{\mathcal{F}}

\newcommand{\bfs}{\mathbf{s}}

\newcommand{\bfu}{\mathbf{u}}
\newcommand{\bfv}{\mathbf{v}}
\newcommand{\bfw}{\mathbf{w}}
\newcommand{\bfx}{\mathbf{x}}
\newcommand{\bfy}{\mathbf{y}}

\newcommand{\conv}{{\rm conv}}

\newcommand{\Int}{{\rm int}}
\newcommand{\reInt}{{\rm reInt}}
\newcommand{\reBd}{{\rm reBd}}
\newcommand{\bd}{{\rm bd}}

\title{A Physics-informed Sheaf Model}
\author{
        Chuan-Shen Hu \\ 
	Division of Mathematical Sciences\\
	School of Physical and Mathematical Sciences\\
        Nanyang Technological University \\
	Singapore 637371 \\
	\texttt{chuanshen.hu@ntu.edu.sg} \\
	\And
	Xiang Liu \\
	Division of Mathematical Sciences\\
	School of Physical and Mathematical Sciences\\
        Nanyang Technological University \\
	Singapore 637371 \\
        \texttt{liuxiangmath@qq.com}\\
	\AND
	Kelin Xia \\
	Division of Mathematical Sciences\\
	School of Physical and Mathematical Sciences\\
        Nanyang Technological University \\
	Singapore 637371 \\
	\texttt{xiakelin@ntu.edu.sg}
}
\date{} 

\begin{document}
\maketitle

\begin{abstract}
Normal mode analysis (NMA) provides a mathematical framework for exploring the intrinsic global dynamics of molecules through the definition of an energy function, where normal modes correspond to the eigenvectors of the Hessian matrix derived from the second derivatives of this function. The energy required to 'trigger' each normal mode is proportional to the square of its eigenvalue, with six zero-eigenvalue modes representing universal translation and rotation, common to all molecular systems. In contrast, modes associated with small non-zero eigenvalues are more easily excited by external forces and are thus closely related to molecular functions. Inspired by the anisotropic network model (ANM), this work establishes a novel connection between normal mode analysis and sheaf theory by introducing a cellular sheaf structure, termed the anisotropic sheaf, defined on undirected, simple graphs, and identifying the conventional Hessian matrix as the sheaf Laplacian. By interpreting the global section space of the anisotropic sheaf as the kernel of the Laplacian matrix, we demonstrate a one-to-one correspondence between the zero-eigenvalue-related normal modes and a basis for the global section space. We further analyze the dimension of this global section space, representing the space of harmonic signals, under conditions typically considered in normal mode analysis.  Additionally, we propose a systematic method to streamline the Delaunay triangulation-based construction for more efficient graph generation while preserving the ideal number of normal modes with zero eigenvalues in ANM analysis.
\end{abstract}

\section{Introduction}
\label{Section: Introduction}

Molecular functions are inherently tied to their dynamic behavior, as Richard Feynman famously stated:  'everything that living things do can be understood in terms of the jigglings and wigglings of atoms.' However, molecular motions can be highly complex. In particular, proteins continuously experience various motions under physiological conditions, including atomic thermal vibrations, sidechain rotations, residue movements, and large-scale domain shifts. Protein functions are closely linked to their motions and fluctuations, significantly influencing processes such as drug binding~\cite{Alvarez-Garcia:2014}, molecular docking~\cite{Fischer:2014}, self-assembly~\cite{marsh2014protein}, allosteric signaling~\cite{bu2011proteins}, and enzyme catalysis~\cite{fraser2009hidden}. The extent of protein motion within a cellular environment is largely determined by the local flexibility of its structure, an inherent characteristic of the protein. This flexibility is commonly assessed using the Debye-Waller factor (B-factor), which represents the atomic mean-square displacement and is measured through techniques such as X-ray crystallography, NMR spectroscopy, or single-molecule force experiments~\cite{dudko2006intrinsic}. However, the B-factor is not an absolute measure of flexibility, as it is also influenced by factors such as the crystal environment, solvent type, data collection conditions, and the procedures used for structural refinement~\cite{kondrashov2007protein,hinsen2008structural}.

Various models have been proposed to assess the flexibility of biomolecules, including molecular dynamics (MD)~\cite{mccammon1977dynamics}, normal mode analysis (NMA)~\cite{brooks1983charmm,go1983dynamics,levitt1985protein,tasumi1982normal}, elastic network models (ENM)~\cite{bahar1997direct, bahar1998vibrational, Atilgan:2001,cui2005normal,hinsen1998analysis,li2002coarse,tama2001conformational}, and approaches based on graph theory~\cite{jacobs2001protein}. In particular, NMA serves as a time-independent MD model~\cite{park2013coarse}, decomposing molecular dynamics using eigenvalues and eigenvectors, leveraging this spectral information to predict the motions of a given biomolecule. Notably, the first few eigenvectors derived from NMA models often capture the collective and global motions of the biomolecule, which potentially play a significant role in its functional behavior.

Among the Elastic Network Models (ENMs), the Gaussian Network Model (GNM)~\cite{bahar1998vibrational,bahar1997direct} and the Anisotropic Network Model (ANM)~\cite{Atilgan:2001} have emerged as widely used tools for studying protein dynamics. Both models have gained significant attention in the scientific community due to their simplicity and ability to accurately describe experimental data on equilibrium protein dynamics. The GNM offers a straightforward approach to investigating isotropic motions, while the ANM focuses on anisotropic motions. Specifically, while GNM is commonly applied in molecular flexibility analysis, such as B-factor prediction~\cite{Opron:2014,park2013coarse,LWYang:2008}, ANM is employed to analyze molecular functions directly related to intrinsic global normal modes.

In molecular dynamics, the eigenvectors, known as \textit{eigenmodes}, of the ANM Hessian matrix describe the intrinsic motion of atoms. Generally, eigenmodes associated with smaller eigenvalues correspond to more global and stable molecular motions, whereas those with larger eigenvalues reflect more pronounced and unstable local movements. Notably, the eigenmodes associated with an eigenvalue of zero, referred to as \textit{trivial modes}, represent rigid body motions, such as the translation and rotation of the entire molecule~\cite{cui2005normal}.

Despite the wide applications and significant success of normal mode analysis (NMA) in studying molecular structure, flexibility, dynamics, and function, a solid mathematical foundation for NMA is largely absent. A longstanding problem, involving the rigorous lower bound for the cutoff distance or a systematic approach for constructing the underlying graph essential to NMA models, remains unsolved. Furthermore, while it is known that normal modes are directly influenced by the molecular ``shape,'' the fundamental connections between normal mode behaviors and the underlying molecular topology have not been thoroughly explored. In summary, traditional graph-spectral-based NMA has inherent limitations in characterizing and analyzing the fundamental relationships between molecular dynamics and underlying topology. Establishing a rigorous mathematical framework for normal mode analysis remains a significant challenge.

\paragraph{Sheaf theory}
Molecular representation and featurization are essential in the application of AI to scientific fields. Mathematically, geometry and topology offer foundational frameworks for physical, chemical, and AI models~\cite{huangstability, nguyen2020review, nguyen2019mathematical, nguyen2020mathdl, peigeom, ye2019curvature, ghrist2014elementary}. Geometry focuses on shape and spatial configurations, while topology examines the underlying connections and relationships. Sheaf theory serves as a powerful bridge between geometry and topology, providing a framework to integrate rich and complex information about topological spaces equipped with ``data'' or ``signals'' in their local regions. Specifically, for an underlying topological space, a sheaf assigns algebraic structures (e.g., groups, vector spaces, rings, etc.) to local regions, captures local coherence, and shapes global information by coherently "gluing" these local data together~\cite{bredon2012sheaf,hartshorne2013algebraic,curry2014sheaves,robinson2014topological,kashiwara2018persistent}. Originally developed to address fixed-point problems in partial differential equations and good cover problems in nerve theory, sheaf theory has become a foundation of modern mathematics. Today, it serves as a common language across various disciplines, facilitating exploration in fields such as algebraic geometry, algebraic topology, number theory, and complex geometry~\cite{bredon2012sheaf,hartshorne2013algebraic, artin2012arithmetic,wells2017differential}. In applications, due to its richer and more flexible mathematical framework, the sheaf theory-based representation has also been incorporated into machine learning and deep learning architectures~\cite{barbero2022sheaf_attention,hansen2020sheaf,caralt2024joint,battiloro2023tangent,he2023sheaf,braithwaite2024heterogeneous,duta2024sheaf}.

Recently, a discretized version of sheaves, known as \textit{cellular sheaves} and their variants, has been proposed for Topological Data Analysis (TDA) on simplicial complexes (or cellular complexes)~\cite{curry2014sheaves,ghrist2014elementary,hansen2021opinion,robinson2014topological,hansRiess2022}. Specifically, unlike sheaves defined on general topological spaces, a cellular sheaf is defined on topological objects with strong combinatorial relationships between local structures, such as graphs, simplicial complexes, cell complexes, and even partially-ordered sets. By assigning algebraic objects (e.g., groups, vector spaces, rings) to combinatorial elements in the underlying topological complexes, such as simplices within a simplicial complex, the sheaf cohomology, sheaf Laplacian, and global and local section spaces can be directly constructed and computed algebraically. This approach provides a practical framework from a computational perspective. 

In particular, as a generalization of the cohomology of simplicial or cell complexes, along with various assignments to the stalk spaces and restriction maps, the cellular sheaf cohomology involves more geometric and physical information to the topology structure. In recent developments, the \textit{force cosheaf} model has been developed, providing a solid mathematical foundation for analyzing truss mechanisms~\cite{cooperband2023towards,cooperband2024equivariant,cooperband2023cosheaf,cooperband2024cellular}. The homology theory of the force cosheaf captures the axial forces along incident members connected by truss joints, offering a novel mathematical framework for studying equilibrium stresses and truss system configurations. In particular, the modern form of Maxwell’s Rule for 3-dimensional truss systems can be understood through Euler characteristics within this homology theory, primarily focusing on homology and its Euler characteristics~\cite{cooperband2023cosheaf,cooperband2024cellular}. By treating the cellular sheaf as a dual structure to the force cosheaf, it has been shown that force cosheaf homology can be regarded as the dual counterpart of cellular sheaf cohomology. From an NMA perspective, the physical and geometric insights gained from the global sections, sheaf cohomology, sheaf Laplacian, and spectral information can be leveraged to explore connections between atomic movements and the underlying graph statics.

\paragraph{Main contributions} 
In this paper, we establish a rigorous sheaf theory-based mathematical framework for normal mode analysis (NMA), presenting a physics-informed sheaf model specifically tailored for the anisotropic network model (ANM). The main results of this paper are presented in three parts.

First, for molecules modeled as graphs, we represent the atomic system as a cellular sheaf, termed the \textit{anisotropic sheaf}, based on the principles of the anisotropic network model (ANM). This sheaf acts as a "dual" counterpart to the force cosheaf model with spring constants on edges~\cite{cooperband2023towards,cooperband2023cosheaf,cooperband2024cellular}. We demonstrate that the corresponding sheaf Laplacian matrix is equivalent to the Hessian matrix used in NMA, with its eigenvectors capturing intrinsic global molecular motions (Theorem \ref{Theorem: Main result 1}). 

Second, based on cellular sheaf modeling, sheaf theory concepts such as cohomology, global sections, and Hodge theory are utilized to provide deeper insight into the intrinsic dynamics of molecules. In particular, the global section space, and cellular sheaf cohomology provide a rigorous interpretation of rigid motions, offering a clear geometric description of these physical phenomena (Theorem \ref{Theorem: Main result 2}). Additionally, the spectral information of the sheaf Laplacian, including its eigenvalues and eigenvectors, provides insights into the dynamics and fluctuations of molecules. Furthermore, we present a sheaf-based proof for the existence of a minimal graph model that induces an ANM Hessian with the desired number of trivial modes (Theorem \ref{Theorem: Main result 3}).

Finally, we analyze the relationships between rigid motions and the underlying graph structure. In particular, drawing on the proofs and discussions in Section \ref{Section: Normal Mode Analysis in Anisotropic Sheaf Models}, we introduce the concept of an \textit{admissible simplicial complex} for building the graph from a given atomic system (Definition \ref{Definition: admissible homogeneous 3-complex}). We prove that any ANM Hessian based on the $1$-skeleton of an admissible simplicial complex induces exactly six trivial modes (Theorem \ref{Theorem: Main result 4-1}). Inspired by previous works employing Delaunay triangulation and related mathematical tools for constructing underlying graphs~\cite{xia2014identifying,zhou2014alpha}, we further examine the use of the sheaf framework in Delaunay triangulation-based construction. Specifically, we prove that every Delaunay triangulation of a 3D point cloud is admissible (Corollary \ref{Corollary: Main result 4-2}), ensuring that its $1$-skeleton induces a Hessian matrix with six trivial modes. Additionally, Algorithm \ref{Algorithm: Main result 5-2} provides a systematic method for constructing a minimal graph for a given point cloud in \( \mathbb{R}^3 \) that induces exactly six trivial modes, and any subgraph obtained by removing edges leads to a Hessian matrix with more than six trivial modes.

\paragraph{Organization of the Paper}
The remainder of this paper is organized into four sections (Sections \ref{Section: Foundations of the Mathematical Framework}--\ref{Section: Graph Construction for Anisotropic Sheaves}). Section \ref{Section: Foundations of the Mathematical Framework} introduces the mathematical foundations necessary for this work, including the theory of cellular sheaves on simplicial complexes, global sections, sheaf cohomology, and Laplacians. In Section \ref{Section: Anisotropic Sheaves}, we revisit the mathematical formulation of the anisotropic network model and present the proposed anisotropic sheaf model. Section \ref{Section: Normal Mode Analysis in Anisotropic Sheaf Models} discusses normal mode analysis using anisotropic sheaves, with a focus on the mathematical interpretation of eigenmodes with zero eigenvalues through global sections and an examination of the dimension of the global section space. Finally, Section \ref{Section: Graph Construction for Anisotropic Sheaves} addresses the construction of underlying graphs for anisotropic analysis, providing mathematical proofs for the construction and an efficient method for implementation.

\section{Foundations of the Mathematical Framework}
\label{Section: Foundations of the Mathematical Framework}

Cellular sheaves defined on simplicial complexes and graphs form the foundation of the proposed work. In this section, we introduce the definition of a cellular sheaf on an abstract simplicial complex, along with the associated sheaf cohomology and sheaf Laplacian, with a particular focus on cellular sheaves defined on finite abstract simplicial complexes and graphs. For a more comprehensive and general treatment—including cellular sheaves of vector spaces on posets or cellular sheaves of Hilbert spaces on cell complexes—refer to~\cite{curry2014sheaves, curry2016discrete, robinson2014topological, hansen2019toward}.

\paragraph{Cellular sheaves on simplicial complexes}
Mathematically, a \textit{cellular sheaf} of $\mathbb{R}$-vector spaces on an abstract simplicial complex $K$ is defined as a functor $\mathcal{F}: (K, \leq) \rightarrow \textup{\textsf{Vect}}_{\mathbb{R}}$, from the poset category $(K, \leq)$ to the category $\textup{\textsf{Vect}}_{\mathbb{R}}$ of vector spaces and linear transformations, where $\leq$ denotes the partial order induced by the face relations of simplices in $K$. Specifically, an \textit{abstract simplicial complex} $K$ over a vertex set $V$ is a collection of non-empty subsets of $V$, called \textit{simplices}, with the following property: if $\sigma \in K$ and $\tau$ is a non-empty subset of $\sigma$, then $\tau \in K$, where $\tau$ is referred to as a \textit{face} of $\sigma$. The subset relation $\subseteq$ on simplices in $K$ forms a partial order, typically denoted by $\leq$ (or $\trianglelefteq$), which defines a category $(K, \leq)$. Additionally, a simplex $\sigma$ in $K$ is called a $q$\textit{-simplex} or is said to have \textit{dimension} $q$, with $q \in \mathbb{Z}_{\geq 0}$, if it consists of $q+1$ vertices. The dimension of $\sigma \in K$ is denoted by $\dim(\sigma)$, and the collection of $q$-simplices within $K$ is denoted by $K_{(q)}$. Under this setup, a cellular sheaf $\mathcal{F}: (K, \leq) \rightarrow \textup{\textsf{Vect}}_{\mathbb{R}}$ over an abstract simplicial complex $K$ consists of the following data:
\begin{itemize}
\item[\rm (a)] a vector space $\mathcal{F}_\sigma$ for each simplex $\sigma \in K$;
\item[\rm (b)] an $\mathbb{R}$-linear transformation $\mathcal{F}_{\sigma, \tau}: \mathcal{F}_\sigma \rightarrow \mathcal{F}_\tau$ for simplices $\sigma \leq \tau$.
\end{itemize}
Moreover, as a functor, the map $\mathcal{F}_{\sigma, \sigma}: \mathcal{F}_\sigma \rightarrow \mathcal{F}_\sigma$ is defined as the identity map $\id_{\mathcal{F}_\sigma}$, and $\mathcal{F}_{\tau, \eta} \circ \mathcal{F}_{\sigma, \tau} = \mathcal{F}_{\sigma, \eta}$ for every triple towered simplices $\sigma \leq \tau \leq \eta$ in $K$. The vector space $\mathcal{F}_\sigma$ is called the \textit{stalk} of $\mathcal{F}$ at $\sigma$, and the linear transformation $\mathcal{F}_{\sigma, \tau}: \mathcal{F}_\sigma \rightarrow \mathcal{F}_\tau$ is called the \textit{restriction map} from $\mathcal{F}_\sigma$ to $\mathcal{F}_\tau$. Elements in the stalk space $\mathcal{F}_\sigma$ are referred to as \textit{local sections} of $\mathcal{F}$ on $\sigma$.

Indeed, the cellular sheaf structure can be defined straightforwardly when considering an undirected graph as a simplicial complex composed of $0$- and $1$-simplices (vertices and edges, respectively). Specifically, For a graph $G = (V,E)$, a cellular sheaf $\mathcal{F}: (G, \leq) \rightarrow \textup{\textsf{Vect}}_{\mathbb{R}}$ can be defined as follows:
\begin{itemize}
\item[\rm (a)] a vector space $\mathcal{F}_{v}$ for each vertex $v \in V$;
\item[\rm (b)] a vector space $\mathcal{F}_{e}$ for each edge $v \in E$;
\item[\rm (c)] a linear transformation $\mathcal{F}_{v,e}: \mathcal{F}_v \rightarrow \mathcal{F}_e$ for any ordered pair $v \leq e$.
\end{itemize}
As a convention, we define $\mathcal{F}_{v,e}: \mathcal{F}_v \rightarrow \mathcal{F}_e$ to be the zero map when $v \nleq e$. This convention simplifies the representation of the sheaf coboundary maps and the Laplacian using matrices (e.g., Equation \ref{Eq. Coboundary matrix}). It is worth noting that in the case of graphs, the composition rule $\mathcal{F}_{\tau, \eta} \circ \mathcal{F}_{\sigma, \tau} = \mathcal{F}_{\sigma, \eta}$ holds naturally for every chain of simplices $\sigma \leq \tau \leq \eta$ in $K$, since there are no simplices in dimension greater than or equal to $2$.
\paragraph{Global section spaces of cellular sheaves}
Global sections play an important role in both cellular and general sheaf theory, as they gather local section information from the underlying space and consistently glue them together~\cite{bredon2012sheaf,curry2014sheaves,hansen2019toward}. To define the global section space of a cellular sheaf $\mathcal{F}: (K, \leq) \rightarrow \textup{\textsf{Vect}}_{\mathbb{R}}$ over an abstract simplicial complex $K$, we begin by introducing the concept of chain spaces. Specifically, for every non-negative integer $q \in \mathbb{Z}_{\geq 0}$, the $q$\textit{-th cochain space} of $\mathcal{F}$ is defined as
\begin{equation*}
C^q(K,\mathcal{F}) = \prod_{\sigma \in K_{(q)}} \mathcal{F}_\sigma.
\end{equation*}
That is, $C^q(K,\mathcal{F})$ is the direct sum of the stalk spaces of $\mathcal{F}$ on $q$-simplices.  Elements in $C^q(K,\mathcal{F})$ are usually expressed by $|K_{(q)}|$-tuples $(\mathbf{x}_\sigma)_{\sigma \in K_q}$ with $\mathbf{x}_\sigma \in \mathcal{F}_\sigma$ for every $\sigma \in K_{(q)}$. In particular, if $K$ is finite, then the cochain spaces can be written by
\begin{equation*}
C^q(K,\mathcal{F}) = \bigoplus_{\sigma \in K_{(q)}} \mathcal{F}_\sigma.    
\end{equation*}
\begin{def.}
Let $\mathcal{F}: (K, \leq) \rightarrow \textup{\textsf{Vect}}_{\mathbb{R}}$ be a cellular sheaf over an abstract simplicial complex $K$ with vertex set $V = K_{(0)}$. A $|K_{(0)}|$-tuple $(\mathbf{x}_v)_{v \in V} \in C^0(K,\mathcal{F})$ is called a \textbf{global section} of $\mathcal{F}$ if $\mathcal{F}_{v,e}(\mathbf{x}_v) = \mathcal{F}_{w,e}(\mathbf{x}_w)$ for every $1$-simplex $e \in K_{(1)}$ such that $v \leq e$ and $w \leq e$. The collection of global sections of $\mathcal{F}$, referred to as the \textbf{global section space} of $\mathcal{F}$, is denoted by $\Gamma(K,\mathcal{F})$. The mathematical representation of the global section space is given by
\begin{equation*}
\Gamma(K,\mathcal{F}) = \{ (\bfx_v)_{v \in K_{(0)}} \in C^0(K,\mathcal{F}) \ | \ \mathcal{F}_{v, e}(\bfx_v) = \mathcal{F}_{w, e}(\bfx_w) \text{ if } v \leq e \text{ and } w \leq e \text{ for some } e \in K_{(1)} \}.    
\end{equation*}
\end{def.}
Because every $\mathcal{F}_{v,e}$ and $\mathcal{F}_{w,e}$ are $\mathbb{R}$-linear maps, the global section space $\Gamma(K,\mathcal{F})$ is an $\mathbb{R}$-valued vector subspace of the $0$-th cochain space $C^0(K,\mathcal{F})$. In sheaf theory, this space collects the \textit{harmonic signals} on the simplices that can be coherently assembled into a global signal of the sheaf~\cite{robinson2014topological,hansen2019toward, bodnar2022neural}.
\paragraph{Restriction of cellular sheaves}
In this paragraph, we introduce a sheaf operation known as the \textit{restriction sheaf}, which functions to produce a cellular sheaf on a subcomplex from a cellular sheaf defined on a simplicial complex. This operation can be used to examine the behavior of the sheaf on the substructures of the underlying simplicial complex, such as the dimension of the global section space on these substructures and other related properties. In particular, in the upcoming sections, this operation will be frequently applied to cellular sheaves on simplicial complexes, serving as an effective tool for studying various properties of the sheaves.  

Let $\mathcal{F}: (K,\leq) \rightarrow \textsf{Vect}_{\mathbb{R}}$ be a cellular sheaf defined on a simplicial complex $K$ and let $L$ be a subcomplex of $K$. Then the face relation $\leq$ on simplices in $L$ inherits the partial order from $K$. The restriction sheaf of $\mathcal{F}$ on $L$, denoted as $\mathcal{F}|_L: (L,\leq) \rightarrow \textsf{Vect}_{\mathbb{R}}$, is defined as follows. For every simplex $\sigma$ and pair $\sigma \leq \tau$ in $L \subseteq K$, the stalk space and restrictions map are defined by $(\mathcal{F}|_L)_\sigma = \mathcal{F}_\sigma$ and $(\mathcal{F}|_L)_{\sigma, \tau} = \mathcal{F}_{\sigma, \tau}$. By inheriting the structure of $\mathcal{F}$, $\mathcal{F}|_L$ satisfies all the sheaf axioms. The following proposition provides a straightforward relationship between the global sections of $\mathcal{F}$ and those of $\mathcal{F}|_L$. 
\begin{prop.}\label{Proposition: global section space and restriction sheaf}
Let $\mathcal{F}: (K,\leq) \rightarrow \textup{\textsf{Vect}}_{\mathbb{R}}$ be a cellular sheaf over a simplicial complex $K$ with vertex set $V = K_{(0)}$, and let $L \subseteq K$ be a subcomplex. Let $(\mathbf{x}_v)_{v \in K_{(0)}} \in C^0(K,\mathcal{F})$ be a $|K_{(0)}|$-tuple of local sections. If $(\mathbf{x}_v)_{v \in K_{(0)}}$ is a global section of $\mathcal{F}$, then the restricted $|L_{(0)}|$-tuple $(\mathbf{x}_v)_{v \in L_{(0)}} \in C^0(L,\mathcal{F}|_L)$ is a global section of $\mathcal{F}|_L$.
\end{prop.}
\begin{proof}
Let $\mathcal{G} = \mathcal{F}|_L$ and $e$ be an edge (i.e., a $1$-simplex) of $L$. Because $L$ is a subcomplex of $K$, $e$ is also a $1$-simplex of $K$. Because $(\mathbf{x}_v)_{v \in K_{(0)}} \in \Gamma(K,\mathcal{F})$, $\mathcal{F}_{v,e}(\mathbf{x}_v) = \mathcal{F}_{w,e}(\mathbf{x}_w)$ whenever $v, w \in K_{(0)}$ with $v \leq e$ and $w \leq e$.  Since $L_{(0)} \subseteq K_{(0)}$, this formula shows that $(\mathbf{x}_v)_{v \in L_{(0)}}$ is a global section of $\mathcal{F}|_L$.
\end{proof}
\paragraph{Sheaf cohomology and sheaf Laplacians}
Similar to the Laplacian matrix of an undirected graph, a cellular sheaf defined on a simplicial complex gives rise to a related structure known as the \textit{sheaf Laplacian} on the underlying simplicial complex. To provide an overview of cellular sheaf theory and to leverage some useful results, we briefly introduce sheaf cohomology and the sheaf Laplacians of cellular sheaves of finite-dimensional vector spaces defined on finite abstract simplicial complexes. 

Let $K$ be a finite abstract simplicial complex with vertex set $V = K_{(0)}$, and let $\mathcal{F}: (K, \leq) \rightarrow \textup{\textsf{Vect}}_{\mathbb{R}}$ be a cellular sheaf. To define the sheaf cohomology and sheaf Laplacian of $\mathcal{F}$, the cochain spaces $C^q(K, \mathcal{F}) = \bigoplus_{\sigma \in K_{(q)}} \mathcal{F}_\sigma$ and the coboundary maps $\delta^q: C^q(K, \mathcal{F}) \rightarrow C^{q+1}(K, \mathcal{F})$ are essential components. Specifically, by defining a total order $<$ on $V$, each $q$-simplex in $K$ is uniquely determined by the oriented sequence $[v_0, v_1, \dots, v_q]$ with $v_0 < v_1 < \cdots < v_q$. Moreover, under these defined orientations, the signed incidence function $[\cdot: \cdot]: K \times K \rightarrow \{ -1, 0, 1 \}$ can be defined as follows.  For every paired $(\sigma, \tau) \in K \times K$ of simplicies in $K$, the signed incidence of the pair is defined by
\begin{equation}
\label{Eq. signed incidence function}
[\sigma:\tau] = \begin{cases} (-1)^i &, \ \text{if } \tau = [v_0, v_1, ..., v_n] \text{ and } \sigma = [v_0, ..., \widehat{v_i}, ..., v_n], \\    
0 &, \ \text{otherwise}. \\
\end{cases}    
\end{equation}
For any simplicial complex $K$, the signed incidence function satisfies the following relation for any $\sigma, \tau \in K$:
\begin{equation}
\label{Eq: Signed incidence relation}
\sum_{\eta \in K} [\sigma : \eta] \cdot [\eta : \tau] = 0,
\end{equation}
where analogous signed incidence functions can be defined in more general settings, such as for regular cell complexes (cf. \cite{curry2014sheaves}). The $q$-\textit{th coboundary map} $\delta^q: C^q(K,\mathcal{F}) \rightarrow C^{q+1}(K,\mathcal{F})$ is thus defined as the $\mathbb{R}$-linear map extended by the following assignment:
\begin{equation*}
\delta^q|_{\mathcal{F}_\sigma} = \sum_{\tau \in K_{(q+1)}} [\sigma:\tau] \cdot \mathcal{F}_{\sigma, \tau}.     
\end{equation*}
In particular, the composition map of the maps $\delta^{q+1} \circ \delta^q$ from $C^q(K,\mathcal{F})$ to $C^{q+1}(K,\mathcal{F})$ satisfies $\delta^{q+1} \circ \delta^q = 0$ for every $q \geq 0$ (cf. Lemma 6.6.2, \cite{curry2014sheaves}), and thus the $q$-th sheaf cohomology is defined as the quotient vector space
\begin{equation*}
H^q(K,\mathcal{F}) = \frac{\ker(\delta^q)}{{\rm im}(\delta^{q-1})}.
\end{equation*}
\begin{remark}
To define sheaf cohomology, a total order is used on the vertex set of the underlying simplicial complex. However, as is well-known in the computation of simplicial homology (e.g., Section 3.1, \cite{curry2015topological}), different total orders on $V$ induce isomorphic simplicial homologies for the simplicial complex. Similarly, as a dual structure of homology, the choice of total order on the vertex set does not affect the sheaf cohomology. For more information, refer to \cite{curry2014sheaves,robinson2014topological,XiaoqiWei2024FODS}. 
\end{remark}
The \textit{sheaf Laplacian} is also defined using the coboundary maps $\delta^q: C^q(K,\mathcal{F}) \rightarrow C^{q+1}(K,\mathcal{F})$. In particular, for a cellular sheaf of finite-dimensional $\mathbb{R}$-spaces on a finite simplicial complex $K$, the $q$-\textit{th sheaf Laplacian} of the cochain complex 
\begin{equation*}
\xymatrix@+0.0em{
\cdots
\ar[r]
& C^{q-1}(K,\mathcal{F})
\ar[r]^{\delta^{q-1}}
& C^{q}(K,\mathcal{F})
\ar[r]^{\delta^{q}}
\ar@/^1pc/[l]^{(\delta^{q-1})^*}
& C^{q+1}(K,\mathcal{F})
\ar@/^1pc/[l]^{(\delta^{q})^*}
\ar[r]
& \dots
}    
\end{equation*}
is defined as the linear operator $\Delta^q = (\delta^{q})^* \circ \delta^{q} + \delta^{q-1} \circ (\delta^{q-1})^*$ from $C^q(K,\mathcal{F})$ to itself, where $(\delta^{q-1})^*$ and $(\delta^{q})^*$ are the unique adjoint $\mathbb{R}$-linear maps of $\delta^{q-1}$ and $\delta^{q}$. Especially, the existence of adjoint maps is ensured by the finite-dimensionality of the vector spaces in the cochain $C^\bullet(K,\mathcal{F})$. 
Indeed, by considering each $\delta^{q}$ as a matrix over $\mathbb{R}$, the adjoint map $(\delta^{q})^*$ is precisely the transpose of $\delta^{q}$. For the construction of the sheaf Laplacian in more general settings, such as for cellular sheaves of Hilbert spaces, refer to~\cite{hansen2019toward} for more information.

Sheaf cohomology, the global section space, and the sheaf Laplacian have intrinsically close relationships. First, for a cellular sheaf $\mathcal{F}: (K, \leq) \rightarrow \textup{\textsf{Vect}}_{\mathbb{R}}$ on a simplicial complex $K$, it can also yield the space of global sections. Notably, by conventionally setting $C^{-1}(K,\mathcal{F}) = 0$ and $\delta^{-1}: C^{-1}(K,\mathcal{F}) \rightarrow C^{0}(K,\mathcal{F})$ as the zero map, one obtains
\begin{equation}
\label{Eq. Global section space representation}
H^0(K,\mathcal{F}) = \ker(\delta^0) \simeq \{ (\bfx_v)_{v \in K_0} \ | \ \mathcal{F}_{v,[v,w]}(\bfx_v) = \mathcal{F}_{w,[v,w]}(\bfx_w) \text{ if } [v,w] \in K_1 \} = \Gamma(K,\mathcal{F}).
\end{equation}
Second, by the Hodge decomposition theorem, each $q$-th sheaf cohomology can be identified as the kernel of the $q$-th sheaf Laplacian. This connection between sheaf cohomology and the sheaf Laplacian is formally stated in the following theorem.
\begin{theorem}
\label{Theorem: Hodge's theorem}
Let $(K,\leq)$ be a finite abstract simplicial complex and $\mathcal{F}: (K,\leq) \rightarrow \textup{\textsf{Vect}}_{\mathbb{R}}$ be a cellular sheaf of finite-dimensional $\mathbb{R}$-vector spaces defined on $K$. Let $\Delta^q: C^{q}(K,\mathcal{F}) \rightarrow C^{q}(K,\mathcal{F})$ be the $q$-th sheaf Laplacian. Then, $H^q(K,\mathcal{F}) \simeq \ker(\Delta^q)$. In particular, 
\begin{equation}
\label{Eq. Global section space and the kernel of 0th Laplacian}
\Gamma(K,\mathcal{F}) = H^0(K,\mathcal{F}) = \ker(\Delta^0) = \ker((\delta^0)^* \cdot \delta^0).
\end{equation}
In particular, the nullity of the sheaf Laplacian is the dimension of the global section space, i.e., $\dim_\mathbb{R} \ker(\Delta^0) =  \dim_\mathbb{R} \Gamma(K,\mathcal{F}) = \dim_\mathbb{R} \ker(\delta^0)$.
\end{theorem}
\begin{proof}
For a proof of the theorem, refer to Theorem 3.1 in~\cite{hansen2019toward}.
\end{proof}
From a molecular dynamics perspective, with the introduction of the proposed anisotropic sheaf model in the following section, we demonstrate how incorporating geometric and physical information into the cellular sheaf framework enables local sections at the vertices to capture localized fluctuations and movements. In contrast, integrating these local sections into the global section offers a cohesive view of the motion of the entire simplicial complex.

\section{Anisotropic Sheaves}
\label{Section: Anisotropic Sheaves}
In this section, we introduce the proposed sheaf modeling for atomic systems, which is defined on $1$-dimensional simplicial complexes—namely, undirected, simple, finite graphs. The anisotropic sheaf model, which will be the primary focus of our analysis, is inspired by the Anisotropic Network Model (ANM), a powerful framework for analyzing the normal modes of proteins and exploring the functional relationships and dynamics of numerous proteins~\cite{Atilgan:2001, doruker2000dynamics, eyal2006anisotropic, xia2015multiscale, xia2018multiscale}. 

\begin{figure}
	\centering
  \includegraphics[width=\linewidth]{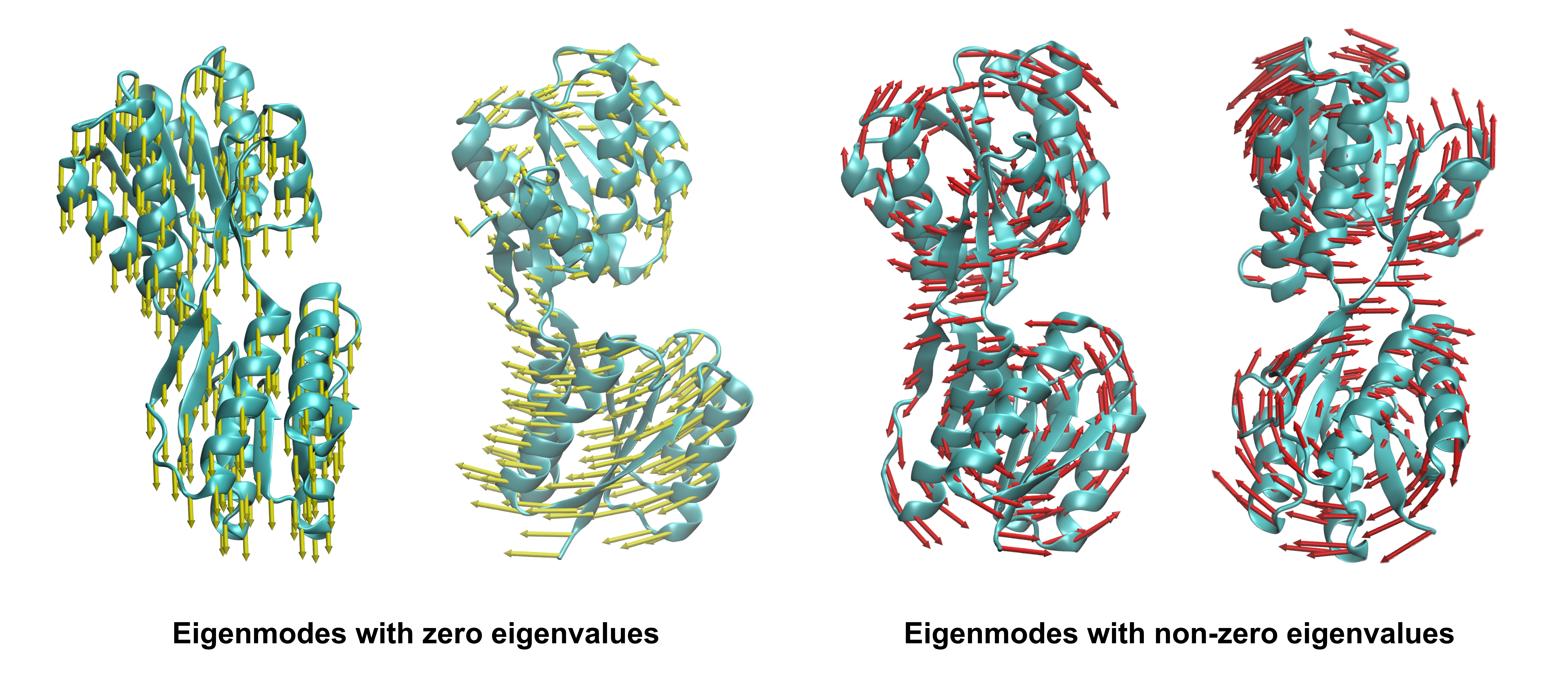}
  \caption{Illustration of the eigenmodes with zero and non-zero eigenvalues for chain A of the protein with ID 1URP from the RCSB Protein Data Bank~\cite{BJORKMAN1998651, Berman:2000}. The visualizations of the eigenmodes were generated using the VMD software~\cite{VMD}.}
\label{fig: Comparison of zero and non-zero eigenmodes}
\end{figure}

\paragraph{Anisotropic network and Hassien matrix}
The mathematical formulation of the anisotropic network and the associated Hessian matrix is presented below.  Refer to~\cite{Atilgan:2001, doruker2000dynamics, eyal2006anisotropic, xia2015multiscale, xia2018multiscale} for more detailed settings, discussions, and formula derivations related to the ANM framework. Given an atomic system with 3D coordinates of atomic positions, particularly for the ${\rm C}_\alpha$ atoms, the anisotropic network is constructed as follows. Let $\{ v_i = (x_i^\circ, y_i^\circ, z_i^\circ) \ | \ i = 1, 2, ..., n \}$ be the coordinates of the equilibrium positions of the atoms. Two distance measurements for two atoms $v_i$ and $v_j$ are considered, the equilibrium distance $s_{ij}^\circ$ and $s_{ij}$. With a spring constant $\gamma$, the harmonic potential between these two atoms is formulated as
\begin{equation*}
V_{ij} = \frac{\gamma}{2} (s_{ij} - s_{ij}^\circ)^2.   
\end{equation*}
Furthermore, by using a fixed cutoff distance \( R_c \) and the Heaviside step function \( H: \mathbb{R} \rightarrow \{ 0, 1 \} \), defined by \( H(x) = 1 \) for \( x \geq 0 \) and \( H(x) = 0 \) otherwise, the potential function \( V_{\text{ANM}} \) for the entire atomic system is given by
\begin{equation}
\label{Eq. ANM energy function}
\begin{split}
V_{\text{ANM}} = \frac{\gamma}{2} \sum_{i,j} (d_{ij} - d_{ij}^\circ)^2 \cdot H(R_c - d_{ij}^\circ),
\end{split}
\end{equation}
where $d_{ij} := \Vert (x_j - x_i, y_j - y_i, z_j - z_i) \Vert$ and $d_{ij}^\circ := \Vert (x_j^\circ - x_i^\circ, y_j^\circ - y_i^\circ, z_j^\circ - z_i^\circ) \Vert$. In particular, $H(R_c - d_{ij}^\circ) = 0$ if the equilibrium distance $d_{ij}^\circ$ is larger than a given cutoff distance $R_c$. Then, for the $i$-th and $j$-th atoms, the ANM Hessian matrix is defined as
\begin{equation}
\label{Eq. small Hessian matrix}
\begin{split}
H_{ij} &= \begin{bmatrix}
\frac{\partial V_{ij}}{\partial x_i \partial x_j} & \frac{\partial V_{ij}}{\partial x_i \partial y_j} & \frac{\partial V_{ij}}{\partial x_i \partial z_j}\\
\frac{\partial V_{ij}}{\partial y_i \partial x_j} & \frac{\partial V_{ij}}{\partial y_i \partial y_j} & \frac{\partial V_{ij}}{\partial y_i \partial z_j}\\
\frac{\partial V_{ij}}{\partial z_i \partial x_j} & \frac{\partial V_{ij}}{\partial z_i \partial y_j} & \frac{\partial V_{ij}}{\partial z_i \partial z_j}\\
\end{bmatrix} \\ 
&= \frac{-\gamma}{(s_{ij}^\circ)^2} \cdot \begin{bmatrix} 
(x_j^\circ-x_i^\circ)(x_j^\circ-x_i^\circ) & (x_j^\circ-x_i^\circ)(y_j^\circ-y_i^\circ) & (x_j^\circ-x_i^\circ)(z_j^\circ-z_i^\circ) \\ (x_j^\circ-x_i^\circ)(y_j^\circ-y_i^\circ) & (y_j^\circ-y_i^\circ)(y_j^\circ-y_i^\circ) & (y_j^\circ-y_i^\circ)(z_j^\circ-z_i^\circ) \\ (x_j^\circ-x_i^\circ)(z_j^\circ-z_i^\circ) & (y_j^\circ-y_i^\circ)(z_j^\circ-z_i^\circ) & (z_j^\circ-z_i^\circ)(z_j^\circ-z_i^\circ)  \\
\end{bmatrix}
\end{split}.
\end{equation}
In particular, according to Equation~\eqref{Eq. ANM energy function}, $H_{ij}$ equals the zero matrix if $d_{ij}^{\circ}$ is larger than the cutoff distance. The force constant of the entire system is described by the following $3n \times 3n$ matrix, known as the \textit{Hessian matrix} of the system. That is,
\begin{equation}
\label{Eq. The big Hessian matrix}
\mathbf{H}_{\rm ANM} = \begin{bmatrix}
H_{11} & H_{12} & \cdots & H_{1n}\\
H_{21} & H_{22} & \cdots & H_{2n}\\
\vdots & \vdots & \ddots & \vdots \\
H_{n1} & H_{n2} & \cdots & H_{nn}\\
\end{bmatrix},
\end{equation}
where each $H_{ij}$ with $i \neq j$ is a $3 \times 3$ matrix defined in Equation \eqref{Eq. small Hessian matrix} and $H_{ii} := -\sum_{j \neq i} H_{ij}$. The Hessian matrix is symmetric and positive semi-definite, and therefore can be diagonalized into \( 3n \) non-negative eigenvalues. The eigenvectors of \( \mathbf{H}_{\rm ANM} \) are referred to as \textit{eigenmodes} of the anisotropic network. 

In physical terms, the eigenvalues of the Hessian matrix represent the energetic cost of displacing the system. Eigenmodes with smaller eigenvalues correspond to low-energy deformations and typically reflect delocalized motions. Specifically, the eigenmodes associated with zero eigenvalues correspond to the rigid motions of the entire biomolecule, such as global translations and rotations. In contrast, eigenmodes with larger eigenvalues represent high-energy modes with higher energetic costs, and they are associated with more significant local deformations of the biomolecule~\cite{cui2005normal}. Figure \ref{fig: Comparison of zero and non-zero eigenmodes} illustrates examples of eigenmodes with zero and non-zero eigenvalues.

With a sufficiently large cutoff distance \( R_c \), the ANM Hessian matrix for the underlying graph structure typically exhibits six zero eigenvalues, corresponding to the trivial modes of translation and rotation for the entire biomolecular complex (Figure \ref{fig: Six non-zero eigenmodes}). However, if \( R_c \) is too small, the number of zero eigenvalue modes can significantly exceed six, creating challenges for accurate ANM analysis. On the other hand, using extremely large cutoff distances complicates the network formed by the atomic system, resulting in significantly higher computational complexity.

\paragraph{Mathematical formulation of anisotropic sheaves}
With an underlying geometric graph as the foundation, the \textit{anisotropic sheaf} is defined as follows. Let $G = (V, E)$ be an undirected, simple, finite graph, where the vertices in $V$ are labeled as $v_1, v_2, \dots, v_n$, each accompanied by coordinates $(x_i^\circ, y_i^\circ, z_i^\circ) \in \mathbb{R}^3$ for $i = 1, 2, \dots, n$. Then, a cellular sheaf $\mathcal{F}: (G, \leq) \rightarrow \textup{\textsf{Vect}}_\mathbb{R}$ is defined by the following assignments. For each vertex $v_i \in V$ and edge $[v_i, v_j] \in E$ with $i < j$, we define $\mathcal{F}_{v_i} = \mathbb{R}^3$ and $\mathcal{F}_{[v_i,v_j]} = \mathbb{R}$. For the linear transformations for ordered pairs $v_i \leq [v_j, v_j]$ and $v_i \leq [v_j, v_j]$, we define
\begin{equation}
\label{Eq. Anisotropic sheaf definition}
\mathcal{F}_{v_i, [v_i,v_j]} = w_{ij} \cdot \begin{bmatrix}
x_j^\circ-x_i^\circ & y_j^\circ-y_i^\circ & z_j^\circ-z_i^\circ \\
\end{bmatrix}
= \mathcal{F}_{v_j, [v_i,v_j]}  
\end{equation}
as a $1 \times 3$ matrix, which represents a linear transformation from $\mathbb{R}^3$ to $\mathbb{R}^1$, and $w_{ij}$ is an assigned \textit{weight value} for the edge $[v_i, v_j]$. Furthermore, as the convention introduced in Section \ref{Section: Foundations of the Mathematical Framework}, we also establish
\begin{equation}
\label{Eq. General cellular sheaf convention}
\mathcal{F}_{v_i, [v_j,v_k]} = \begin{bmatrix}
0 & 0 & 0
\end{bmatrix}
\end{equation}
for $i \notin \{ j, k \}$, denoting the zero map from $\mathcal{F}_{v_i}$ to $\mathcal{F}_{[v_j,v_k]}$ when $v_i$ is not a face of $[v_j,v_k]$. Then, for every $[v_i, v_j] \in E$ with $i < j$, we have 
\begin{equation}
\label{Eq. Local 3 times 3 matrix of the ANM sheaf}
\begin{split}
\mathcal{F}_{v_j, [v_i,v_j]}^T\mathcal{F}_{v_i, [v_i,v_j]} &= \mathcal{F}_{v_i, [v_i,v_j]}^T\mathcal{F}_{v_j, [v_i,v_j]} = w_{ij}^2 \cdot\begin{bmatrix}
x_j^\circ-x_i^\circ \\ y_j^\circ-y_i^\circ \\ z_j^\circ-z_i^\circ \\
\end{bmatrix}  \cdot \begin{bmatrix}
x_j^\circ-x_i^\circ & y_j^\circ-y_i^\circ & z_j^\circ-z_i^\circ \\
\end{bmatrix} 
\\ 
& = w_{ij}^2 \cdot \begin{bmatrix}
(x_j^\circ-x_i^\circ)^2 & (x_j^\circ-x_i^\circ)(y_j^\circ-y_i^\circ) & (x_j^\circ-x_i^\circ)(z_j^\circ-z_i^\circ) \\ 
(x_j^\circ-x_i^\circ)(y_j^\circ-y_i^\circ) & (y_j^\circ-y_i^\circ)^2 & (y_j^\circ-y_i^\circ)(z_j^\circ-z_i^\circ) \\ 
(x_j^\circ-x_i^\circ)(z_j^\circ-z_i^\circ) & (y_j^\circ-y_i^\circ)(z_j^\circ-z_i^\circ) & (z_j^\circ-z_i^\circ)^2 \\
\end{bmatrix}.   
\end{split}
\end{equation}
In particular, compared to Equation~\eqref{Eq. small Hessian matrix},  the matrix $\mathcal{F}_{v_j, [v_i,v_j]}^T\mathcal{F}_{v_i, [v_i,v_j]}$ corresponds to the Hessian matrix for the $i$-th and $j$-th atoms in the ANM model, with the only difference being the scaling factor $\frac{-\gamma}{(s_{ij}^\circ)^2}$. We term the cellular sheaf $\mathcal{F}$ as an anisotropic sheaf defined on the geometric graph $G = (V,E)$.

Moreover, to establish the sheaf cohomology of an anisotropic sheaf, we consider the following setup. By \eqref{Eq. signed incidence function}, the signed incidence is defined by $[v_i:[v_i,v_j]] := -1$, $[v_j:[v_i,v_j]] := 1$, and $[v_i:[v_j,v_k]] = 0$ for $i \notin \{ j, k \}$. Let $E = \{ e_1, ..., e_m \}$, the set of edges in $G$, then the $0$-th and $1$-st cochain spaces of $\mathcal{F}$ can be explicitly expressed by
\begin{equation*}
C^0(G,\mathcal{F}) = \bigoplus_{k = 1}^n \mathcal{F}_{v_k} \simeq \mathbb{R}^{3n} \text{ and } C^1(G,\mathcal{F}) = \bigoplus_{l = 1}^m \mathcal{F}_{e_l} \simeq \mathbb{R}^{m}.   
\end{equation*}
Representing elements in $C^0(G, \mathcal{F})$ and $C^1(G, \mathcal{F})$ as elements in $\mathbb{R}^{3n}$ and $\mathbb{R}^{m}$ in column form, the $0$-th coboundary map becomes an $\mathbb{R}$-linear map $\mathbf{C}: C^0(G, \mathcal{F}) \longrightarrow C^1(G, \mathcal{F})$, represented by the following $m \times 3n$ matrix: 
\begin{equation}
\label{Eq. Coboundary matrix}
\mathbf{C} = \begin{bmatrix}
[v_1:e_1] \cdot \mathcal{F}_{v_1, e_1} & [v_2:e_1] \cdot \mathcal{F}_{v_2, e_1} & \cdots & [v_n:e_1] \cdot \mathcal{F}_{v_n, e_1}\\
[v_1:e_2] \cdot \mathcal{F}_{v_1, e_2} & [v_2:e_2] \cdot \mathcal{F}_{v_2, e_2} & \cdots & [v_n:e_2] \cdot \mathcal{F}_{v_n, e_2}\\
\vdots & \vdots & \ddots & \vdots \\
[v_1:e_m] \cdot \mathcal{F}_{v_1, e_m} & [v_2:e_m] \cdot \mathcal{F}_{v_2, e_m} & \cdots & [v_n:e_m] \cdot \mathcal{F}_{v_n, e_m}\\
\end{bmatrix},
\end{equation}
where $\mathbf{C}$ is composed of $m \times n$ blocks, and the $(l,k)$-th block is the $1 \times 3$ matrix $[v_k:e_l] \cdot \mathcal{F}_{v_k, e_l}$. Then, the $0$-th sheaf Laplacian $L_\mathcal{F} := \Delta^0$ is defined as the $3n \times 3n$ matrix
\begin{equation}
\label{Eq. Sheaf Laplacian}
L_\mathcal{F} = \mathbf{C}^T \cdot \mathbf{C} = \begin{bmatrix}
L_{11} & L_{12} & \cdots & L_{1n}\\
L_{21} & L_{22} & \cdots & L_{2n}\\
\vdots & \vdots & \ddots & \vdots \\
L_{n1} & L_{n2} & \cdots & L_{nn}\\
\end{bmatrix},   
\end{equation}
where the $(i,j)$-block matrix, with $i, j \in \{ 1, 2, \dots, n \}$, of the Laplacian matrix $L_\mathcal{F}$ is given by
\begin{equation*}
L_{ij} = \sum_{l = 1}^m \ [v_j : e_l] \cdot [v_i : e_l] \cdot \mathcal{F}_{v_i,e_l}^T\mathcal{F}_{v_j,e_l}.
\end{equation*}
Since $G$ is a simple graph, i.e., a $1$-dimensional abstract simplicial complex, any two vertices $v_i$ and $v_j$ with $i < j$ can have at most one edge between them. In particular, the $(i,j)$-block matrix of $L_\mathcal{F}$ satisfies the equation
\begin{equation*}
L_{ij} = \begin{cases} [v_j : [v_i,v_j]] \cdot [v_i : [v_i,v_j]] \cdot \mathcal{F}_{v_i,[v_i,v_j]}^T\mathcal{F}_{v_j,[v_i,v_j]} = -\mathcal{F}_{v_i,[v_i,v_j]}^T\mathcal{F}_{v_j,[v_i,v_j]} &, \ \text{if } [v_i,v_j] \in E, \\    
0 &, \ \text{otherwise}. \\
\end{cases}
\end{equation*}
More precisely, by Equation \eqref{Eq. Local 3 times 3 matrix of the ANM sheaf}, for $[v_i,v_j] \in E$, matrix $L_{ij}$ can be represented by
\begin{equation*}
L_{ij} = -\mathcal{F}_{v_i,[v_i,v_j]}^T\mathcal{F}_{v_j,[v_i,v_j]} = -w_{ij}^2 \cdot \begin{bmatrix} 
(x_j^\circ-x_i^\circ)(x_j^\circ-x_i^\circ) & (x_j^\circ-x_i^\circ)(y_j^\circ-y_i^\circ) & (x_j^\circ-x_i^\circ)(z_j^\circ-z_i^\circ) \\ (x_j^\circ-x_i^\circ)(y_j^\circ-y_i^\circ) & (y_j^\circ-y_i^\circ)(y_j^\circ-y_i^\circ) & (y_j^\circ-y_i^\circ)(z_j^\circ-z_i^\circ) \\ (x_j^\circ-x_i^\circ)(z_j^\circ-z_i^\circ) & (y_j^\circ-y_i^\circ)(z_j^\circ-z_i^\circ) & (z_j^\circ-z_i^\circ)(z_j^\circ-z_i^\circ)  \\
\end{bmatrix}.   
\end{equation*}
On the other hand, by adapting the convention in Equation~\eqref{Eq. General cellular sheaf convention} and the definition of $\mathcal{F}_{v_i, [v_i,v_j]} = \mathcal{F}_{v_j, [v_i,v_j]}$,
\begin{equation*}
L_{ii} = \sum_{l = 1}^m \mathcal{F}_{v_i, e_l}^T\mathcal{F}_{v_i, e_l} = \sum_{j = 1}^n \mathcal{F}_{v_i, [v_i,v_j]}^T\mathcal{F}_{v_i, [v_i,v_j]} = \sum_{j = 1}^n \mathcal{F}_{v_i, [v_i,v_j]}^T\mathcal{F}_{v_j, [v_i,v_j]} = \sum_{j \neq i} -L_{ij}
\end{equation*}
for every $i \in \{ 1, 2, ..., n \}$. In particular, when setting each weight $w_{ij}$ to be $\gamma^{1/2}/s_{ij}^\circ$, the induced sheaf Laplacian $L$ is exactly the Hessian matrix from the ANM model (cf. Equations \eqref{Eq. The big Hessian matrix} and \eqref{Eq. small Hessian matrix}), i.e., $L_{ij} = H_{ij}$. Furthermore, by Theorem \ref{Theorem: Hodge's theorem}, the space $\ker(L_\mathcal{F})$ equals the $0$-th cohomology $H^0(G,\calF)$; that is, the space of global sections of $\calF$ over the graph $G$. According to the ANM framework, the global section space corresponds to the eigenmodes associated with the zero eigenvalue of the Hessian matrix. We summarize the above inferences in the following theorem.

\begin{theorem}
\label{Theorem: Main result 1}
Let $G = (V, E)$ be an undirected, simple, finite graph with vertex coordinates $(x_i^\circ, y_i^\circ, z_i^\circ)$, where each vertex has distinct coordinates. Let $\mathcal{F}: (G,\leq) \rightarrow \textup{\textsf{Vect}}_\mathbb{R}$ be the anisotropic sheaf defined as in~\eqref{Eq. Anisotropic sheaf definition}. If $w_{ij} = \gamma^{1/2}/s_{ij}^\circ$, where $\gamma$ denotes the spring constant and $s_{ij}^\circ$ represents the equilibrium distance between the $i$-th and $j$-th vertices, then the sheaf Laplacian $L_\mathcal{F} := \Delta^0$ equals the Hessian matrix $\mathbf{H}_{\rm ANM}$ based on the underlying graph $G = (V, E)$.
\end{theorem}
\begin{exam.}
\label{Example: anisotropic sheaf defined on K3}
Let $G = K_3$ be the complete graph with three vertices.  Furthermore, we represent $V = \{ 1, 2, 3\}$ and $E = \{ [1,2], [1,3], [2,3] \}$. Let $\mathcal{F}: (G,\leq) \rightarrow \textup{\textsf{Vect}}_\mathbb{R}$ be an anisotropic sheaf. Then the $0$-th coboundary matrix is
\begin{equation*}
\mathbf{C} = \begin{bmatrix}
-\mathcal{F}_{1, [1,2]} & \mathcal{F}_{2, [1,2]} &  \mathbf{0} \\
-\mathcal{F}_{1, [1,3]} & \mathbf{0} & \mathcal{F}_{3, [1,3]} \\
 \mathbf{0} & -\mathcal{F}_{2, [2,3]} & \mathcal{F}_{3, [2,3]} \\
\end{bmatrix},
\end{equation*}
where each block is a $1 \times d$ matrix. On the other hand, the sheaf Laplacian $L_\mathcal{F}$ is the matrix
\begin{equation*}
\begin{split}
L_\mathcal{F} = \mathbf{C}^T \cdot \mathbf{C}  &= \begin{bmatrix}
-\mathcal{F}_{1, [1,2]}^T & -\mathcal{F}_{1, [1,3]}^T &  \mathbf{0} \\
\mathcal{F}_{2, [1,2]}^T &  \mathbf{0} & -\mathcal{F}_{2, [2,3]}^T \\
 \mathbf{0} & \mathcal{F}_{3, [1,3]}^T & \mathcal{F}_{3, [2,3]}^T \\
\end{bmatrix} \cdot \begin{bmatrix}
-\mathcal{F}_{1, [1,2]} & \mathcal{F}_{2, [1,2]} &  \mathbf{0} \\
-\mathcal{F}_{1, [1,3]} & \mathbf{0} & \mathcal{F}_{3, [1,3]} \\
\mathbf{0} & -\mathcal{F}_{2, [2,3]} & \mathcal{F}_{3, [2,3]} \\
\end{bmatrix} \\
&= \begin{bmatrix}
L_{11} & -\mathcal{F}_{1, [1,2]}^T\mathcal{F}_{2, [1,2]} & -\mathcal{F}_{1, [1,3]}^T\mathcal{F}_{3, [1,3]}\\
-\mathcal{F}_{2, [1,2]}^T \mathcal{F}_{1, [1,2]} & L_{22} & -\mathcal{F}_{2,[2,3]}^T\mathcal{F}_{3,[2,3]} \\
-\mathcal{F}_{3, [1,3]}^T\mathcal{F}_{1, [1,3]}  & -\mathcal{F}_{3, [2,3]}^T\mathcal{F}_{2, [2,3]} & L_{33} \\
\end{bmatrix},
\end{split}    
\end{equation*}
where the diagonal block matrices are 
\begin{equation*}
\begin{split}
L_{11} &= \mathcal{F}_{1, [1,2]}^T \mathcal{F}_{1, [1,2]} + \mathcal{F}_{1, [1,3]}^T \mathcal{F}_{1, [1,3]}, \\    
L_{22} &= \mathcal{F}_{2, [1,2]}^T\mathcal{F}_{2, [1,2]} + \mathcal{F}_{2, [2,3]}^T\mathcal{F}_{2, [2,3]}, \\
L_{33} &= \mathcal{F}_{3, [1,3]}^T\mathcal{F}_{3, [1,3]} + \mathcal{F}_{3, [2,3]}^T\mathcal{F}_{3, [2,3]}.
\end{split}     
\end{equation*}
In particular, the rank of $\mathbf{C}$ is at most 3. Let $\mathbf{C}_1$, $\mathbf{C}_2$, and $\mathbf{C}_3$ denote the first, second, and third rows of $\mathbf{C}$, respectively. Suppose $\lambda_1, \lambda_2, \lambda_3 \in \mathbb{R}$ satisfy $\lambda_1 \mathbf{C}_1 + \lambda_2 \mathbf{C}_2 + \lambda_3 \mathbf{C}_3 = 0$. Then, it follows that $\lambda_1 \mathcal{F}_{1, [1,2]} + \lambda_2 \mathcal{F}_{1, [1,3]} = \mathbf{0}$ and $\lambda_2 \mathcal{F}_{3, [1,3]} + \lambda_3 \mathcal{F}_{3, [2,3]} = \mathbf{0}$. Furthermore, if $\mathcal{F}_{1, [1,2]}$ and $\mathcal{F}_{1, [1,3]}$ are linearly independent and $\mathcal{F}_{3, [2,3]} \neq \mathbf{0}$, then $\lambda_1 = \lambda_2 = \lambda_3 = 0$, implying that ${\rm rank}(\mathbf{C}) = 3$.
\end{exam.}

If \( w_{ij} = 1 \) for every \( i < j \), the anisotropic sheaf can be regarded as a dual structure to the force cosheaf defined on the graph~\cite{cooperband2023towards,cooperband2024equivariant,cooperband2023cosheaf,cooperband2024cellular}. Specifically, the restriction maps of the anisotropic sheaf from vertices to edges correspond to the transpose matrices of the force cosheaf’s maps from edges to vertices. In particular, the force cosheaf can be represented as a functor $\mathcal{H}: (G,\leq)^{\rm op} \rightarrow \textsf{Vect}_{\mathbb{R}}$ from the opposite category of $(G,\leq)$ to $\textsf{Vect}_{\mathbb{R}}$, and the induced boundary matrix $\mathbf{B}: C_1(G,\mathcal{H}) \rightarrow C_0(G,\mathcal{H})$ is exactly the transpose of the boundary matrix $\mathbf{C}: C^0(G,\mathcal{F}) \rightarrow C^1(G,\mathcal{F})$, where $C_0(G,\mathcal{H}) = C^0(G,\mathcal{F})$, $C_1(G,\mathcal{H}) = C^1(G,\mathcal{F})$. In particular,
the $1$-st cosheaf homology $H_1(G,\mathcal{H})$, with its dimension representing the $|S|$, satisfies
\begin{equation}
\begin{split}
\dim H_1(G,\mathcal{H}) &= \dim \ker(\mathbf{B}) = \dim C_1(G,\mathcal{H}) - {\rm rank}(\mathbf{B}) = \dim C_1(G,\mathcal{H}) - {\rm rank}(\mathbf{B}^T) \\ &= \dim C_1(G,\mathcal{H}) - {\rm rank}(\mathbf{C}) = |E| - {\rm rank}(\mathbf{C}) = \dim_\mathbb{R} H^1(G,\mathcal{F}).
\end{split}   
\end{equation}
On the other hand, the dimension of $0$-th force cosheaf homology $H_0(G,\mathcal{H})$ is given by
\begin{equation}
\begin{split}
\dim H_0(G,\mathcal{H}) &= \dim \frac{C_0(G,\mathcal{H})}{{\rm im}(\mathbf{B})} = \dim C_0(G,\mathcal{H}) - \dim {\rm im}(\mathbf{B}) \\
&= \dim C_0(G,\mathcal{H}) - {\rm rank}(\mathbf{B}) = \dim C_0(G,\mathcal{H}) - {\rm rank}(\mathbf{B}^T) \\ 
&= \dim C_0(G,\mathcal{H}) - {\rm rank}(\mathbf{C}) = \dim \ker(\mathbf{C}) = \dim H^0(G,\mathcal{F}).
\end{split}
\end{equation}
In other words, the dimension of the $0$-th force cosheaf equals the dimensional of the global section space of the anisotropic sheaf $\mathcal{F}: (G,\leq) \rightarrow \textsf{Vect}_{\mathbb{R}}$. More precisely, the $3$-dimensional Maxwell's rule states that
\begin{equation}
\label{Eq. Maxwell's rule}
3|V| - |E| = \dim H_0(G,\mathcal{H}) - \dim H_1(G,\mathcal{H}) = 6 + |M| - |S|,
\end{equation}
where $|M| = \dim H_0(G,\mathcal{H}) - 6 = \dim H_0(G,\mathcal{H}) - \binom{4}{2}$ is the number of \textit{linkage mechanisms} and $|S| = \dim H_1(G,\mathcal{F})$ is the number of \textit{self-stresses} $|S|$ of the truss mechanism. For further details and a more general version of the $n$-dimensional Maxwell's rule, refer to~\cite{cooperband2023towards,cooperband2024equivariant,cooperband2023cosheaf,cooperband2024cellular}.

\paragraph{Ranks of anisotropic sheaves}
At the end of the section, the concept of the \textit{rank} of an anisotropic sheaf on a graph is introduced, which is crucial for determining the dimension of the global section space of the sheaf. In general, for a cellular sheaf $\mathcal{F}$ on a graph $G = (V, E)$, the rank of $\mathcal{F}$ reflects the linear independence of the linear transformations $\mathcal{F}_{v, e}$ considered as vectors in $\mathbb{R}^3$. As illustrated in Example \ref{Example: anisotropic sheaf defined on K3}, the matrix $\mathbf{C}$ achieves a rank of $3$ when $\mathcal{F}_{1, [1,2]}$ and $\mathcal{F}_{1, [1,3]}$ are linearly independent and $\mathcal{F}_{3, [2,3]} \neq \mathbf{0}$. In other words, counting the number of linearly independent vectors $\mathcal{F}_{v, e} \in \mathbb{R}^3$ is useful for approximating the rank of the coboundary matrix and, consequently, for estimating the dimension of the global section space. The formal definition of the rank of an anisotropic sheaf is provided below.
\begin{def.}
Let $G = (V,E)$ be an undirected, simple, and finite graph, and let $\mathcal{F}: (G,\leq) \rightarrow \textup{\textsf{Vect}}_\mathbb{R}$ be an anisotropic sheaf. The \textbf{rank} of $\mathcal{F}$ is defined as follows.
\begin{equation*}
{\rm rank}(\mathcal{F}) = \dim_\mathbb{R} \bigg( {\rm span}_\mathbb{R} \bigg\{ \mathcal{F}_{v_i, [v_i,v_j]} \ \bigg| \ i < j \bigg\}  \bigg).    
\end{equation*}
An anisotropic sheaf $\mathcal{F}: (G,\leq) \rightarrow \textup{\textsf{Vect}}_\mathbb{R}$ is said to have \textbf{full rank} if ${\rm rank}(\mathcal{F}) = 3$.
\end{def.}
It is evident that ${\rm rank}(\mathcal{F}) \leq 3$, since all vectors $\mathcal{F}_{v, e}$ lie in the $3$-dimensional Euclidean space. Excluding cases where an anisotropic sheaf $\mathcal{F}: (G, \leq) \rightarrow \textup{\textsf{Vect}}_\mathbb{R}$ has $w_{ij} = 0$ for some $i < j$, the following proposition simplifies the definition of the rank of an anisotropic sheaf to facilitate a more efficient analysis.
\begin{prop.}
Let $G = (V, E)$ be a graph with $n$ vertices, each associated with 3D coordinates $(x_i^\circ, y_i^\circ, z_i^\circ)$. The vertices are denoted by $v_1, \ldots, v_n \in V$, and the edges by $[v_i, v_j]$ with $i < j$. Let $\mathcal{F}: (G, \leq) \rightarrow \textup{\textsf{Vect}}_\mathbb{R}$ be an anisotropic sheaf defined as in~\eqref{Eq. Anisotropic sheaf definition}. If $w_{ij} \neq 0$ for every $i < j$, and $[v_1, v_j]$ for all $j \in \{ 2, ..., n\}$, then
\begin{equation*}
{\rm rank}(\mathcal{F}) = \dim_\mathbb{R} \bigg( {\rm span}_\mathbb{R} \bigg\{ \mathcal{F}_{v_1, [v_1,v_j]} \ \bigg| \ j \in \{ 2, ..., n \} \bigg\}  \bigg),    
\end{equation*}
which admits a basis for the space spanned by the vectors $\mathcal{F}_{v_i, [v_i,v_j]}$ with $i < j$. In particular, if $G = K_n$ is a complete graph with $n$ vertices and $k$ is a fixed number in $\{ 1, 2, ..., n \}$, then the set $\{ \mathcal{F}_{v_k, [v_k,v_l]} \ | \ l \in \{ 1, 2, ..., n \} \setminus \{ k \} \}$ spans all the vectors $\mathcal{F}_{v_i, [v_i,v_j]}$ with $i < j$. 
\end{prop.}
\begin{proof}
For every $i < j$ in $\{ 1, 2, ..., n \}$, the weight $w_{ij}$ is non-zero. Then,
\begin{equation}
\label{Eq. ij vector spanned by 1j and 1i}
\begin{split}
\frac{1}{w_{ij}} \cdot \mathcal{F}_{v_i, [v_i,v_j]} &= \begin{bmatrix}
x_j^\circ-x_i^\circ & y_j^\circ-y_i^\circ & z_j^\circ-z_i^\circ \\
\end{bmatrix} = \begin{bmatrix}
x_j^\circ & y_j^\circ & z_j^\circ \\
\end{bmatrix} - \begin{bmatrix}
x_i^\circ & y_i^\circ & z_i^\circ \\
\end{bmatrix} \\
&= \begin{bmatrix}
x_j^\circ - x_1^\circ & y_j^\circ - y_1^\circ & z_j^\circ - z_1^\circ \\
\end{bmatrix} - \begin{bmatrix}
x_i^\circ - x_1^\circ & y_i^\circ - y_1^\circ & z_i^\circ - z_1^\circ \\
\end{bmatrix} \\ &= \frac{1}{w_{1j}} \cdot \mathcal{F}_{v_1, [v_1,v_j]} - \frac{1}{w_{1i}} \cdot \mathcal{F}_{v_1, [v_1,v_i]} \in {\rm span}_\mathbb{R} \{ \mathcal{F}_{v_1, [v_1,v_j]}, \mathcal{F}_{v_1, [v_1,v_i]} \}
\end{split}    
\end{equation}
if $[v_1, v_i]$ and $[v_1, v_j]$ also belong to the edge set $E$ of $G$. This shows that $\{ \mathcal{F}_{v_1, [v_1,v_2]}, ...,  \mathcal{F}_{v_1, [v_1,v_{n}]} \}$ spans all the vectors $\mathcal{F}_{v_i, [v_i,v_j]}$ with $i < j$, as desired. Because $\{ \mathcal{F}_{v_1, [v_1,v_j]} \ | \ j \in \{ 2, ..., n \}\}$ is a subset of $\{ \mathcal{F}_{v_i, [v_i,v_j]} \ | \ i < j \}$, the proposition follows.
\end{proof}
{\color{red}

}

\section{Normal Mode Analysis in Anisotropic Sheaf Models}
\label{Section: Normal Mode Analysis in Anisotropic Sheaf Models}
In this section, we explore the physical significance of the global sections and the global section space of the anisotropic sheaf defined on an atomic system. The section is organized into two main parts. 

First, we establish the connection between the global sections of the anisotropic sheaf and the normal modes of the atomic system. Specifically, as demonstrated in Theorem \ref{Theorem: Main result 2}, each global section corresponds to a normal mode involving global translation or rotation in the anisotropic network model associated with the graph. 

Second, we examine the dimension of the global section space of a given anisotropic sheaf. By analyzing the anisotropic sheaf on a complete graph under specific geometric conditions, such as the coordinates being in general position, we provide a mathematical proof for the existence of a graph model of an anisotropic sheaf with a global section space of dimension exactly 6—generally considered an ideal condition in ANM analysis (cf. Theorems \ref{Theorem: Main result 3-} and \ref{Theorem: Main result 3}).
\paragraph{Global Section Space and Normal Modes}
\begin{figure}
	\centering
  \includegraphics[width=\linewidth]{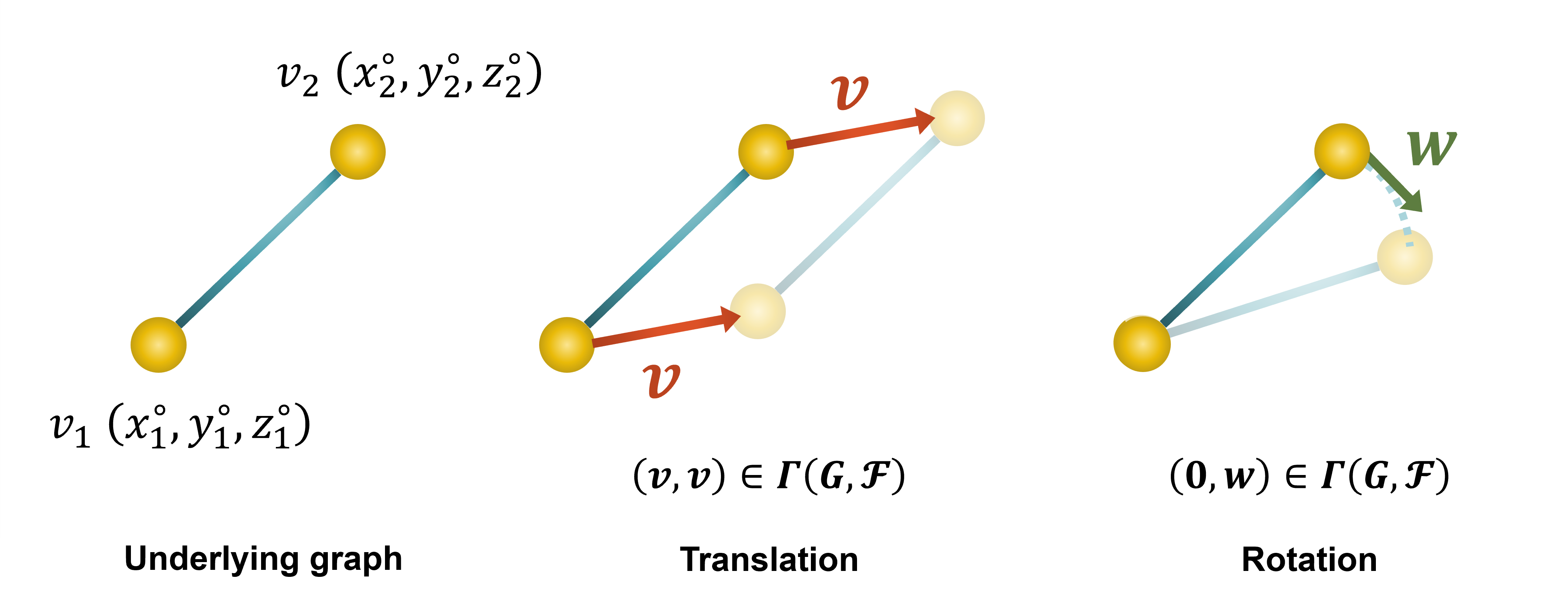}
  \caption{An illustration of two global sections of an anisotropic sheaf defined on a graph $G$, consisting of two vertices $v_1$ and $v_2$ with coordinates $(x_1^\circ, y_1^\circ, z_1^\circ)$ and $(x_2^\circ, y_2^\circ, z_2^\circ)$, connected by a single edge. According to the definition of anisotropic sheaves, the stalk spaces are $\mathcal{F}_{v_1} = \mathbb{R}^3$, $\mathcal{F}_{v_2} = \mathbb{R}^3$, $\mathcal{F}_{[v_1,v_2]} = \mathbb{R}$. The restriction maps $\mathcal{F}_{v_1, [v_1, v_2]}$ and $ = \mathcal{F}_{v_2, [v_1, v_2]}$ are defined as the $1 \times 3$ matrix $\begin{bmatrix}
x_2^\circ-x_1^\circ & y_2^\circ-y_1^\circ & z_2^\circ-z_1^\circ \\
\end{bmatrix}$, which sends each vector $\mathbf{x} \in \mathbb{R}^3$ to the inner product value $\langle \mathbf{u}, \mathbf{x} \rangle$, where $\mathbf{u} = (x_2^\circ-x_1^\circ, y_2^\circ-y_1^\circ, z_2^\circ-z_1^\circ)$. Then, for any $\mathbf{v} \in \mathbb{R}^3$, $(\mathbf{v}, \mathbf{v})$ is a global section since $\mathcal{F}_{v_1,[v_1,v_2]}(\mathbf{v}) =  \mathcal{F}_{v_2,[v_1,v_2]}(\mathbf{v})$. On the other hand, if $\mathbf{w} \in \mathbb{R}^3$ is perpendicular to $\mathbf{u}$, then $(\mathbf{0}, \mathbf{w})$ with $\mathbf{0}$ is a global section since $\mathcal{F}_{v_2, [v_1,v_2]}(\mathbf{w}) = \langle \mathbf{u}, \mathbf{w} \rangle = 0 = \langle \mathbf{u}, \mathbf{0} \rangle = \mathcal{F}_{v_1, [v_1,v_2]}(\mathbf{0})$.}
\label{fig: demo example of global sections}
\end{figure}

Global sections of an anisotropic sheaf have direct geometric meanings in normal mode analysis within molecular dynamics. Starting with a simple example of a graph consisting of two vertices and one edge with 3D coordinate information, Figure \ref{fig: demo example of global sections} illustrates the geometric visualization of two types of global sections for any anisotropic sheaf defined on this graph. In this example, the tuples $(\mathbf{v}, \mathbf{v})$ and $(\mathbf{0}, \mathbf{w})$, where $\mathbf{v}, \mathbf{w}, \mathbf{0} = (0,0,0) \in \mathbb{R}^3$, are global sections since $\mathcal{F}_{v_1,[v_1,v_2]}(\mathbf{v}) = \mathcal{F}_{v_2,[v_1,v_2]}(\mathbf{v})$ and $\mathcal{F}_{v_2,[v_1,v_2]}(\mathbf{w}) = \mathcal{F}_{v_1,[v_1,v_2]}(\mathbf{0})$. Specifically, $(\mathbf{v}, \mathbf{v})$ corresponds to a translation operation where both vertices move consistently along the 3D direction of $\mathbf{v}$, resulting in the translation of the entire graph. In contrast, $\mathbf{w}$ is set perpendicular to the vector $(x_2^\circ - x_1^\circ, y_2^\circ - y_1^\circ, z_2^\circ - z_1^\circ)$, implying that $(\mathbf{0}, \mathbf{w})$ is a global section. This corresponds to a movement where vertex $v_1$ remains fixed while vertex $v_2$ rotates, with $\mathbf{w}$ as the velocity of the circular motion centered at $v_1$ with radius $s_{12}^\circ$. 

\begin{figure}
	\centering
  \includegraphics[width=\linewidth]{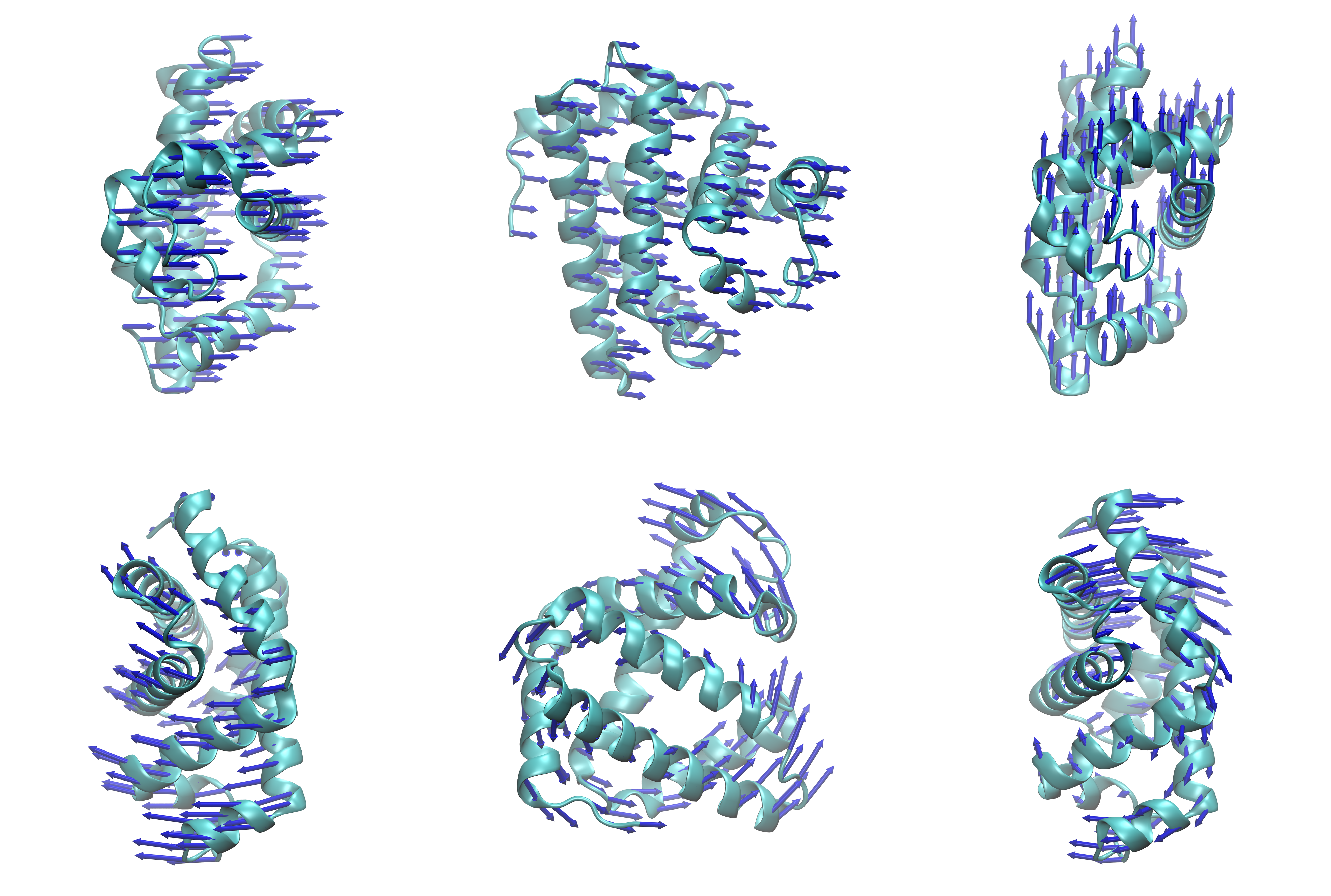}
  \caption{Illustration of six linearly independent eigenmodes with zero eigenvalues for the protein with ID 103M from the RCSB Protein Data Bank~\cite{smith1999correlations, Berman:2000}. Three of the eigenmodes correspond to translations of the entire protein in three directions, while the other three correspond to rotations of the entire protein around three axes. The visualizations of the eigenmodes were generated using the VMD software~\cite{VMD}.}
\label{fig: Six non-zero eigenmodes}
\end{figure}

The following theorem shows that the dimension of the global section space of any anisotropic sheaf must be at least $6$. This space has a linearly independent set consisting of three global sections corresponding to translations in three directions, and three more corresponding to rotations around the three principal axes—operations that act on the entire structure as a whole (Figure \ref{fig: Six non-zero eigenmodes}). This phenomenon is summarized in the following theorem.

\begin{theorem}\label{Theorem: Main result 2}
Let $G = (V,E)$ be an undirected, simple, and finite graph, where each vertex is annotated with a $3$-dimensional coordinate $(x_i^\circ, y_i^\circ, z_i^\circ)$. Let $\mathcal{F}: (G,\leq) \rightarrow \textup{\textsf{Vect}}_\mathbb{R}$ be an anisotropic sheaf. If there are three affinely independent coordinates, then $\dim_\mathbb{R} \ker(L_{\calF}) \geq 6$. In particular, this space contains a linearly independent set $S$ with $|S| = 6$, comprising three global sections corresponding to translations in three directions and three more corresponding to rotations around the three principal axes.
\end{theorem}
\begin{proof}
Suppose $V = \{ v_1, v_2, ..., v_n \}$ consists of $n$ vertices. Then, the $0$-th cochain space $C^0(G,\mathcal{F})$ of $\mathcal{F}$ is defined as the direct sum $\mathbb{R}^3 \oplus \mathbb{R}^3 \oplus \cdots \oplus \mathbb{R}^3 \simeq \mathbb{R}^{3n}$ of $n$ $\mathbb{R}^3$ spaces. Elements in $C^0(G,\mathcal{F})$ are $n$-tuples $\alpha = (\mathbf{x}_1, \mathbf{x}_2, ..., \mathbf{x}_n)$ of vectors in $\mathbb{R}^3$, recording local sections on vertices $v_1, v_2, ..., v_n$. Let $\{ \mathbf{e}_1, \mathbf{e}_2, \mathbf{e}_3 \}$ be the standard basis for $\mathbb{R}^3$, and let $\alpha_k = (\mathbf{e}_k, \mathbf{e}_k, ..., \mathbf{e}_k)$ be the $n$-tuple consisting of $n$ $\mathbf{e}_k$ vectors. Then, $\{ \alpha_1, \alpha_2, \alpha_3 \}$ is a linearly independent subset of $C^0(G,\mathcal{F})$. Therefore, it is sufficient to show that, in addition to the global sections corresponding to translations, there are three linearly independent global sections arising from rotational operations. To simplify the proof, we focus on rotations around the $x$-, $y$-, and $z$-axes and claim that these rotational operations correspond to global sections of any anisotropic sheaf. Let $R_z: \mathbb{R}^3 \rightarrow \mathbb{R}^3$ denote the rotation around the $z$-axis.  Let $(x_i^\circ, y_i^\circ, z_i^\circ)$ be the coordinates of the $i$-th vertex of $G$.  Then the velocity vector due to the rotation $R_z$ at this vertex is $(-y_i^\circ, x_i^\circ, 0)$. Therefore, 
\begin{equation*}
\begin{split}
\mathcal{F}_{v_i,[v_i,v_j]}(-y_i^\circ, x_i^\circ, 0) &= \langle (x_j^\circ - x_i^\circ, y_j^\circ - y_i^\circ, z_j^\circ - z_i^\circ), (-y_i^\circ, x_i^\circ, 0) \rangle \\
&= -x_j^\circ y_i^\circ + x_i^\circ y_i^\circ + x_i^\circ y_j^\circ - x_i^\circ y_i^\circ = -x_j^\circ y_i^\circ + x_i^\circ y_j^\circ, \\
\mathcal{F}_{v_i,[v_i,v_j]}(-y_j^\circ, x_j^\circ, 0) &= \langle (x_j^\circ - x_i^\circ, y_j^\circ - y_i^\circ, z_j^\circ - z_i^\circ), (-y_j^\circ, x_j^\circ, 0) \rangle \\
&= -x_j^\circ y_j^\circ + x_i^\circ y_j^\circ + x_j^\circ y_j^\circ - x_j^\circ y_i^\circ = x_i^\circ y_j^\circ - x_j^\circ y_i^\circ.
\end{split}    
\end{equation*}
for every $v_i, v_j \in V$.  By setting $\alpha_z = (\bfx_i)_{i = 1}^n \in C^0(G,\mathcal{F}) = \bigoplus_{i = 1}^n \mathcal{F}_{v_i}$ with $\bfx_i = (-y_i^\circ, x_i^\circ, 0) \in \mathcal{F}_{v_i} = \mathbb{R}^3$ for each $i$, we deduce that $\alpha_z \in \Gamma(G,\mathcal{F})$.  By symmetric arguments, the associated $\alpha_x$ and $\alpha_y$ are also global sections of $\mathcal{F}$. 

Second, we claim that if the point cloud $\{ (x_i^\circ, y_i^\circ, z_i^\circ) \}_{i = 1}^n$ contains three affinely independent coordinates, then the set $\{ \alpha_1, \alpha_2, \alpha_3, \alpha_x, \alpha_y, \alpha_z \}$ is linearly independent. Without loss of generality, we assume that $\{ (x_i^\circ, y_i^\circ, z_i^\circ) \}_{i = 1}^3$ is affinely independent in $\mathbb{R}^3$, and it is sufficient to prove that the matrix
\begin{equation*} 
\begin{bmatrix}
1 & 0 & 0 & 0 & z_1^\circ & y_1^\circ \\
0 & 1 & 0 & z_1^\circ & 0 & -x_1^\circ \\
0 & 0 & 1 & -y_1^\circ & -x_1^\circ & 0 \\
1 & 0 & 0 & 0 & z_2^\circ & y_2^\circ \\
0 & 1 & 0 & z_2^\circ & 0 & -x_2^\circ \\
0 & 0 & 1 & -y_2^\circ & -x_2^\circ & 0 \\
1 & 0 & 0 & 0 & z_3^\circ & y_3^\circ \\
0 & 1 & 0 & z_3^\circ & 0 & -x_3^\circ \\
0 & 0 & 1 & -y_3^\circ & -x_3^\circ & 0 \\
\end{bmatrix} \sim
\begin{bmatrix}
1 & 0 & 0 & 0 & 0 & 0 \\
0 & 1 & 0 & 0 & 0 & 0 \\
0 & 0 & 1 & 0 & 0 & 0 \\
1 & 0 & 0 & 0 & z_2^\circ - z_1^\circ & y_2^\circ - y_1^\circ \\
0 & 1 & 0 & z_2^\circ - z_1^\circ & 0 & x_1^\circ - x_2^\circ \\
0 & 0 & 1 & y_1^\circ - y_2^\circ & x_1^\circ - x_2^\circ & 0 \\
1 & 0 & 0 & 0 & z_3^\circ - z_1^\circ & y_3^\circ - y_1^\circ \\
0 & 1 & 0 & z_3^\circ - z_1^\circ & 0 & x_1^\circ - x_3^\circ \\
0 & 0 & 1 & y_1^\circ - y_3^\circ & x_1^\circ - x_3^\circ & 0 \\
\end{bmatrix}
\end{equation*}
has rank $6$, where the second matrix is obtained through column elimination. Since the triangle formed by the three coordinates is non-degenerate, the submatrix consisting of the last three columns of the right-hand-side matrix has a rank of $3$. Consequently, the whole matrix has a rank of $6$, as desired.
\end{proof}
\begin{remark}
Using cosheaf language, Theorem \ref{Theorem: Main result 2} can be interpreted that for any non-degenerate $1$-simplicial complex in \( \mathbb{R}^3 \), the cosheaf homology space, representing the linkage kinematic degrees of freedom, has dimension at least \( \binom{4}{2} \)~\cite{cooperband2024cellular}. Dually, within the framework of the proposed anisotropic sheaf model, we explicitly construct the global sections corresponding to the $3$-dimensional translations and rotations, which are directly connected to the geometric significance of atomic local movements.
\end{remark}

Theorems \ref{Theorem: Main result 2} and \ref{Theorem: Main result 1} together offer a mathematical interpretation of the geometric significance of the eigenmodes associated with a zero eigenvalue by analyzing the global sections of the proposed anisotropic sheaf. Specifically, Theorem \ref{Theorem: Main result 1} establishes that each eigenmode of the anisotropic Hessian matrix $\mathbf{H}_{\rm ANM}$ with a zero eigenvalue corresponds to a global section of the sheaf. Importantly, by Theorem \ref{Theorem: Main result 2}, this correspondence holds independently of the cutoff distance $R_c$ used in network construction for Equation~\eqref{Eq. ANM energy function}, ensuring that all underlying anisotropic networks consistently exhibit the same six eigenmodes—three corresponding to translations and three to rotations. In particular, as demonstrated in Theorem \ref{Theorem: Main result 2}, the Hessian matrix $\mathbf{H}_{\rm ANM}$ associated with any atomic system must exhibit more than six trivial eigenvalues if the system contains at least three affinely independent coordinates.

Next, we present two propositions as preparation for a further analysis of global section spaces. The first proposition provides a formula relating the dimension of the global section space of an anisotropic sheaf to the rank of the 0-coboundary map. This allows one to compute the dimension of $H^0(G, \mathcal{F})$ by calculating the rank of the coboundary matrix.
\begin{prop.}
\label{Proposition: Dimension of the global section space of an anisotropic sheaf}
Let $G = (V,E)$ be an undirected, simple, and finite graph. Let $\mathcal{F}: (G,\leq) \rightarrow \textup{\textsf{Vect}}_\mathbb{R}$ be an anisotropic sheaf. Let $\mathbf{C}: C^0(G,\calF) \rightarrow C^1(G,\calF)$ be the $0$-th coboundary matrix. Then
\begin{equation*}
\dim_\mathbb{R} \ker(L_\calF) = \dim_\mathbb{R} H^0(G,\calF) = 3 \cdot |V| - {\rm rank}(\mathbf{C}).    
\end{equation*}
\end{prop.}
\begin{proof}
As presented in \eqref{Eq. Coboundary matrix}, the coboundary map $\mathbf{C}$ is an $m \times dn$, where $n = |V|$ and $m = |E|$ are the numbers of vertices and edges of $G$, respectively. In particular, it is a linear transformation from $\mathbb{R}^{3n} \simeq C^0(G,\calF)$ to $\mathbb{R}^m \simeq C^1(G,\calF)$. By Theorem \ref{Theorem: Hodge's theorem} and the dimension theorem, we have $\dim_\mathbb{R} \ker(L_\calF) = \dim_\mathbb{R} H^0(G,\calF) = \dim_\mathbb{R} \ker(\mathbf{C}) = \dim_\mathbb{R} C^0(G,\calF) - {\rm rank}(\mathbf{C}) = 3 \cdot |V| - {\rm rank}(\mathbf{C})$, as desired.    
\end{proof}
The second proposition plays a crucial role in investigating the dimensions of the global section spaces of an anisotropic sheaf and its restriction sheaves. This proposition will be highly valuable in exploring the kernel of the Laplacian matrix of anisotropic sheaves.
\begin{prop.}
\label{Proposition: Contravariant relation of global section spaces}
Let $V$ be a finite set and $G_1 = (V,E_1), G_2 = (V,E_2)$ be undirected simple graphs defined on the same vertex set. Suppose $E_1 \subseteq E_2$ and $\calF_1, \calF_2$ are anisotropic sheaves defined on $G_1$ and $G_2$, respectively. If $\mathcal{F}_1 = \mathcal{F}_2|_{G_1}$, then $\dim_\mathbb{R} \ker(L_{\calF_2}) \leq \dim_\mathbb{R} \ker(L_{\calF_1})$.   
\end{prop.}
\begin{proof}
Suppose $V = \{ v_1, ..., v_n \}$ and $E_1 = \{ e_1, ..., e_m \}$ consist of $n$ vertices and $m$ edges, respectively. Since $E_1 \subseteq E_2$, $E_2 = \{ e_1, ..., e_m, e_{m+1}, ..., e_{m+d} \}$, where $e_{m+1}, ..., e_{m+d} \in E_2 \setminus E_1$ are the additional edges of $G_2$. Then, the $0$-th coboundary matrix $\mathbf{C}_2: C^0(G_2,\calF_2) \rightarrow C^1(G_2,\calF_2)$ can be represented by
\begin{equation*}
\mathbf{C}_2 = \begin{bmatrix}
[v_1:e_1] \cdot (\mathcal{F}_1)_{v_1, e_1} & [v_2:e_1] \cdot (\mathcal{F}_1)_{v_2, e_1} & \cdots & [v_n:e_1] \cdot (\mathcal{F}_1)_{v_n, e_1}\\
[v_1:e_2] \cdot (\mathcal{F}_1)_{v_1, e_2} & [v_2:e_2] \cdot (\mathcal{F}_1)_{v_2, e_2} & \cdots & [v_n:e_2] \cdot (\mathcal{F}_1)_{v_n, e_2}\\
\vdots & \vdots & \ddots & \vdots \\
[v_1:e_m] \cdot (\mathcal{F}_1)_{v_1, e_m} & [v_2:e_m] \cdot (\mathcal{F}_1)_{v_2, e_m} & \cdots & [v_n:e_m] \cdot (\mathcal{F}_1)_{v_n, e_m}\\
[v_1:e_{m+1}] \cdot (\mathcal{F}_2)_{v_1, e_{m+1}} & [v_2:e_{m+1}] \cdot (\mathcal{F}_2)_{v_2, e_{m+1}} & \cdots & [v_n:e_{m+1}] \cdot (\mathcal{F}_2)_{v_n, e_{m+1}}\\
[v_1:e_{m+2}] \cdot (\mathcal{F}_2)_{v_1, e_{m+2}} & [v_2:e_{m+2}] \cdot (\mathcal{F}_2)_{v_2, e_{m+2}} & \cdots & [v_n:e_{m+2}] \cdot (\mathcal{F}_2)_{v_n, e_{m+2}}\\
\vdots & \vdots & \ddots & \vdots \\
[v_1:e_{m+d}] \cdot (\mathcal{F}_2)_{v_1, e_{m+d}} & [v_2:e_{m+d}] \cdot (\mathcal{F}_2)_{v_2, e_{m+d}} & \cdots & [v_n:e_{m+d}] \cdot (\mathcal{F}_2)_{v_n, e_{m+d}}\\
\end{bmatrix},
\end{equation*}
where the submatrix of the first $m$ rows of $\mathbf{C}_2$ is the $0$-th coboundary matrix of the sheaf $\mathcal{F}_1: (G_1,\leq) \rightarrow \textup{\textsf{Vect}}_\mathbb{R}$, denoted by $\mathbf{C}_1$. In particular, ${\rm rank}(\mathbf{C}_1) \leq {\rm rank}(\mathbf{C}_2)$. By Proposition \ref{Proposition: Dimension of the global section space of an anisotropic sheaf}, $\dim_\mathbb{R} \ker(L_{\calF_2}) \leq \dim_\mathbb{R} \ker(L_{\calF_1})$.
\end{proof}


\paragraph{Dimension Analysis of Global Section Spaces}
In the second part of this section, we examine the dimension of the global section space of the proposed anisotropic sheaf model. Generally, by selecting an appropriate cutoff distance $R_c$ (see Equation~\eqref{Eq. ANM energy function}), the associated ANM Hessian matrix typically exhibits six zero eigenvalues, corresponding to the trivial eigenmodes of translation and rotation for the entire biomolecular complex (Figure \ref{fig: Six non-zero eigenmodes}). However, if $R_c$ is not sufficiently large, the number of zero eigenvalue eigenmodes may significantly exceed six, complicating the ANM analysis. 

To analyze the dimension of the global section space of an anisotropic sheaf $\mathcal{F}: (G, \leq) \rightarrow \textup{\textsf{Vect}}_\mathbb{R}$ with non-zero weight values $w_{ij}$ (refer to Equation \eqref{Eq. Anisotropic sheaf definition}), we focus on the case where the anisotropic sheaf is of the form: 
\begin{equation}
\label{Eq. Anisotropic sheaf definition-form 2}
\mathcal{F}_{v_i, [v_i,v_j]} = \begin{bmatrix}
x_j^\circ-x_i^\circ & y_j^\circ-y_i^\circ & z_j^\circ-z_i^\circ \\
\end{bmatrix}
= \mathcal{F}_{v_j, [v_i,v_j]}  
\end{equation}
That is, we set $w_{ij} := 1$ for every edge $[e_i,e_j] \in E$. In particular, since each row of the induced coboundary matrix, $\mathbf{C}$ only involves two $1 \times 3$ matrices $\mathcal{F}_{e_i,[e_i,e_j]}$ and $\mathcal{F}_{e_j,[e_i,e_j]}$, dividing the scalar $w_{ij}$ to this row doesn't change the rank of the matrix. That is, to investigate the dimension of the global section space of an anisotropic sheaf with non-zero edge weights, it is sufficient to study the anisotropic sheaf of the form in Equation~\eqref{Eq. Anisotropic sheaf definition-form 2}. In particular, the linear relation depicted in \eqref{Eq. ij vector spanned by 1j and 1i} becomes
\begin{equation}
\label{Eq. ij vector spanned by 1j and 1i-simple}
\begin{split}
\mathcal{F}_{v_i, [v_i,v_j]} &= \mathcal{F}_{v_1, [v_1,v_j]} - \mathcal{F}_{v_1, [v_1,v_i]} \in {\rm span}_\mathbb{R} \{ \mathcal{F}_{v_1, [v_1,v_j]}, \mathcal{F}_{v_1, [v_1,v_i]} \}.
\end{split}    
\end{equation}
Evidently, by verifying the local consistency of the local sections on the stalk spaces of vertices in the underlying graph, sheaves defined as in Equations~\eqref{Eq. Anisotropic sheaf definition} and \eqref{Eq. Anisotropic sheaf definition-form 2} share the same global section space. In particular, the associated sheaf Laplacians have identical nullity.
\begin{theorem}\label{Theorem: weighted and non-weighted anisotropic sheaves}
Let $G = (V, E)$ be an undirected, simple, and finite graph. Define an anisotropic sheaf $\mathcal{F}: (G, \leq) \rightarrow \textup{\textsf{Vect}}_\mathbb{R}$ as in Equation~\eqref{Eq. Anisotropic sheaf definition}, with non-zero weights $w_{ij}$. For every edge $[v_i, v_j] \in E$ with endpoints $v_i, v_j \in V$, the sheaf map is given by:
\begin{equation*}
\mathcal{F}_{v_i, [v_i, v_j]} = w_{ij} \cdot 
\begin{bmatrix}
x_j^\circ - x_i^\circ & y_j^\circ - y_i^\circ & z_j^\circ - z_i^\circ \\
\end{bmatrix}
= \mathcal{F}_{v_j, [v_i, v_j]}.
\end{equation*}
On the other hand, let $\mathcal{G}: (G, \leq) \rightarrow \textup{\textsf{Vect}}_\mathbb{R}$ be the anisotropic sheaf defined as in Equation~\eqref{Eq. Anisotropic sheaf definition-form 2}. Then, $\mathcal{F}$ and $\mathcal{G}$ have the same global section space, i.e., $H^0(G, \mathcal{F}) = H^0(G, \mathcal{G})$.
\end{theorem}
\begin{proof}
Let $v_i$ and $v_j$ be two vertices in $G$, forming an edge $[v_i, v_j] \in E$. Let $\mathbf{x}_{v_i} \in \mathcal{F}_{v_i} = \mathbb{R}^3 = \mathcal{G}_{v_i}$ and $\mathbf{x}_{v_j} \in \mathcal{F}_{v_j} = \mathbb{R}^3 = \mathcal{G}_{v_j}$ be local sections on the stalk spaces on $v_i$ and $v_j$, respectively. Because $w_{ij} \neq 0$, equation $\mathcal{F}_{v_i,[v_i,v_j]}(\mathbf{x}_{v_i}) = \mathcal{F}_{v_j,[v_i,v_j]}(\mathbf{x}_{v_j})$ if and only if $\mathcal{G}_{v_i,[v_i,v_j]}(\mathbf{x}_{v_i}) = \mathcal{G}_{v_j,[v_i,v_j]}(\mathbf{x}_{v_j})$. Thus, $\mathcal{F}$ and $\mathcal{G}$ have the same global section space, i.e., $H^0(G, \mathcal{F}) = H^0(G, \mathcal{G})$. Furthermore, since the global sections determine the null space of the associated sheaf Laplacian, both sheaves also share the same nullity. This completes the proof.      
\end{proof}
\begin{remark}
In applications, the weight values $w_{ij}$ for the edges are set as $\gamma^{1/2}/s_{ij}^{\circ}$ (refer to Equation \eqref{Eq. small Hessian matrix} and Theorem \ref{Theorem: Main result 1}). Since the spring constant $\gamma$  is generally non-zero, anisotropic sheaves with non-zero weight values align with the assumptions used in ANM analysis.
\end{remark}

Using the anisotropic sheaf model, the second main result of this section provides a sheaf-based perspective for proving the existence of underlying graphs that model a given atomic system, ensuring that the number of trivial eigenmodes is exactly six. Specifically, as demonstrated in Theorem \ref{Theorem: Main result 3-}, for an ANM Hessian matrix constructed on the complete graph of atoms in general position, the number of trivial modes is precisely six. 

Moreover, by Proposition \ref{Proposition: Contravariant relation of global section spaces}, adding additional edges to the underlying graph of the atomic system reduces the number of trivial modes. Consequently, it follows from Proposition \ref{Proposition: Contravariant relation of global section spaces} that there exists a graph with the minimal number of edges that ensures $\dim_{\mathbb{R}} \ker(\mathbf{H}_{\rm ANM}) = 6$. This result is formalized in the following theorem.


\begin{theorem}\label{Theorem: Main result 3}
Let $V \subseteq \mathbb{R}^3$ be a finite point cloud in general position, where the cardinality of $|V|$ is larger than $3$. Then there is a graph $G = (V, E)$ with edges of endpoints in $V$ that satisfies the following two properties:
\begin{itemize}
\item[\rm (a)] For the anisotropic sheaf $\mathcal{F}: (G, \leq) \rightarrow \textup{\textsf{Vect}}_\mathbb{R}$ defined as in Equation~\eqref{Eq. Anisotropic sheaf definition-form 2}, $\dim_{\mathbb{R}} H^0(G, \mathcal{F}) = 6$.
\item[\rm (b)] Let $G' = (V,E')$ be the graph obtained by removing an arbitrary edge in $E$, and let $\mathcal{G}: (G, \leq) \rightarrow \textup{\textsf{Vect}}_\mathbb{R}$ be the anisotropic sheaf defined as in Equation~\eqref{Eq. Anisotropic sheaf definition-form 2}, then $\dim_{\mathbb{R}} H^0(G, \mathcal{G}) > 6$.
\end{itemize}
\end{theorem}

For a point cloud $V \subseteq \mathbb{R}^3$ in general position, a graph $G = (V,E)$ with edges of endpoints in $V$ that satisfies properties (a) and (b) is referred to a \textit{minimal graph} for desired nullity of the induced Hessian matrix.

\section{Graph Construction for Anisotropic Sheaves}
\label{Section: Graph Construction for Anisotropic Sheaves}
In practical applications, modeling a network from a point cloud with a complete graph is often inefficient due to the excessive number of edges involved. In the following discussion, we propose an alternative approach by constructing an anisotropic sheaf $\mathcal{F}$ on a graph $G$ using $3$-dimensional homogeneous simplicial complex to represent the point cloud data in general position, such as the ${\rm C}_\alpha$ point cloud extracted from a protein structure. In particular, by employing 3D Delaunay triangulation to connect the data points~\cite{delaunay1934bulletin}, the resulting anisotropic sheaf consistently maintains the global section space at a dimension of $6$. Compared to complete graphs, this method offers a more streamlined network model that ensures the constructed anisotropic sheaf achieves the minimal dimension of the global section space.
\paragraph{Homogeneous simplicial $3$-complex construction}
In this paper, we construct the underlying graph of a given point cloud in $\mathbb{R}^3$ by considering simplicial complexes composed of $3$-dimensional tetrahedrons. Specifically, we focus on \textit{homogeneous} (or \textit{pure}) simplicial $3$-complexes in $\mathbb{R}^3$, built upon the point cloud to obtain the underlying graph. These complexes play a significant role in computer graphics, particularly in generating network structures from discrete objects embedded in 2D or 3D Euclidean spaces~\cite{thompson1998handbook}. We briefly recall the definition of homogeneous simplicial $3$-complexes below.
\begin{def.}
\label{Definition: Homogeneous simplicial 3-complex}
A \textbf{pure} or \textbf{homogeneous} simplicial $3$-complex $K$ in the $3$-dimensional Euclidean space $\mathbb{R}^3$ is a simplicial complex consisting of geometric simplices in $\mathbb{R}^3$, where every simplex of dimension less than $3$ serves as a face of at least one simplex $\sigma$ in $K$ of dimension exactly $3$. For every $2$-simplex $\tau$ in $K$, the \textbf{degree of upper adjacency}, denoted by $\deg_{K,u}(\tau)$, is defined as the number of $3$-simplices $\sigma$ in $K$ such that $\tau \leq \sigma$.  
\end{def.}
In other words, a homogeneous simplicial $3$-complex $K$ comprises a collection of non-overlapping $3$-simplices, with the meaning of ``non-overlapping'' that two $3$-simplices intersect only at a shared $q$-face with $0 \leq q < 3$. Notably, to explore the face intersection behaviors between two $3$-simplices within a homogeneous simplicial $3$-complex, we utilize the notation $\sigma \sim_q \tau$ to denote the relation that $\sigma \cap \tau$ is a $q$-simplex with $0 \leq q < 3$. The following two propositions serve as useful tools for investigating the face relations of simplices in $\mathbb{R}^3$.
\begin{prop.}
\label{Proposition: |K| = |L| with homogeneous K, L, then K=L}
Let $K$ be a finite homogeneous simplicial $3$-complex in $\mathbb{R}^3$ and let $L$ be the simplicial complex generated by a collection of $3$-simplices in $K$. Then, $|K| = |L|$ if and only if $K = L$.  
\end{prop.}
\begin{proof}
It is clear that $|K| = |L|$ if $K = L$. Therefore, it is sufficient to prove the `only if' direction. Since $L$ is generated by simplices in $K$, we must have $L \subseteq K$. Suppose $K \neq L$, say $\tau \in K \setminus L$. In particular, $\tau \subseteq |K| = |L|$.  Because $K$ is a homogeneous simplicial $3$-complex, there is a $3$-simplex $\sigma \in K$ such that $\tau \subseteq \sigma$. Because $\tau \notin L$ and $L$ is a simplicial complex, $\sigma \notin L$. Because $\dim(\sigma) = 3$, the interior of $\sigma$ is non-empty (Proposition \ref{Proposition: Elementary Munkres properties}(b)). Let $\mathbf{x}$ be an interior point in $\sigma$. Because $K$ is a simplicial complex in $\mathbb{R}^3$ and $\dim(\sigma) = 3$, $\sigma$ is the only simplex in $K$ containing $\mathbf{x}$ (Proposition \ref{Proposition: Elementary Munkres properties}(e)). In particular, $\mathbf{x} \notin \tau$ for every $\tau \in L$. This shows that $\mathbf{x} \in |K| \setminus |L|$, which is a contradiction.
\end{proof}
\begin{prop.}\label{Proposition: Upper adjacency of a 2-simplex}
Let $K$ be a homogeneous simplicial $3$-complex in $\mathbb{R}^3$, then either $\deg_{K,u}(\tau) = 1$ or $\deg_{K,u}(\tau) = 2$ for every $2$-simplex $\tau \in K$.
\end{prop.}
\begin{proof}
Let $\tau \in K$ be a $2$-simplex. Because $K$ is a homogeneous simplicial $3$-complex, there is a $3$-simplex $\sigma \in K$ such that $\tau \leq \sigma$.  In particular, $\deg_{K,u}(\tau) \geq 1$.  Let $H$ be the hyperplane spanned by $\tau$, such that $\sigma$ is contained in one of the closed half-spaces separated by $H$. Then, by Proposition \ref{Proposition: Supporting hyperplane of simplex on face}(d), $\deg_{K,u}(\tau) > 1$ only if there is a unique $3$-simplex $\sigma' \in K$ with $\tau \leq \sigma'$ such that $\sigma'$ lies in the opposite closed half-space of $H$, as desired.
\end{proof}
In particular, building on a finite point cloud in $\mathbb{R}^3$, such as a collection of atom coordinates, we consider certain homogeneous $3$-simplices, referred to as \textit{admissible simplicial complexes}, which are constructed based on the point cloud. The properties required for selecting admissible simplicial complexes are formally defined below.
\begin{def.}
\label{Definition: admissible homogeneous 3-complex}
Let $V$ be a finite point cloud in $\mathbb{R}^3$. A simplicial complex $K = K(V)$ of simplicies in $\mathbb{R}^3$ is said to be \textbf{admissible} for $V$ if it satisfies the following three properties:
\begin{itemize}
\item[\rm (a)] $V$ is the vertex set of $K$;
\item[\rm (b)] $K$ is a homogeneous simplicial $3$-complex;
\item[\rm (c)] For every two $3$-simplicies $\tau$ and $\rho$ in $K$, there is a sequence $\sigma_1, ..., \sigma_n$ of $3$-simplicies in $K$ such that $\sigma_1 = \tau$, $\sigma_n = \rho$, and $\sigma_i \sim_2 \sigma_{i+1}$ for every $i \in \{ 1, ..., n-1 \}$.
\end{itemize}
\end{def.}
Specifically, the $1$-skeleton $G$ of an admissible simplicial complex $K = K(V)$, constructed from a given point cloud $V \subseteq \mathbb{R}^3$, is primarily considered as the underlying graph of the ANM model in this study. In algebraic topology, for any simplicial complex $K$, the $1$-skeleton of $K$ is defined as the subcomplex generated by all its $1$-simplices~\cite{munkres2018elements}. Furthermore, for a homogeneous simplicial $3$-complex $K$ in $\mathbb{R}^3$, the $1$-skeleton is a homogeneous $1$-complex, resulting in a simple, undirected, geometric graph embedded in $\mathbb{R}^3$, with the $0$-simplices of $K$ as its vertices. Notably, as dictated by property (c), $G$ is a connected graph and every anisotropic sheaf defined on $G$ (cf. \eqref{Eq. Anisotropic sheaf definition}) has rank $3$ since $K$ is a homogeneous simplicial $3$-complex. In particular, the following proposition shows that the homogeneous simplicial $3$-complex produced through the 3D Delaunay triangulation of any point cloud $V \subseteq \mathbb{R}^3$ in general position is admissible in the sense of Definition \ref{Definition: admissible homogeneous 3-complex}.
\begin{prop.}
\label{Proposition: DL is a 3D Tetrahedral mesh}
Let $V \subseteq \mathbb{R}^3$ be a finite point cloud in general position. Let $K = K(V)$ be the 3D Delaunay triangulation of $V$. Then, $K$ is admissible, i.e., $K$ satisfies (a), (b), and (c) in Definition \ref{Definition: admissible homogeneous 3-complex}.
\end{prop.}
\begin{proof}
Since a 3D Delaunay triangulation is a homogeneous $3$-simplicial complex over the point cloud as its vertex set, properties (a) and (b) hold. It remains to prove that $K$ satisfies property (c). Let $|K| \subseteq \mathbb{R}^3$ be the geometric realization of $K$. By the definition of 3D Delaunay triangulation, $|K|$ is the convex hull of $V$.  Pick an arbitrary $3$-simplex $\sigma_0 \in K$, and define $L \subseteq K$ as the subcomplex generated by the collection $\{ \sigma_0 \} \cup C$, where $C$ is the collection of $3$-simplices $\sigma \in K$ such that there exists a sequence of $3$-simplices $\sigma_1, \dots, \sigma_n$ in $K$ with $\sigma_n = \sigma$ and $\sigma_{i} \sim_2 \sigma_{i+1}$ for $i = 0, 1, \dots, n-1$. In particular, $L$ is also a homogeneous simplicial $3$-complex. We claim that $|L| = |K|$, and this shows that $L = K$ by Proposition \ref{Proposition: |K| = |L| with homogeneous K, L, then K=L}. 

Suppose $|L| \neq |K|$. Then, there exists a point $\mathbf{x}_0 \in |K| \setminus |L|$. We claim that there is a $3$-simplex $\sigma \in K$ and a hyperplane $H$ spanned by a $2$-face $\tau \leq \sigma$, with $\deg_{L,u}(\tau) = 1$ (see Definition \ref{Definition: Homogeneous simplicial 3-complex} and Proposition \ref{Proposition: Upper adjacency of a 2-simplex}), such that $H$ separates $\mathbf{x}_0$ and $\sigma$ into disjoint open and closed half-spaces. To prove this fact by induction, we proceed as follows. Pick an arbitrary $3$-simplex $\sigma \in L$. By Corollary \ref{Corollary: A point outside the sigma and supporting hyperplane thm}, there is a $2$-face $\tau \leq \sigma$ such that the spanned hyperplane $H$ separates $\mathbf{x}_0$ and $\sigma$ into disjoint open and closed half-spaces. We have done if $\deg_{L,u}(\tau) = 1$. Otherwise, $\deg_{L,u}(\tau) = 2$ by Proposition \ref{Proposition: Upper adjacency of a 2-simplex}. Therefore, there is a $3$-simplex $\sigma' \in L$ such that $\tau = \sigma \cap \sigma'$ and $\{ \mathbf{x}_0 \} \cup \sigma'$ lies in the opposite closed half-space of $H$. By the same arguments, there is a $2$-face $\tau'$ of $\sigma'$ such that the hyperplane $H'$ spanned by $\tau'$ separates $\mathbf{x}_0$ and $\sigma'$. In particular, we have $\deg_{L,u}(\tau') = 1$ or $2$. This process terminates because $L$ is a finite simplicial complex, and the claim is proved by induction.

Let $\sigma \in K$, along with $H$ and $\tau$, be selected with the properties described in the previous paragraph. Say $\tau = \conv(\bfx_1, \bfx_2, \bfx_3)$ for some $\bfx_1, \bfx_2, \bfx_3 \in V$. Because $|K|$ is convex, the $3$-simplex $\omega := \conv(\bfx_0, \bfx_1, \bfx_2, \bfx_3)$ is a closed subset of $|K|$. In particular, let $\mathbf{z} \in \reInt(\tau)$, then the line segment $\conv(\mathbf{z}, \mathbf{x}_0)$ is a closed subset of $|K|$. Because $\deg_{L,u}(\tau) = 1$, there is a $3$-simplex $\sigma' \in K \setminus L$ such that $\mathbf{z} \in \sigma \cap \sigma'$. Because $K$ is a simplicial complex, it forces that $\tau = \sigma \cap \sigma'$; in particular, $\sigma \sim_2 \sigma'$. However, by the definition of $L$, this implies that $\sigma' \in L$ since $\sigma \in L$.  It leads to a contradiction, and we conclude that $|L|$ and $|K|$ must be equal, as desired.
\end{proof} 
The $1$\textit{-skeleton} $G$ of an abstract simplicial complex $K$ is the collection of $0$- and $1$-simplices within $K$, forming a simplicial complex of dimension $1$, which can be viewed as a graph $G = (V, E)$ consisting of vertices and edges. Notably, let $K$ be an admissible simplicial complex defined as in Definition \ref{Definition: admissible homogeneous 3-complex}, and let $G$ be the $1$-skeleton of $K$, then the property (c) implies $G$ is a connected geometric graph embedded in $\mathbb{R}^3$. The following theorem demonstrates that every $1$-skeleton $G$ derived in this manner supports anisotropic sheaves to achieve the minimum dimension of the global section space.
\begin{theorem}\label{Theorem: Main result 4-1}
Let $V = \{ v_1, v_2, ..., v_n \} \subseteq \mathbb{R}^3$ be a finite point cloud in general position, and let $K = K(V)$ be an admissible simplicial complex for $V$, as defined in Definition \ref{Definition: admissible homogeneous 3-complex}. If $G$ is the $1$-skeleton of $K$, and $\mathcal{F}: (G, \leq) \rightarrow \textup{\textsf{Vect}}_\mathbb{R}$ is an anisotropic sheaf with $w_{ij} \neq 0$ whenever $i < j$, then $\dim_\mathbb{R} \Gamma(G,\mathcal{F}) = 6$.
\end{theorem}
\begin{proof}
Let $\mathbf{C}: C^0(G, \mathcal{F}) \rightarrow C^1(G, \mathcal{F})$ be the associated $0$-coboundary matrix of sheaf $\mathcal{F}$. For every non-empty collection $S$ of $3$-simplicies in $K$, denote $N(S)$ as the set $N(S) = \{ \tau \in K_{(3)} \ | \ \sigma \sim_2 \tau \text{ for some } \sigma \in S\}$. We begin with the case of the subcomplex generated by a single $3$-simplex and prove the theorem by induction. First, pick an arbitrary $3$-simplex $\sigma_1$ in $K$ and define $L_1$ as the subcomplex generated by $S_1 = \{ \sigma_1 \}$, i.e., $L_1 = \{ \tau \in K \ | \ \tau \leq \sigma_1 \}$. Let $G_1 = (V_1, E_1)$ denote the $1$-skeleton of $L_1$ and let $\mathcal{F}_1 = \mathcal{F}|_{G_1}$. By extending zeros, each row in the $0$-th coboundary matrix of sheaf $\mathcal{F}_1$, denoted as $\mathbf{C}_{(1)}: C^0(G_1, \mathcal{F}_1) \rightarrow C^1(G_1, \mathcal{F}_1)$, can be identified as a row of the $0$-th coboundary matrix $\mathbf{C}_{(1)}: C^0(G, \mathcal{F}) \rightarrow C^1(G, \mathcal{F})$ of sheaf $\mathcal{F}$. Because $\sigma_1$ is a non-degenerate $3$-simplex, the rank of $\mathbf{C}_{(1)}$ equals $6$ by Example \ref{Example: K4 example}. In particular,
\begin{equation*}
{\rm rank}(\mathbf{C}_{(1)}) = 6 = 3 \cdot 4 - 6 = 3 \cdot |V_1| - 6.
\end{equation*}
Second, let $L_2$ be the simplicial complex generated by \( S_1 \cup N(S_1) \), i.e., $L_2 = \{ \tau \in K \ | \ \tau \leq \sigma \text{ for some } \sigma \in S_1 \cup N(S_1) \}$, then the restriction sheaf $\mathcal{F}_2 := \mathcal{F}|_{G_2}$, and the $0$-th coboundary matrix $\mathbf{C}_{(2)}: C^0(G_2, \mathcal{F}_2) \rightarrow C^1(G_2, \mathcal{F}_2)$ are defined. In particular, we have $L_1 \subseteq L_2$, $V_1 \subseteq V_2$, $E_1 \subseteq E_2$, and $\mathcal{F}_2|_{G_1} = \mathcal{F}_1$. Again, $\mathbf{C}_{(2)}$ is a submatrix of $\mathbf{C}$, where the corresponding rows are extended to match those in $\mathbf{C}$ by appending zeros. By the definition of $L_2$, each vertex belongs to $V_2 \setminus V_1$ must be a vertex of a member in $N(S_1)$, and hence has at least $3$ edges in $E_2$ that connect to vertices in $V_1$. Since every tetrahedron in $N(S_1)$ is non-degenerate, each vertex in $V_2 \setminus V_1$ contributes at least three additional linearly independent rows to the extension matrix  $\mathbf{C}_{(2)}$ from matrix $\mathbf{C}_{(1)}$. Then,
\begin{equation*}
{\rm rank}(\mathbf{C}_{(2)}) \geq {\rm rank}(\mathbf{C}_{(1)}) + 3 \cdot | V_2 \setminus V_1 | = 3 \cdot |V_1| - 6 + 3 \cdot | V_2 \setminus V_1 | = 3 \cdot |V_2| - 6.  
\end{equation*}
On the other hand, since ${\rm rank}(\mathbf{C}_{(2)}) = 3 \cdot |V_2| - \dim_{\mathbb{R}}(\ker(L_{\mathcal{F}_2})) \leq 3 \cdot |V_2| - 6$ by Theorem \ref{Theorem: Main result 2}, we obtain
\begin{equation*}
{\rm rank}(\mathbf{C}_{(2)}) = 3 \cdot |V_2| - 6.    
\end{equation*}
If $V_2 = V_1$, then ${\rm rank}(\mathbf{C}_{(2)}) = {\rm rank}(\mathbf{C}_{(1)})$ since $3 \cdot |V_2| - 6 = 3 \cdot |V_1| - 6 = {\rm rank}(\mathbf{C}_{(1)}) \leq {\rm rank}(\mathbf{C}_{(2)}) \leq 3 \cdot |V_2| - 6$ by Theorem \ref{Theorem: Main result 2}. Similarly, we define $S_2$ as the collection of all $3$-simplices in $L_2$ and the same process can be continued by defining $L_3$ as the simplicial complex generated by $S_2 \cup N(S_2)$. By continuing this process, we will have the following filtration of subcomplexes of $K$:
\begin{equation*}
L_1 \subseteq L_2 \subseteq L_3 \subseteq \cdots    
\end{equation*}
with $1$-skeletons $G_{n} = (V_n, E_n)$, sets $S_n$ and $N(S_n)$, sheaves $\mathcal{F}_n: (G_n, \leq) \rightarrow \textup{\textsf{Vect}}_{\mathbb{R}}$, and the $0$-th coboundary matrices $\mathbf{C}_{(n)}: C^0(G_n, \mathcal{F}_n) \rightarrow C^1(G_n, \mathcal{F}_n)$ with $n = 1, 2, ...$, satisfying 
\begin{equation*}
{\rm rank}(\mathbf{C}_{(n)}) = 3 \cdot |V_n| - 6.    
\end{equation*}
Because $V$ is finite, this process must terminate. Moreover, every two tetrahedrons $\tau$ and $\rho$ in $K$ admit a sequence $\tau_1, ..., \tau_m$ of $3$-simplicies such that $\tau_1 = \tau$, $\tau_m = \rho$, and $\tau_i \sim_2 \tau_{i+1}$ for each $i$ since $K$ is admissible according to Definition \ref{Definition: admissible homogeneous 3-complex}. Eventually, this process encompasses all the $3$-simplices in $K$ and associates an $N \in \mathbb{N}$ such that $L_N = K$, $G_N = G$, and $\mathbf{C}_{(N)} = \mathbf{C}$. By Proposition \ref{Proposition: Dimension of the global section space of an anisotropic sheaf}, we conclude that $\dim_\mathbb{R} \ker(L_\calF) = 3 \cdot |V_N| - {\rm rank}(\mathbf{C}_{(N)}) = 3 \cdot |V| - {\rm rank}(\mathbf{C}) = 3 \cdot |V| - (3 \cdot |V| - 6) = 6$.
\end{proof}
The following corollary follows from Proposition \ref{Proposition: DL is a 3D Tetrahedral mesh} and Theorem \ref{Theorem: Main result 4-1}. Specifically, for any 3D atomic system in general position, the $1$-skeleton of the 3D Delaunay triangulation of the system guarantees that the associated ANM Hessian matrix has exactly six eigenmodes with zero eigenvalues.

\begin{coro}\label{Corollary: Main result 4-2}
Let $V = \{ v_1, v_2, ..., v_n \} \subseteq \mathbb{R}^3$ be a finite point cloud in general position, and let $K = K(V)$ be the 3D Delaunay triangulation of $V$. If $G$ is the $1$-skeleton of $K$, and $\mathcal{F}: (G, \leq) \rightarrow \textup{\textsf{Vect}}_\mathbb{R}$ is an anisotropic sheaf with $w_{ij} \neq 0$ whenever $i < j$, then $\dim_\mathbb{R} \Gamma(G,\mathcal{F}) = 6$.    
\end{coro}
\paragraph{Minimal Graph Construction}

At the end of this section, we propose a systematic method for constructing $3$-simplices and their faces in an admissible simplicial complex, as defined in Definition \ref{Definition: admissible homogeneous 3-complex}. The $1$-skeleton of this simplicial complex induces the desired Hessian matrix with nullity exactly $6$. Furthermore, this graph is minimal in the sense of Theorem \ref{Theorem: Main result 3}; that is, removing any edge from the constructed graph would result in a Hessian matrix with nullity larger than $6$. The construction is presented in the following algorithm.
\begin{algorithm}[H]
\caption{Algorithm for the minimal graph construction}\label{Algorithm: Main result 5-2}
\begin{algorithmic}[1]
\Require A finite point cloud $V \subseteq \mathbb{R}^3$ in general position in $\mathbb{R}^3$ ($|V| \geq 4$).
\Ensure A minimal graph $G = (V, E)$ that satisfies the properties (a) and (b) in Theorem \ref{Theorem: Main result 3}.
\State Pick an arbitrary point in $V$, say $\mathbf{v}_1$.
\State Pick another three points that are nearest to $\mathbf{v}_1$, say $\mathbf{v}_2$, $\mathbf{v}_3$, $\mathbf{v}_4$.
\State Let $K_1$ be the homogeneous simplicial complex generated by ${\rm conv}(\mathbf{v}_1, \mathbf{v}_2, \mathbf{v}_3, \mathbf{v}_4)$.
\For{$i = 5$ to $|V|$}
    \State Pick a point $\mathbf{v}_i \in V \setminus \{\mathbf{v}_1, \mathbf{v}_2, \ldots, \mathbf{v}_{i-1}\}$ that has the smallest distance to $|K_{i-1}|$.
    \State Pick a 2-simplex $\sigma$ in $K_{i-1}$, say $\sigma = \mathrm{conv}(\mathbf{v}_{j_1}, \mathbf{v}_{j_2}, \mathbf{v}_{j_3})$, such that 
    \begin{equation}\label{Equation: Crucial step in the Algorithm}
    \mathrm{conv}(\mathbf{v}_{j_1}, \mathbf{v}_{j_2}, \mathbf{v}_{j_3}, \mathbf{v}_i) \cap K_{i-1} = \{ \mathbf{v}_{j_1}, \mathbf{v}_{j_2}, \mathbf{v}_{j_3}, \mathbf{v}_i \}.    
    \end{equation}
    \State Add $\mathbf{v}_i$ and the edges $\{\mathbf{v}_i, \mathbf{v}_{j_1}\}$, $\{\mathbf{v}_i, \mathbf{v}_{j_2}\}$, $\{\mathbf{v}_i, \mathbf{v}_{j_3}\}$ to $K_i$.
\EndFor
\State Let $K_N$ denote the final simplicial complex after all vertices in $V$ have been processed.
\State Let $G = (V, E)$ be the $1$-skeleton of the simplicial complex $K_N$.
\Return $G$.
\end{algorithmic}
\end{algorithm}

For step 3 in Algorithm \ref{Algorithm: Main result 5-2}, ${\rm conv}(\mathbf{v}_1, \mathbf{v}_2, \mathbf{v}_3, \mathbf{v}_4) \cap V = {\mathbf{v}_1, \mathbf{v}_2, \mathbf{v}_3, \mathbf{v}_4}$, since $\mathbf{v}_2$, $\mathbf{v}_3$, and $\mathbf{v}_4 \in V$ are the nearest points to $\mathbf{v}_1$. For step 6, such a $2$-simplex exists by the proof of the existence of a $2$-face in the homogeneous simplicial complex using Corollary \ref{Corollary: A point outside the sigma and supporting hyperplane thm}. Furthermore, Equation~\eqref{Equation: Crucial step in the Algorithm} holds due to the minimal distance property of $\mathbf{v}_i$.

Due to the construction, the resulting $K_N$ is an admissible simplicial complex as defined in Definition \ref{Definition: admissible homogeneous 3-complex}. Let $\mathcal{F}: (G, \leq) \rightarrow \textup{\textsf{Vect}}_\mathbb{R}$ be the anisotropic sheaf defined as in~\eqref{Eq. Anisotropic sheaf definition-form 2}, i.e., 
\begin{equation*} 
\mathcal{F}_{v_i, [v_i,v_j]} = \begin{bmatrix} x_j^\circ - x_i^\circ & y_j^\circ - y_i^\circ & z_j^\circ - z_i^\circ \ \end{bmatrix} = \mathcal{F}_{v_j, [v_i,v_j]}. \end{equation*} 
By Theorem \ref{Theorem: Main result 4-1}, the global section space of the anisotropic sheaf $\mathcal{F}$ has rank $6$, i.e., the nullity of the induced ANM Hessian matrix is exactly $6$.

Finally, for the minimality of the generated $1$-skeleton, note that there are exactly $6 + 3 \cdot (|V| - 4) = 3 \cdot |V| - 6$ edges. Let $G = (V, E)$ be the output graph, and let $G' = (V, E')$ be obtained by removing an edge from $E$. Let $\mathcal{F}': (G', \leq) \rightarrow \textup{\textsf{Vect}}_\mathbb{R}$ be the induced anisotropic sheaf. The coboundary matrix $\mathbf{C}': C^0(G', \mathcal{F}') \rightarrow C^1(G', \mathcal{F}')$ then has dimensions $|E'| \times 3|V| = (3|V| - 7) \times 3|V|$. Consequently: 
\begin{equation*} 
\dim_{\mathbb{R}} \ker(\mathbf{C}') = 3|V| - \mathrm{rank}(\mathbf{C}') \geq 3|V| - (3|V| - 7) = 7. 
\end{equation*} 
This confirms that removing any edge increases the nullity of the Hessian matrix beyond $6$, ensuring the minimality of the constructed graph.

\bibliography{refs_2}

\begin{thebibliography}{10}

\bibitem{Alvarez-Garcia:2014}
D.~Alvarez-Garcia and X.~Barril.
\newblock Relationship between protein flexibility and binding: Lessons for structure-based drug design.
\newblock {\em Journal of Chemical Theory and Computation}, 10(6):2608--2614, 2014.

\bibitem{artin2012arithmetic}
M.~Artin, G.~Cornell, C.~Chai, J.~Silverman, C.~Chinburg, G.~Faltings, B.~Gross, F.~McGuiness, J.~Milne, M.~Rosen, et~al.
\newblock {\em Arithmetic Geometry}.
\newblock Springer New York, 2012.

\bibitem{Atilgan:2001}
A.~R. Atilgan, S.~R. Durell, R.~L. Jernigan, M.~C. Demirel, O.~Keskin, and I.~Bahar.
\newblock Anisotropy of fluctuation dynamics of proteins with an elastic network model.
\newblock {\em Biophysical journal}, 80(1):505--515, 2001.

\bibitem{bahar1998vibrational}
I.~Bahar, A.~R. Atilgan, M.~C. Demirel, and B.~Erman.
\newblock Vibrational dynamics of folded proteins: significance of slow and fast motions in relation to function and stability.
\newblock {\em Physical review letters}, 80(12):2733--2736, 1998.

\bibitem{bahar1997direct}
I.~Bahar, A.~R. Atilgan, and B.~Erman.
\newblock Direct evaluation of thermal fluctuations in proteins using a single-parameter harmonic potential.
\newblock {\em Folding and Design}, 2(3):173--181, 1997.

\bibitem{barbero2022sheaf_attention}
F.~Barbero, C.~Bodnar, H.~S. de~Oc{\'a}riz~Borde, and P.~Lio.
\newblock Sheaf attention networks.
\newblock In {\em NeurIPS 2022 Workshop on Symmetry and Geometry in Neural Representations}, 2022.

\bibitem{battiloro2023tangent}
C.~Battiloro, Z.~Wang, H.~Riess, P.~Di~Lorenzo, and A.~Ribeiro.
\newblock Tangent bundle filters and neural networks: From manifolds to cellular sheaves and back.
\newblock In {\em ICASSP 2023-2023 IEEE International Conference on Acoustics, Speech and Signal Processing (ICASSP)}, pages 1--5. IEEE, 2023.

\bibitem{Berman:2000}
H.~M. Berman, J.~Westbrook, Z.~Feng, G.~Gilliland, T.~N. Bhat, H.~Weissig, I.~N. Shindyalov, and P.~E. Bourne.
\newblock The protein data bank.
\newblock {\em Nucleic acids research}, 28(1):35--242, 2000.

\bibitem{BJORKMAN1998651}
A.~Björkman and S.~L. Mowbray.
\newblock Multiple open forms of ribose-binding protein trace the path of its conformational change.
\newblock {\em Journal of Molecular Biology}, 279(3):651--664, 1998.

\bibitem{bodnar2022neural}
C.~Bodnar, F.~Di~Giovanni, B.~Chamberlain, P.~Lio, and M.~Bronstein.
\newblock Neural sheaf diffusion: A topological perspective on heterophily and oversmoothing in {G}{N}{N}s.
\newblock {\em Advances in Neural Information Processing Systems}, 35:18527--18541, 2022.

\bibitem{braithwaite2024heterogeneous}
L.~Braithwaite, I.~Duta, and P.~Li{\`o}.
\newblock Heterogeneous sheaf neural networks.
\newblock {\em arXiv preprint arXiv:2409.08036}, 2024.

\bibitem{bredon2012sheaf}
G.~E. Bredon.
\newblock {\em Sheaf theory}, volume 170.
\newblock Springer Science \& Business Media, 2012.

\bibitem{brooks1983charmm}
B.~R. Brooks, R.~E. Bruccoleri, B.~D. Olafson, D.~J. States, S.~a. Swaminathan, and M.~Karplus.
\newblock {C}{H}{A}{R}{M}{M}: a program for macromolecular energy, minimization, and dynamics calculations.
\newblock {\em Journal of computational chemistry}, 4(2):187--217, 1983.

\bibitem{bu2011proteins}
Z.~Bu and D.~J. Callaway.
\newblock Proteins move! {P}rotein dynamics and long-range allostery in cell signaling.
\newblock {\em Advances in protein chemistry and structural biology}, 83:163--221, 2011.

\bibitem{caralt2024joint}
F.~H. Caralt, G.~Bernardez, I.~Duta, E.~Alarcon, and P.~Lio.
\newblock Joint diffusion processes as an inductive bias in sheaf neural networks.
\newblock In {\em ICML 2024 Workshop on Geometry-grounded Representation Learning and Generative Modeling}, 2024.

\bibitem{cooperband2024cellular}
Z.~Cooperband.
\newblock {\em Cellular Cosheaves, Graphic Statics, and Mechanics}.
\newblock PhD thesis, University of Pennsylvania, 2024.

\bibitem{cooperband2023towards}
Z.~Cooperband and R.~Ghrist.
\newblock Towards homological methods in graphic statics.
\newblock {\em Journal of the International Association for Shell and Spatial Structures}, 64(4):266--277, 2023.

\bibitem{cooperband2023cosheaf}
Z.~Cooperband, R.~Ghrist, and J.~Hansen.
\newblock A cosheaf theory of reciprocal figures: Planar and higher genus graphic statics.
\newblock {\em arXiv preprint arXiv:2311.12946}, 2023.

\bibitem{cooperband2024equivariant}
Z.~Cooperband, M.~Lopez, and B.~Schulze.
\newblock Equivariant cosheaves and finite group representations in graphic statics.
\newblock {\em arXiv preprint arXiv:2401.09392}, 2024.

\bibitem{cui2005normal}
Q.~Cui and I.~Bahar.
\newblock {\em Normal Mode Analysis: Theory and Applications to Biological and Chemical Systems}.
\newblock Chapman \& Hall/CRC Mathematical Biology Series. CRC Press, 2005.

\bibitem{curry2016discrete}
J.~Curry, R.~Ghrist, and V.~Nanda.
\newblock Discrete morse theory for computing cellular sheaf cohomology.
\newblock {\em Foundations of Computational Mathematics}, 16:875--897, 2016.

\bibitem{curry2014sheaves}
J.~M. Curry.
\newblock {\em Sheaves, cosheaves and applications}.
\newblock PhD thesis, University of Pennsylvania, 2014.

\bibitem{curry2015topological}
J.~M. Curry.
\newblock Topological data analysis and cosheaves.
\newblock {\em Japan Journal of Industrial and Applied Mathematics}, 32:333--371, 2015.

\bibitem{delaunay1934bulletin}
B.~Delaunay, S.~Vide, A.~Lam{\'e}moire, and V.~De~Georges.
\newblock Bulletin de l'{A}cademie des {S}ciences de l'{URSS}.
\newblock {\em Classe des sciences math{\'e}matiques et naturelles}, 6:793--800, 1934.

\bibitem{doruker2000dynamics}
P.~Doruker, A.~R. Atilgan, and I.~Bahar.
\newblock Dynamics of proteins predicted by molecular dynamics simulations and analytical approaches: Application to $\alpha$-amylase inhibitor.
\newblock {\em Proteins: Structure, Function, and Bioinformatics}, 40(3):512--524, 2000.

\bibitem{dudko2006intrinsic}
O.~K. Dudko, G.~Hummer, and A.~Szabo.
\newblock Intrinsic rates and activation free energies from single-molecule pulling experiments.
\newblock {\em Physical review letters}, 96(10):108101, 2006.

\bibitem{duta2024sheaf}
I.~Duta, G.~Cassar{\`a}, F.~Silvestri, and P.~Li{\`o}.
\newblock Sheaf hypergraph networks.
\newblock {\em Advances in Neural Information Processing Systems}, 36, 2024.

\bibitem{ewald1996combinatorial}
G.~Ewald.
\newblock {\em Combinatorial Convexity and Algebraic Geometry}.
\newblock Graduate Texts in Mathematics. Springer New York, 1996.

\bibitem{eyal2006anisotropic}
E.~Eyal, L.-W. Yang, and I.~Bahar.
\newblock Anisotropic network model: systematic evaluation and a new web interface.
\newblock {\em Bioinformatics}, 22(21):2619--2627, 2006.

\bibitem{Fischer:2014}
M.~Fischer, R.~G. Coleman, J.~S. Fraser, and B.~K. Shoichet.
\newblock Incorporation of protein flexibility and conformational energy penalties in docking screens to improve ligand discovery.
\newblock {\em Nature Chemistry}, 6:575--583, 2014.

\bibitem{fraser2009hidden}
J.~S. Fraser, M.~W. Clarkson, S.~C. Degnan, R.~Erion, D.~Kern, and T.~Alber.
\newblock Hidden alternative structures of proline isomerase essential for catalysis.
\newblock {\em Nature}, 462(7273):669--673, 2009.

\bibitem{hansRiess2022}
R.~Ghrist and H.~Riess.
\newblock Cellular sheaves of lattices and the {T}arski {L}aplacian.
\newblock {\em Homology, Homotopy and Applications}, 24(1):325--345, 2022.

\bibitem{ghrist2014elementary}
R.~W. Ghrist.
\newblock {\em Elementary applied topology}, volume~1.
\newblock Createspace Seattle, 2014.

\bibitem{go1983dynamics}
N.~Go, T.~Noguti, and T.~Nishikawa.
\newblock Dynamics of a small globular protein in terms of low-frequency vibrational modes.
\newblock {\em Proceedings of the National Academy of Sciences}, 80(12):3696--3700, 1983.

\bibitem{hansen2020sheaf}
J.~Hansen and T.~Gebhart.
\newblock Sheaf neural networks.
\newblock {\em arXiv preprint arXiv:2012.06333}, 2020.

\bibitem{hansen2019toward}
J.~Hansen and R.~Ghrist.
\newblock Toward a spectral theory of cellular sheaves.
\newblock {\em Journal of Applied and Computational Topology}, 3(4):315--358, 2019.

\bibitem{hansen2021opinion}
J.~Hansen and R.~Ghrist.
\newblock Opinion dynamics on discourse sheaves.
\newblock {\em SIAM Journal on Applied Mathematics}, 81(5):2033--2060, 2021.

\bibitem{hartshorne2013algebraic}
R.~Hartshorne.
\newblock {\em Algebraic geometry}, volume~52.
\newblock Springer Science \& Business Media, 2013.

\bibitem{he2023sheaf}
Y.~He, C.~Bodnar, and P.~Lio.
\newblock Sheaf-based positional encodings for graph neural networks.
\newblock In {\em NeurIPS 2023 Workshop on Symmetry and Geometry in Neural Representations}, 2023.

\bibitem{hinsen1998analysis}
K.~Hinsen.
\newblock Analysis of domain motions by approximate normal mode calculations.
\newblock {\em Proteins: Structure, Function, and Bioinformatics}, 33(3):417--429, 1998.

\bibitem{hinsen2008structural}
K.~Hinsen.
\newblock Structural flexibility in proteins: impact of the crystal environment.
\newblock {\em Bioinformatics}, 24(4):521--528, 2008.

\bibitem{huangstability}
Y.~Huang, W.~Lu, J.~Robinson, Y.~Yang, M.~Zhang, S.~Jegelka, and P.~Li.
\newblock On the stability of expressive positional encodings for graphs.
\newblock In {\em The Twelfth International Conference on Learning Representations}, 2024.

\bibitem{VMD}
W.~Humphrey, A.~Dalke, and K.~Schulten.
\newblock {VMD} -- visual molecular dynamics.
\newblock {\em Journal of Molecular Graphics}, 14(1):33--38, 1996.

\bibitem{jacobs2001protein}
D.~J. Jacobs, A.~J. Rader, L.~A. Kuhn, and M.~F. Thorpe.
\newblock Protein flexibility predictions using graph theory.
\newblock {\em Proteins: Structure, Function, and Bioinformatics}, 44(2):150--165, 2001.

\bibitem{kashiwara2018persistent}
M.~Kashiwara and P.~Schapira.
\newblock Persistent homology and microlocal sheaf theory.
\newblock {\em Journal of Applied and Computational Topology}, 2(1):83--113, 2018.

\bibitem{kondrashov2007protein}
D.~A. Kondrashov, A.~W. Van~Wynsberghe, R.~M. Bannen, Q.~Cui, and G.~N. Phillips.
\newblock Protein structural variation in computational models and crystallographic data.
\newblock {\em Structure}, 15(2):169--177, 2007.

\bibitem{levitt1985protein}
M.~Levitt, C.~Sander, and P.~S. Stern.
\newblock Protein normal-mode dynamics: trypsin inhibitor, crambin, ribonuclease and lysozyme.
\newblock {\em Journal of molecular biology}, 181(3):423--447, 1985.

\bibitem{li2002coarse}
G.~Li and Q.~Cui.
\newblock A coarse-grained normal mode approach for macromolecules: an efficient implementation and application to {C}a$^{2+}$-{A}{T}{P}ase.
\newblock {\em Biophysical Journal}, 83(5):2457--2474, 2002.

\bibitem{marsh2014protein}
J.~A. Marsh and S.~A. Teichmann.
\newblock Protein flexibility facilitates quaternary structure assembly and evolution.
\newblock {\em PLoS biology}, 12(5):e1001870, 2014.

\bibitem{mccammon1977dynamics}
J.~A. McCammon, B.~R. Gelin, and M.~Karplus.
\newblock Dynamics of folded proteins.
\newblock {\em nature}, 267(5612):585--590, 1977.

\bibitem{munkres2018elements}
J.~R. Munkres.
\newblock {\em Elements of algebraic topology}.
\newblock CRC Press, 2018.

\bibitem{nguyen2020review}
D.~D. Nguyen, Z.~Cang, and G.-W. Wei.
\newblock A review of mathematical representations of biomolecular data.
\newblock {\em Physical Chemistry Chemical Physics}, 22(8):4343--4367, 2020.

\bibitem{nguyen2019mathematical}
D.~D. Nguyen, Z.~Cang, K.~Wu, M.~Wang, Y.~Cao, and G.-W. Wei.
\newblock Mathematical deep learning for pose and binding affinity prediction and ranking in {D}3{R} grand challenges.
\newblock {\em Journal of computer-aided molecular design}, 33:71--82, 2019.

\bibitem{nguyen2020mathdl}
D.~D. Nguyen, K.~Gao, M.~Wang, and G.-W. Wei.
\newblock {M}ath{D}l: mathematical deep learning for {D}3{R} grand challenge 4.
\newblock {\em Journal of computer-aided molecular design}, 34:131--147, 2020.

\bibitem{Opron:2014}
K.~Opron, K.~L. Xia, and G.~W. Wei.
\newblock Fast and anisotropic flexibility-rigidity index for protein flexibility and fluctuation analysis.
\newblock {\em Journal of Chemical Physics}, 140:234105, 2014.

\bibitem{park2013coarse}
J.-K. Park, R.~Jernigan, and Z.~Wu.
\newblock Coarse grained normal mode analysis vs. refined gaussian network model for protein residue-level structural fluctuations.
\newblock {\em Bulletin of mathematical biology}, 75:124--160, 2013.

\bibitem{peigeom}
H.~Pei, B.~Wei, K.~C.-C. Chang, Y.~Lei, and B.~Yang.
\newblock Geom-{G}{C}{N}: Geometric graph convolutional networks.
\newblock In {\em International Conference on Learning Representations}, 2020.

\bibitem{robinson2014topological}
M.~Robinson.
\newblock {\em Topological signal processing}.
\newblock Mathematical Engineering. Springer Berlin Heidelberg, 2014.

\bibitem{smith1999correlations}
R.~D. Smith.
\newblock {\em Correlations between bound N-alkyl isocyanide orientations and pathways for ligand binding in recombinant myoglobins}.
\newblock Rice University, 1999.

\bibitem{tama2001conformational}
F.~Tama and Y.-H. Sanejouand.
\newblock Conformational change of proteins arising from normal mode calculations.
\newblock {\em Protein engineering}, 14(1):1--6, 2001.

\bibitem{tasumi1982normal}
M.~Tasumi, H.~Takeuchi, S.~Ataka, A.~Dwivedi, and S.~Krimm.
\newblock Normal vibrations of proteins: {G}lucagon.
\newblock {\em Biopolymers: Original Research on Biomolecules}, 21(3):711--714, 1982.

\bibitem{thompson1998handbook}
J.~F. Thompson, B.~K. Soni, and N.~P. Weatherill.
\newblock {\em Handbook of grid generation}.
\newblock CRC press, 1998.

\bibitem{XiaoqiWei2024FODS}
X.~Wei and G.-W. Wei.
\newblock Persistent sheaf laplacians.
\newblock {\em Foundations of Data Science}, 2024.

\bibitem{wells2017differential}
R.~O. Wells and O.~RAYMOND.
\newblock {\em Differential and complex geometry: origins, abstractions and embeddings}.
\newblock Springer, 2017.

\bibitem{xia2014identifying}
F.~Xia, D.~Tong, L.~Yang, D.~Wang, S.~C. Hoi, P.~Koehl, and L.~Lu.
\newblock Identifying essential pairwise interactions in elastic network model using the alpha shape theory.
\newblock {\em Journal of Computational Chemistry}, 35(15):1111--1121, 2014.

\bibitem{xia2018multiscale}
K.~Xia.
\newblock Multiscale virtual particle based elastic network model (mvp-enm) for normal mode analysis of large-sized biomolecules.
\newblock {\em Physical Chemistry Chemical Physics}, 20(1):658--669, 2018.

\bibitem{xia2015multiscale}
K.~Xia, K.~Opron, and G.-W. Wei.
\newblock Multiscale gaussian network model (m{G}{N}{M}) and multiscale anisotropic network model (m{A}{N}{M}).
\newblock {\em The Journal of chemical physics}, 143(20), 2015.

\bibitem{LWYang:2008}
L.~W. Yang and C.~P. Chng.
\newblock Coarse-grained models reveal functional dynamics--{I}. elastic network models--theories, comparisons and perspectives.
\newblock {\em Bioinformatics and Biology Insights}, 2:25 -- 45, 2008.

\bibitem{ye2019curvature}
Z.~Ye, K.~S. Liu, T.~Ma, J.~Gao, and C.~Chen.
\newblock Curvature graph network.
\newblock In {\em International conference on learning representations}, 2019.

\bibitem{zhou2014alpha}
W.~Zhou and H.~Yan.
\newblock Alpha shape and delaunay triangulation in studies of protein-related interactions.
\newblock {\em Briefings in bioinformatics}, 15(1):54--64, 2014.

\end{thebibliography}
\bibliographystyle{abbrv}

\appendix
\section{Anisotropic Sheaves on Complete Graphs}
\label{Section: Anisotropic Sheaves on Complete Graphs}
To prove that any anisotropic sheaf $\mathcal{F}: (K_n, \leq) \rightarrow \textup{\textsf{Vect}}_\mathbb{R}$ of rank $3$ over a complete graph $K_n$ ($n \geq 3$) must satisfy the condition $\dim_\mathbb{R} \Gamma(K_n, \mathcal{F}) = 6$, we begin by examining the cases where $n = 4$, $5$, and $6$.
\begin{exam.}
\label{Example: K4 example}
Let $G = K_4$ be the complete graph with vertex set $V = \{ 1,2,3,4 \}$ accompanied with 3D coordinates, and let $\mathcal{F}: (G,\leq) \rightarrow \textup{\textsf{Vect}}_\mathbb{R}$ be the anisotropic sheaf defined in Equation~\eqref{Eq. Anisotropic sheaf definition-form 2}, then the coboundary matrix $\mathbf{C}: C^0(G,\mathcal{F}) \rightarrow C^1(G,\mathcal{F})$ is a $6 \times 12$ matrix represented by
\begin{equation*}
\mathbf{C} = \left[\begin{array}{cccc}
-\mathcal{F}_{1, [1,2]} & \mathcal{F}_{2, [1,2]} & \mathbf{0} & \mathbf{0} \\
-\mathcal{F}_{1, [1,3]} & \mathbf{0} & \mathcal{F}_{3, [1,3]} & \mathbf{0} \\
-\mathcal{F}_{1, [1,4]} & \mathbf{0} & \mathbf{0} & \mathcal{F}_{4, [1,4]} \\
\hline
\mathbf{0} & -\mathcal{F}_{2, [2,3]} & \mathcal{F}_{3, [2,3]} & \mathbf{0} \\
\mathbf{0} & -\mathcal{F}_{2, [2,4]} & \mathbf{0} & \mathcal{F}_{4, [2,4]} \\
\hline
\mathbf{0} & \mathbf{0} & -\mathcal{F}_{3, [3,4]} & \mathcal{F}_{4, [3,4]} \\
\end{array}
\right] =
\left[\begin{array}{cccc}
-\mathcal{F}_{1, [1,2]} & \mathcal{F}_{1, [1,2]} & \mathbf{0} & \mathbf{0} \\
-\mathcal{F}_{1, [1,3]} & \mathbf{0} & \mathcal{F}_{1, [1,3]} & \mathbf{0} \\
-\mathcal{F}_{1, [1,4]} & \mathbf{0} & \mathbf{0} & \mathcal{F}_{1, [1,4]} \\
\hline
\mathbf{0} & -\mathcal{F}_{2, [2,3]} & \mathcal{F}_{2, [2,3]} & \mathbf{0} \\
\mathbf{0} & -\mathcal{F}_{2, [2,4]} & \mathbf{0} & \mathcal{F}_{2, [2,4]} \\
\hline
\mathbf{0} & \mathbf{0} & -\mathcal{F}_{3, [3,4]} & \mathcal{F}_{3, [3,4]} \\
\end{array}
\right].    
\end{equation*}
Then, ${\rm rank}(\mathbf{C}) \leq 6$ and hence $\dim_\mathbb{R}(\ker(L_\mathcal{F})) = 3 \cdot 4 - {\rm rank}(\mathbf{C}) \geq  6$. Furthermore, by applying column operations with block matrices, the coboundary matrix becomes
\begin{equation}
\label{Eq. Column and Row operations for K4-1}
\begin{split}
\mathbf{C} &=
\left[\begin{array}{cccc}
-\mathcal{F}_{1, [1,2]} & \mathcal{F}_{1, [1,2]} & \mathbf{0} & \mathbf{0} \\
-\mathcal{F}_{1, [1,3]} & \mathbf{0} & \mathcal{F}_{1, [1,3]} & \mathbf{0} \\
-\mathcal{F}_{1, [1,4]} & \mathbf{0} & \mathbf{0} & \mathcal{F}_{1, [1,4]} \\
\hline
\mathbf{0} & -\mathcal{F}_{2, [2,3]} & \mathcal{F}_{2, [2,3]} & \mathbf{0} \\
\mathbf{0} & -\mathcal{F}_{2, [2,4]} & \mathbf{0} & \mathcal{F}_{2, [2,4]} \\
\hline
\mathbf{0} & \mathbf{0} & -\mathcal{F}_{3, [3,4]} & \mathcal{F}_{3, [3,4]} \\
\end{array} \right]
\longrightarrow  
\left[\begin{array}{cccc}
\mathbf{0} & \mathcal{F}_{1, [1,2]} & \mathbf{0} & \mathbf{0} \\
\mathbf{0} & \mathbf{0} & \mathcal{F}_{1, [1,3]} & \mathbf{0} \\
\mathbf{0} & \mathbf{0} & \mathbf{0} & \mathcal{F}_{1, [1,4]} \\
\hline
\mathbf{0} & -\mathcal{F}_{2, [2,3]} & \mathcal{F}_{2, [2,3]} & \mathbf{0} \\
\mathbf{0} & -\mathcal{F}_{2, [2,4]} & \mathbf{0} & \mathcal{F}_{2, [2,4]} \\
\hline
\mathbf{0} & \mathbf{0} & -\mathcal{F}_{3, [3,4]} & \mathcal{F}_{3, [3,4]} \\
\end{array} \right] \\
&\longrightarrow
\left[\begin{array}{cccc}
\mathbf{0} & \mathcal{F}_{1, [1,2]} & \mathbf{0} & \mathbf{0} \\
\mathbf{0} & \mathcal{F}_{1, [1,3]} & \mathcal{F}_{1, [1,3]} & \mathbf{0} \\
\mathbf{0} & \mathcal{F}_{1, [1,4]} & \mathbf{0} & \mathcal{F}_{1, [1,4]} \\
\hline
\mathbf{0} & \mathbf{0} & \mathcal{F}_{2, [2,3]} & \mathbf{0} \\
\mathbf{0} & \mathbf{0} & \mathbf{0} & \mathcal{F}_{2, [2,4]} \\
\hline
\mathbf{0} & \mathbf{0} & -\mathcal{F}_{3, [3,4]} & \mathcal{F}_{3, [3,4]} \\
\end{array} \right] 
\longrightarrow  
\left[\begin{array}{cccc}
\mathbf{0} & \mathcal{F}_{1, [1,2]} & \mathbf{0} & \mathbf{0} \\
\mathbf{0} & \mathcal{F}_{1, [1,3]} & \mathcal{F}_{1, [1,3]} & \mathbf{0} \\
\mathbf{0} & \mathcal{F}_{1, [1,4]} & \mathcal{F}_{1, [1,4]} & \mathcal{F}_{1, [1,4]} \\
\hline
\mathbf{0} & \mathbf{0} & \mathcal{F}_{2, [2,3]} & \mathbf{0} \\
\mathbf{0} & \mathbf{0} & \mathcal{F}_{2, [2,4]} & \mathcal{F}_{2, [2,4]} \\
\hline
\mathbf{0} & \mathbf{0} & \mathbf{0} & \mathcal{F}_{3, [3,4]} \\
\end{array} \right] = \widetilde{\mathbf{C}}.
\end{split}
\end{equation}
In particular, ${\rm rank}(\mathbf{C}) = {\rm rank}(\widetilde{\mathbf{C}})$. Moreover, by investigating the rank of matrix $\widetilde{\mathbf{C}}$, the following assertions hold:
\begin{itemize}
\item[\rm (a)] If ${\rm rank}(\mathcal{F}) = 1$, then ${\rm rank}(\mathbf{C}) = 3$ and $\dim_\mathbb{R}(\ker(L_\mathcal{F})) = 9$. 
\item[\rm (b)] If ${\rm rank}(\mathcal{F}) = 2$, then ${\rm rank}(\mathbf{C}) = 5$ and $\dim_\mathbb{R}(\ker(L_\mathcal{F})) = 7$.
\item[\rm (c)] If ${\rm rank}(\mathcal{F}) = 3$, then ${\rm rank}(\mathbf{C}) = 6$ and $\dim_\mathbb{R}(\ker(L_\mathcal{F})) = 6$.
\end{itemize}
\end{exam.}
\begin{proof}
Suppose the assumption of (a) holds. Without loss of generality, we may assume that all the vectors $\mathcal{F}_{i,[i,j]}$ are generated by $\mathcal{F}_{1,[1,2]} \neq [ 0 \ 0 \ 0 ]$. Especially, say $\mathcal{F}_{1,[1,3]} = \lambda \cdot \mathcal{F}_{1,[1,2]}$ and $\mathcal{F}_{1,[1,4]} = \mu \cdot \mathcal{F}_{1,[1,2]}$ for some $\lambda, \mu \in \mathbb{R}$. In particular, by column operations,
\begin{equation*}
\mathbf{C} \longrightarrow
\left[\begin{array}{cccc}
\mathbf{0} & \mathcal{F}_{1, [1,2]} & \mathbf{0} & \mathbf{0} \\
\mathbf{0} & \mathbf{0} & \lambda \cdot \mathcal{F}_{1, [1,2]} & \mathbf{0} \\
\mathbf{0} & \mathbf{0} & \mathbf{0} & \mu \cdot \mathcal{F}_{1, [1,2]} \\
\hline
\mathbf{0} & -(\lambda - 1) \cdot \mathcal{F}_{1, [1,2]} & (\lambda - 1) \cdot \mathcal{F}_{1, [1,2]} & \mathbf{0} \\
\mathbf{0} & -(\mu - 1) \cdot \mathcal{F}_{1, [1,2]} & \mathbf{0} & (\mu - 1) \cdot \mathcal{F}_{1, [1,2]} \\
\hline
\mathbf{0} & \mathbf{0} & -(\mu - \lambda) \cdot \mathcal{F}_{1, [1,2]} & (\mu - \lambda) \cdot \mathcal{F}_{1, [1,2]} \\
\end{array} \right].
\end{equation*}
Evidently, the first three rows of the right-hand-side matrix are linearly independent and generate all the rows. This shows that ${\rm rank}(\mathbf{C}) = 3$. By Proposition \ref{Proposition: Dimension of the global section space of an anisotropic sheaf}, $\dim_\mathbb{R}(\ker(L_\mathcal{F})) = 3 \times 4 - 3 = 9$.

Suppose the assumption of (b) holds. Without loss of generality, we may assume that the collection $\{ \mathcal{F}_{1, [1,2]}, \mathcal{F}_{1, [1,3]} \}$ generates all vectors $\mathcal{F}_{i,[i,j]}$ with $i < j$. Say $\mathcal{F}_{1,[1,4]} = \alpha \mathcal{F}_{1,[1,2]} + \beta \mathcal{F}_{1,[1,3]}$ for some $\alpha, \beta \in \mathbb{R}$. As in the previous modification to the matrix $\mathbf{C}$, the column and row operations result in the following procedure:
\begin{equation*}
\begin{split}
\mathbf{C} &\longrightarrow  
\left[\begin{array}{cccc}
\mathbf{0} & \mathcal{F}_{1, [1,2]} & \mathbf{0} & \mathbf{0} \\
\mathbf{0} & \mathbf{0} & \mathcal{F}_{1, [1,3]} & \mathbf{0} \\
\mathbf{0} & \mathbf{0} & \mathbf{0} & \mathcal{F}_{1, [1,4]} \\
\hline
\mathbf{0} & -\mathcal{F}_{2, [2,3]} & \mathcal{F}_{2, [2,3]} & \mathbf{0} \\
\mathbf{0} & -\mathcal{F}_{2, [2,4]} & \mathbf{0} & \mathcal{F}_{2, [2,4]} \\
\hline
\mathbf{0} & \mathbf{0} & -\mathcal{F}_{3, [3,4]} & \mathcal{F}_{3, [3,4]} \\
\end{array} \right] \longrightarrow
\left[\begin{array}{cccc}
\mathbf{0} & \mathcal{F}_{1, [1,2]} & \mathbf{0} & \mathbf{0} \\
\mathbf{0} & \mathbf{0} & \mathcal{F}_{1, [1,3]} & \mathbf{0} \\
\mathbf{0} & \mathbf{0} & \mathbf{0} & \alpha \mathcal{F}_{1,[1,2]} + \beta \mathcal{F}_{1,[1,3]} \\
\hline
\mathbf{0} & -\mathcal{F}_{1, [1,3]} & -\mathcal{F}_{1, [1,2]} & \mathbf{0} \\
\mathbf{0} & -\mathcal{F}_{1, [1,4]} & \mathbf{0} & -\mathcal{F}_{1, [1,2]} \\
\hline
\mathbf{0} & \mathbf{0} & -\mathcal{F}_{1, [1,4]} & -\mathcal{F}_{1, [1,3]} \\
\end{array} \right] \\
&\longrightarrow \left[\begin{array}{cccc}
\mathbf{0} & \mathcal{F}_{1, [1,2]} & \mathbf{0} & \mathbf{0} \\
\mathbf{0} & \mathbf{0} & \mathcal{F}_{1, [1,3]} & \mathbf{0} \\
\mathbf{0} & -\alpha \cdot (\alpha \mathcal{F}_{1,[1,2]} + \beta \mathcal{F}_{1,[1,3]}) & -\beta\cdot (\alpha \mathcal{F}_{1,[1,2]} + \beta \mathcal{F}_{1,[1,3]}) & \mathbf{0} \\
\hline
\mathbf{0} & -\mathcal{F}_{1, [1,3]} & -\mathcal{F}_{1, [1,2]} & \mathbf{0} \\
\mathbf{0} & -\mathcal{F}_{1, [1,4]} & \mathbf{0} & -\mathcal{F}_{1, [1,2]} \\
\hline
\mathbf{0} & \mathbf{0} & -\mathcal{F}_{1, [1,4]} & -\mathcal{F}_{1, [1,3]} \\
\end{array} \right] \\
&\longrightarrow
\left[\begin{array}{cccc}
\mathbf{0} & \mathcal{F}_{1, [1,2]} & \mathbf{0} & \mathbf{0} \\
\mathbf{0} & \mathbf{0} & \mathcal{F}_{1, [1,3]} & \mathbf{0} \\
\mathbf{0} & -\alpha\beta\mathcal{F}_{1,[1,3]} & -\alpha\beta\mathcal{F}_{1,[1,2]} & \mathbf{0} \\
\hline
\mathbf{0} & -\mathcal{F}_{1, [1,3]} & -\mathcal{F}_{1, [1,2]} & \mathbf{0} \\
\mathbf{0} & -\mathcal{F}_{1, [1,4]} & \mathbf{0} & -\mathcal{F}_{1, [1,2]} \\
\hline
\mathbf{0} & \mathbf{0} & -\mathcal{F}_{1, [1,4]} & -\mathcal{F}_{1, [1,3]} \\
\end{array} \right] \longrightarrow
\left[\begin{array}{cccc}
\mathbf{0} & \mathcal{F}_{1, [1,2]} & \mathbf{0} & \mathbf{0} \\
\mathbf{0} & \mathbf{0} & \mathcal{F}_{1, [1,3]} & \mathbf{0} \\
\mathbf{0} & \mathbf{0} & \mathbf{0} & \mathbf{0} \\
\hline
\mathbf{0} & -\mathcal{F}_{1, [1,3]} & -\mathcal{F}_{1, [1,2]} & \mathbf{0} \\
\mathbf{0} & -\mathcal{F}_{1, [1,4]} & \mathbf{0} & -\mathcal{F}_{1, [1,2]} \\
\hline
\mathbf{0} & \mathbf{0} & -\mathcal{F}_{1, [1,4]} & -\mathcal{F}_{1, [1,3]} \\
\end{array} \right]. \\
\end{split}  
\end{equation*}
Because $\mathcal{F}_{1, [1,2]}$ and $\mathcal{F}_{1, [1,3]}$ are linearly independent row vectors, the matrix $\mathbf{C}$ has rank $5$. In particular, the dimension of the kernel of $L_\mathcal{F}$ is $\dim_\mathbb{R}(\ker(L_\mathcal{F})) = 3 \cdot 4 - 5 = 7$.

Finally, suppose the assumption of (c) holds. Because the rank of $\mathcal{F}$ equals 3, the sets $\{ \mathcal{F}_{1, [1,2]},  \mathcal{F}_{1, [1,3]},  \mathcal{F}_{1, [1,4]} \}$, $\{ \mathcal{F}_{2, [2,3]},  \mathcal{F}_{2, [2,4]} \}$, and $\{\mathcal{F}_{3, [3,4]} \}$ depicted in~\eqref{Eq. Column and Row operations for K4-1} are sets of linearly independent row vectors. This shows that ${\rm rank}(\mathbf{C}) = {\rm rank}(\widetilde{\mathbf{C}}) = 6$ and $\dim_\mathbb{R}(L_\mathcal{F}) = 6$.
\end{proof}
\begin{exam.}
\label{Example: K5 example}
Let $G = K_5$ be the complete graph with vertex set $V = \{ 1,2,3,4,5\}$ accompanied with 3D coordinates, and let $\mathcal{F}: (G,\leq) \rightarrow \textup{\textsf{Vect}}_\mathbb{R}$ be the anisotropic sheaf defined in Equation~\eqref{Eq. Anisotropic sheaf definition-form 2}, then the coboundary matrix $\mathbf{C}: C^0(G,\mathcal{F}) \rightarrow C^1(G,\mathcal{F})$ is a $10 \times 15$ matrix represented by
\begin{equation*}
\mathbf{C} = \left[\begin{array}{ccccc}
-\mathcal{F}_{1, [1,2]} & \mathcal{F}_{1, [1,2]} & \mathbf{0} & \mathbf{0} & \mathbf{0}  \\
-\mathcal{F}_{1, [1,3]} & \mathbf{0} & \mathcal{F}_{1, [1,3]} & \mathbf{0} & \mathbf{0}  \\
-\mathcal{F}_{1, [1,4]} & \mathbf{0} & \mathbf{0} & \mathcal{F}_{1, [1,4]} & \mathbf{0}  \\
-\mathcal{F}_{1, [1,5]} & \mathbf{0} & \mathbf{0} & \mathbf{0}  & \mathcal{F}_{1, [1,5]} \\
\hline
\mathbf{0} & -\mathcal{F}_{2, [2,3]} & \mathcal{F}_{2, [2,3]} & \mathbf{0} & \mathbf{0}  \\
\mathbf{0} & -\mathcal{F}_{2, [2,4]} & \mathbf{0} & \mathcal{F}_{2, [2,4]} & \mathbf{0}  \\
\mathbf{0} & -\mathcal{F}_{2, [2,5]} & \mathbf{0} & \mathbf{0} & \mathcal{F}_{2, [2,5]} \\
\hline
\mathbf{0} & \mathbf{0} & -\mathcal{F}_{3, [3,4]} & \mathcal{F}_{3, [3,4]} & \mathbf{0}  \\
\mathbf{0} & \mathbf{0} & -\mathcal{F}_{3, [3,5]} & \mathbf{0} & \mathcal{F}_{3, [3,5]}  \\
\hline
\mathbf{0} & \mathbf{0} & \mathbf{0} & -\mathcal{F}_{4, [4,5]} & \mathcal{F}_{4, [4,5]}  \\
\end{array} \right].   
\end{equation*}
In particular, ${\rm rank}(\mathbf{C}) \leq 10$ and hence $\dim_\mathbb{R}(\ker(L_\mathcal{F})) = 3 \cdot 5 - {\rm rank}(\mathbf{C}) \geq 5$. Furthermore, matrix $\mathbf{C}$ contains the following $6 \times 12$ matrix $\mathbf{D}$ that corresponds to the $0$-th coboundary matrix of the restriction sheaf $\mathcal{F}|_H$, where $H$ is the complete graph with vertices $1, 2, 3$, and $4$:
\begin{equation*}
\mathbf{D} = \left[\begin{array}{ccccc}
-\mathcal{F}_{1, [1,2]} & \mathcal{F}_{1, [1,2]} & \mathbf{0} & \mathbf{0} \\
-\mathcal{F}_{1, [1,3]} & \mathbf{0} & \mathcal{F}_{1, [1,3]} & \mathbf{0} \\
-\mathcal{F}_{1, [1,4]} & \mathbf{0} & \mathbf{0} & \mathcal{F}_{1, [1,4]} \\
\hline
\mathbf{0} & -\mathcal{F}_{2, [2,3]} & \mathcal{F}_{2, [2,3]} & \mathbf{0} \\
\mathbf{0} & -\mathcal{F}_{2, [2,4]} & \mathbf{0} & \mathcal{F}_{2, [2,4]} \\
\hline
\mathbf{0} & \mathbf{0} & -\mathcal{F}_{3, [3,4]} & \mathcal{F}_{3, [3,4]} \\
\end{array} \right].   
\end{equation*}
More precisely, matrix $\mathbf{D}$ is defined by collecting the $1$-st, $2$-nd, $3$-th, $5$-th, $6$-th, and $7$-th rows of matrix $\mathbf{C}$ and reduce the last column of $\mathbf{C}$. Furthermore, suppose $\mathcal{F}_{1,[1,5]} = \alpha_2 \mathcal{F}_{1,[1,2]} + \alpha_3 \mathcal{F}_{1,[1,3]} + \alpha_4 \mathcal{F}_{1,[1,4]}$ with real numbers $\alpha_2, \alpha_3, \alpha_4 \in \mathbb{R}$, then ${\rm rank}(\mathbf{C}) \leq {\rm rank}(\mathbf{D}) + 3$ with equality if $\mathcal{F}_{1, [1,2]}, \mathcal{F}_{1, [1,3]}, \mathcal{F}_{1, [1,4]}$ are linearly independent.
\end{exam.}
\begin{proof}
By applying column operations with block matrices, the coboundary matrix $\mathbf{C}$ becomes
\begin{equation*}
\widetilde{\mathbf{C}} = \left[\begin{array}{ccccc}
\mathbf{0} & \mathcal{F}_{1, [1,2]} & \mathbf{0} & \mathbf{0} & \mathbf{0}  \\
\mathbf{0} & \mathbf{0} & \mathcal{F}_{1, [1,3]} & \mathbf{0} & \mathbf{0}  \\
\mathbf{0} & \mathbf{0} & \mathbf{0} & \mathcal{F}_{1, [1,4]} & \mathbf{0}  \\
\mathbf{0} & \mathbf{0} & \mathbf{0} & \mathbf{0}  & \mathcal{F}_{1, [1,5]} \\
\hline
\mathbf{0} & -\mathcal{F}_{2, [2,3]} & \mathcal{F}_{2, [2,3]} & \mathbf{0} & \mathbf{0}  \\
\mathbf{0} & -\mathcal{F}_{2, [2,4]} & \mathbf{0} & \mathcal{F}_{2, [2,4]} & \mathbf{0}  \\
\mathbf{0} & -\mathcal{F}_{2, [2,5]} & \mathbf{0} & \mathbf{0} & \mathcal{F}_{2, [2,5]} \\
\hline
\mathbf{0} & \mathbf{0} & -\mathcal{F}_{3, [3,4]} & \mathcal{F}_{3, [3,4]} & \mathbf{0}  \\
\mathbf{0} & \mathbf{0} & -\mathcal{F}_{3, [3,5]} & \mathbf{0} & \mathcal{F}_{3, [3,5]}  \\
\hline
\mathbf{0} & \mathbf{0} & \mathbf{0} & -\mathcal{F}_{4, [4,5]} & \mathcal{F}_{4, [4,5]}  \\
\end{array} \right]
\end{equation*}
with ${\rm rank}(\mathbf{C}) = {\rm rank}(\widetilde{\mathbf{C}})$. These column operations also affect the matrix $\mathbf{D}$ as
\begin{equation*}
\widetilde{\mathbf{D}} = \left[\begin{array}{ccccc}
\mathbf{0} & \mathcal{F}_{1, [1,2]} & \mathbf{0} & \mathbf{0} \\
\mathbf{0} & \mathbf{0} & \mathcal{F}_{1, [1,3]} & \mathbf{0} \\
\mathbf{0} & \mathbf{0} & \mathbf{0} & \mathcal{F}_{1, [1,4]} \\
\hline
\mathbf{0} & -\mathcal{F}_{2, [2,3]} & \mathcal{F}_{2, [2,3]} & \mathbf{0} \\
\mathbf{0} & -\mathcal{F}_{2, [2,4]} & \mathbf{0} & \mathcal{F}_{2, [2,4]} \\
\hline
\mathbf{0} & \mathbf{0} & -\mathcal{F}_{3, [3,4]} & \mathcal{F}_{3, [3,4]} \\
\end{array} \right].   
\end{equation*}
Similarly, we have ${\rm rank}(\mathbf{D}) = {\rm rank}(\widetilde{\mathbf{D}})$. In particular, based on the linear relation described in~\eqref{Eq. ij vector spanned by 1j and 1i-simple}, the matrix $\widetilde{\mathbf{C}}$ can be written by
\begin{equation*}
\begin{split}
\widetilde{\mathbf{C}} &= \left[\begin{array}{ccccc}
\mathbf{0} & \mathcal{F}_{1, [1,2]} & \mathbf{0} & \mathbf{0} & \mathbf{0}  \\
\mathbf{0} & \mathbf{0} & \mathcal{F}_{1, [1,3]} & \mathbf{0} & \mathbf{0}  \\
\mathbf{0} & \mathbf{0} & \mathbf{0} & \mathcal{F}_{1, [1,4]} & \mathbf{0}  \\
\mathbf{0} & \mathbf{0} & \mathbf{0} & \mathbf{0}  & \mathcal{F}_{1, [1,5]} \\
\hline
\mathbf{0} & -(\mathcal{F}_{1, [1,3]} - \mathcal{F}_{1, [1,2]}) & \mathcal{F}_{1, [1,3]} - \mathcal{F}_{1, [1,2]} & \mathbf{0} & \mathbf{0}  \\
\mathbf{0} & -(\mathcal{F}_{1, [1,4]} - \mathcal{F}_{1, [1,2]}) & \mathbf{0} & \mathcal{F}_{1, [1,4]} - \mathcal{F}_{1, [1,2]} & \mathbf{0}  \\
\mathbf{0} & -(\mathcal{F}_{1, [1,5]} - \mathcal{F}_{1, [1,2]}) & \mathbf{0} & \mathbf{0} & \mathcal{F}_{1, [1,5]} - \mathcal{F}_{1, [1,2]} \\
\hline
\mathbf{0} & \mathbf{0} & -(\mathcal{F}_{1, [1,4]} - \mathcal{F}_{1, [1,3]}) & \mathcal{F}_{1, [1,4]} - \mathcal{F}_{1, [1,3]} & \mathbf{0}  \\
\mathbf{0} & \mathbf{0} & -(\mathcal{F}_{1, [1,5]} - \mathcal{F}_{1, [1,3]}) & \mathbf{0} & \mathcal{F}_{1, [1,5]} - \mathcal{F}_{1, [1,3]} \\
\hline
\mathbf{0} & \mathbf{0} & \mathbf{0} & -(\mathcal{F}_{1, [1,5]} - \mathcal{F}_{1, [1,4]}) & \mathcal{F}_{1, [1,5]} - \mathcal{F}_{1, [1,4]}  \\
\end{array} \right].
\end{split}    
\end{equation*}
Next, the following process is validated by performing row operations for elimination.
\begin{equation*}
\begin{split}
\widetilde{\mathbf{C}} &\longrightarrow \left[\begin{array}{ccccc}
\mathbf{0} & \mathcal{F}_{1, [1,2]} & \mathbf{0} & \mathbf{0} & \mathbf{0}  \\
\mathbf{0} & \mathbf{0} & \mathcal{F}_{1, [1,3]} & \mathbf{0} & \mathbf{0}  \\
\mathbf{0} & \mathbf{0} & \mathbf{0} & \mathcal{F}_{1, [1,4]} & \mathbf{0}  \\
\mathbf{0} & \mathbf{0} & \mathbf{0} & \mathbf{0}  & \mathcal{F}_{1, [1,5]} \\
\hline
\mathbf{0} & -\mathcal{F}_{1, [1,3]} & - \mathcal{F}_{1, [1,2]} & \mathbf{0} & \mathbf{0}  \\
\mathbf{0} & -\mathcal{F}_{1, [1,4]} & \mathbf{0} & - \mathcal{F}_{1, [1,2]} & \mathbf{0}  \\
\mathbf{0} & -\mathcal{F}_{1, [1,5]} & \mathbf{0} & \mathbf{0} & - \mathcal{F}_{1, [1,2]} \\
\hline
\mathbf{0} & \mathbf{0} & -\mathcal{F}_{1, [1,4]} & - \mathcal{F}_{1, [1,3]} & \mathbf{0}  \\
\mathbf{0} & \mathbf{0} & -\mathcal{F}_{1, [1,5]} & \mathbf{0} &  - \mathcal{F}_{1, [1,3]} \\
\hline
\mathbf{0} & \mathbf{0} & \mathbf{0} & -\mathcal{F}_{1, [1,5]} & - \mathcal{F}_{1, [1,4]}  \\
\end{array} \right]\\
&\longrightarrow \left[\begin{array}{ccccc}
\mathbf{0} & \mathcal{F}_{1, [1,2]} & \mathbf{0} & \mathbf{0} & \mathbf{0}  \\
\mathbf{0} & \mathbf{0} & \mathcal{F}_{1, [1,3]} & \mathbf{0} & \mathbf{0}  \\
\mathbf{0} & \mathbf{0} & \mathbf{0} & \mathcal{F}_{1, [1,4]} & \mathbf{0}  \\
\mathbf{0} & -\alpha_2 \cdot \sum_{i = 2}^4 \alpha_i \mathcal{F}_{1,[1,i]} & -\alpha_3 \cdot \sum_{i = 2}^4 \alpha_i \mathcal{F}_{1,[1,i]} & -\alpha_4 \cdot \sum_{i = 2}^4 \alpha_i \mathcal{F}_{1,[1,i]}  & \mathbf{0} \\
\hline
\mathbf{0} & -\mathcal{F}_{1, [1,3]} & - \mathcal{F}_{1, [1,2]} & \mathbf{0} & \mathbf{0}  \\
\mathbf{0} & -\mathcal{F}_{1, [1,4]} & \mathbf{0} & - \mathcal{F}_{1, [1,2]} & \mathbf{0}  \\
\mathbf{0} & -\sum_{i = 2}^4 \alpha_i \mathcal{F}_{1,[1,i]} & \mathbf{0} & \mathbf{0} & - \mathcal{F}_{1, [1,2]} \\
\hline
\mathbf{0} & \mathbf{0} & -\mathcal{F}_{1, [1,4]} & - \mathcal{F}_{1, [1,3]} & \mathbf{0}  \\
\mathbf{0} & \mathbf{0} & -\sum_{i = 2}^4 \alpha_i \mathcal{F}_{1,[1,i]} & \mathbf{0} &  - \mathcal{F}_{1, [1,3]} \\
\hline
\mathbf{0} & \mathbf{0} & \mathbf{0} & -\sum_{i = 2}^4 \alpha_i \mathcal{F}_{1,[1,i]} & - \mathcal{F}_{1, [1,4]}  \\
\end{array} \right] = \overline{\mathbf{C}}.
\end{split}    
\end{equation*}
Let $\overline{\mathbf{C}}_k$ with $k \in \{ 1,2, ..., 10 \}$ be the $k$-th rows of $\overline{\mathbf{C}}$. By adding the row vector combination
\begin{equation*}
\alpha_2^2 \overline{\mathbf{C}}_1 + \alpha_3^2 \overline{\mathbf{C}}_2 + \alpha_4^2 \overline{\mathbf{C}}_3 - \alpha_2\alpha_3 \overline{\mathbf{C}}_5 - \alpha_2\alpha_4 \overline{\mathbf{C}}_6 - \alpha_3\alpha_4 \overline{\mathbf{C}}_8 
\end{equation*}
to the $4$th row of $\overline{\mathbf{C}}$. Then, the matrix $\overline{\mathbf{C}}$ becomes
\begin{equation*}
\overline{\overline{\mathbf{C}}} = 
\left[\begin{array}{ccccc}
\mathbf{0} & \mathcal{F}_{1, [1,2]} & \mathbf{0} & \mathbf{0} & \mathbf{0}  \\
\mathbf{0} & \mathbf{0} & \mathcal{F}_{1, [1,3]} & \mathbf{0} & \mathbf{0}  \\
\mathbf{0} & \mathbf{0} & \mathbf{0} & \mathcal{F}_{1, [1,4]} & \mathbf{0}  \\
\mathbf{0} & \mathbf{0} & \mathbf{0} & \mathbf{0} & \mathbf{0} \\
\hline
\mathbf{0} & -\mathcal{F}_{1, [1,3]} & - \mathcal{F}_{1, [1,2]} & \mathbf{0} & \mathbf{0}  \\
\mathbf{0} & -\mathcal{F}_{1, [1,4]} & \mathbf{0} & - \mathcal{F}_{1, [1,2]} & \mathbf{0}  \\
\mathbf{0} & -\sum_{i = 2}^4 \alpha_i \mathcal{F}_{1,[1,i]} & \mathbf{0} & \mathbf{0} & - \mathcal{F}_{1, [1,2]} \\
\hline
\mathbf{0} & \mathbf{0} & -\mathcal{F}_{1, [1,4]} & - \mathcal{F}_{1, [1,3]} & \mathbf{0}  \\
\mathbf{0} & \mathbf{0} & -\sum_{i = 2}^4 \alpha_i \mathcal{F}_{1,[1,i]} & \mathbf{0} &  - \mathcal{F}_{1, [1,3]} \\
\hline
\mathbf{0} & \mathbf{0} & \mathbf{0} & -\sum_{i = 2}^4 \alpha_i \mathcal{F}_{1,[1,i]} & - \mathcal{F}_{1, [1,4]}  \\
\end{array} \right].    
\end{equation*}
During the row operation process, the submatrix $\widetilde{\mathbf{D}}$ of $\widetilde{\mathbf{C}}$ becomes
\begin{equation*}
\overline{\overline{\mathbf{D}}} = \left[\begin{array}{ccccc}
\mathbf{0} & \mathcal{F}_{1, [1,2]} & \mathbf{0} & \mathbf{0} \\
\mathbf{0} & \mathbf{0} & \mathcal{F}_{1, [1,3]} & \mathbf{0} \\
\mathbf{0} & \mathbf{0} & \mathbf{0} & \mathcal{F}_{1, [1,4]} \\
\hline
\mathbf{0} & -\mathcal{F}_{1, [1,3]} & -\mathcal{F}_{1, [1,2]} & \mathbf{0} \\
\mathbf{0} & -\mathcal{F}_{1, [1,4]} & \mathbf{0} & -\mathcal{F}_{1, [1,2]} \\
\hline
\mathbf{0} & \mathbf{0} & -\mathcal{F}_{1, [1,4]} & -\mathcal{F}_{1, [1,3]} \\
\end{array} \right],   
\end{equation*}
corresponding to the $7$-th, $9$-th, and $10$-th rows of $\overline{\overline{\mathbf{C}}}$. In particular, matrices $\overline{\overline{\mathbf{D}}}$ and $\mathbf{D}$ have the same rank. Because rows $\overline{\overline{\mathbf{D}}}_7, \overline{\overline{\mathbf{D}}}_9$, and $\overline{\overline{\mathbf{D}}}_{10}$ are all non-zero rows that do not correspond to the rows in $\overline{\overline{\mathbf{D}}}$, we deduce that
\begin{equation*}
{\rm rank}(\mathbf{C}) = {\rm rank}(\overline{\overline{\mathbf{C}}}) \leq {\rm rank}(\overline{\overline{\mathbf{D}}}) + 3 = {\rm rank}(\mathbf{D}) + 3.
\end{equation*}
In addition, if the $1 \times 3$ row vectors $\mathcal{F}_{1, [1,2]}, \mathcal{F}_{1, [1,3]}, \mathcal{F}_{1, [1,4]}$ are linearly independent, then ${\rm rank}(\mathbf{C}) = {\rm rank}(\mathbf{D}) + 3$.
\end{proof}
Next, in the following example, we further explore the anisotropic sheaf defined on $K_6$. Especially, we will build on the results from Example \ref{Example: K5 example} to establish that ${\rm rank}(\mathbf{C}) = {\rm rank}(\mathbf{D}) + 3 = 9 + 3 = 12$ for the anisotropic sheaf of rank $3$ on $K_6$. Moreover, we will show that by systematically applying the methods outlined in Examples \ref{Example: K5 example} and \ref{Example: K6 example}, similar rank properties are preserved for rank-$3$ sheaves $\mathcal{F}: (K_n,\leq) \rightarrow \textup{\textsf{Vect}}_\mathbb{R}$. Specifically, based on the demonstrations in Examples \ref{Example: K5 example} and \ref{Example: K6 example}, we can confirm that ${\rm rank}(\mathbf{C}) = 3 \cdot (|V| - 2) = 3 \cdot (n - 2) = 3n - 6$. Notably, $\dim_\mathbb{R}(L_\mathcal{F}) = 3 \cdot |V| - {\rm rank}(\mathbf{C}) = 3n - (3n - 6) = 6$ remains a constant.
\begin{exam.}
\label{Example: K6 example}
Let $G = K_6$ be the complete graph with vertex set $V = \{ 1,2,3,4,5,6\}$ accompanied with 3D coordinates, and let $\mathcal{F}: (G,\leq) \rightarrow \textup{\textsf{Vect}}_\mathbb{R}$ be the anisotropic sheaf defined in Equation~\eqref{Eq. Anisotropic sheaf definition-form 2}, then the coboundary matrix $\mathbf{C}: C^0(G,\mathcal{F}) \rightarrow C^1(G,\mathcal{F})$ is a $15 \times 18$ matrix represented by
\begin{equation*}
\mathbf{C} = \left[\begin{array}{cccccc}
-\mathcal{F}_{1, [1,2]} & \mathcal{F}_{1, [1,2]} & \mathbf{0} & \mathbf{0} & \mathbf{0}  & \mathbf{0}  \\
-\mathcal{F}_{1, [1,3]} & \mathbf{0} & \mathcal{F}_{1, [1,3]} & \mathbf{0} & \mathbf{0}  & \mathbf{0}  \\
-\mathcal{F}_{1, [1,4]} & \mathbf{0} & \mathbf{0} & \mathcal{F}_{1, [1,4]} & \mathbf{0}  & \mathbf{0}  \\
-\mathcal{F}_{1, [1,5]} & \mathbf{0} & \mathbf{0} & \mathbf{0}  & \mathcal{F}_{1, [1,5]} & \mathbf{0}  \\
-\mathcal{F}_{1, [1,6]} & \mathbf{0} & \mathbf{0} & \mathbf{0}  & \mathbf{0} & \mathcal{F}_{1, [1,6]}  \\
\hline
\mathbf{0} & -\mathcal{F}_{2, [2,3]} & \mathcal{F}_{2, [2,3]} & \mathbf{0} & \mathbf{0} & \mathbf{0} \\
\mathbf{0} & -\mathcal{F}_{2, [2,4]} & \mathbf{0} & \mathcal{F}_{2, [2,4]} & \mathbf{0} & \mathbf{0} \\
\mathbf{0} & -\mathcal{F}_{2, [2,5]} & \mathbf{0} & \mathbf{0} & \mathcal{F}_{2, [2,5]} & \mathbf{0} \\
\mathbf{0} & -\mathcal{F}_{2, [2,6]} & \mathbf{0} & \mathbf{0} & \mathbf{0} & \mathcal{F}_{2, [2,6]} \\
\hline
\mathbf{0} & \mathbf{0} & -\mathcal{F}_{3, [3,4]} & \mathcal{F}_{3, [3,4]} & \mathbf{0} & \mathbf{0} \\
\mathbf{0} & \mathbf{0} & -\mathcal{F}_{3, [3,5]} & \mathbf{0} & \mathcal{F}_{3, [3,5]} & \mathbf{0} \\
\mathbf{0} & \mathbf{0} & -\mathcal{F}_{3, [3,6]} & \mathbf{0} & \mathbf{0} & \mathcal{F}_{3, [3,6]} \\
\hline
\mathbf{0} & \mathbf{0} & \mathbf{0} & -\mathcal{F}_{4, [4,5]} & \mathcal{F}_{4, [4,5]} & \mathbf{0} \\
\mathbf{0} & \mathbf{0} & \mathbf{0} & -\mathcal{F}_{4, [4,6]} & \mathbf{0} & \mathcal{F}_{4, [4,6]} \\
\hline
\mathbf{0} & \mathbf{0} & \mathbf{0} & \mathbf{0} & -\mathcal{F}_{5, [5,6]} & \mathcal{F}_{5, [5,6]} \\
\end{array}\right].   
\end{equation*}
If ${\rm rank}(\mathcal{F}) = 3$, then ${\rm rank}(\mathbf{C}) = {\rm rank}(\mathbf{D}) + 3 = 9 + 3 = 12$, where
\begin{equation*}
\mathbf{D} = \left[\begin{array}{ccccc}
-\mathcal{F}_{1, [1,2]} & \mathcal{F}_{1, [1,2]} & \mathbf{0} & \mathbf{0} & \mathbf{0} \\
-\mathcal{F}_{1, [1,3]} & \mathbf{0} & \mathcal{F}_{1, [1,3]} & \mathbf{0} & \mathbf{0} \\
-\mathcal{F}_{1, [1,4]} & \mathbf{0} & \mathbf{0} & \mathcal{F}_{1, [1,4]} & \mathbf{0} \\
-\mathcal{F}_{1, [1,5]} & \mathbf{0} & \mathbf{0} & \mathbf{0}  & \mathcal{F}_{1, [1,5]} \\
\hline
\mathbf{0} & -\mathcal{F}_{2, [2,3]} & \mathcal{F}_{2, [2,3]} & \mathbf{0} & \mathbf{0} \\
\mathbf{0} & -\mathcal{F}_{2, [2,4]} & \mathbf{0} & \mathcal{F}_{2, [2,4]} & \mathbf{0} \\
\mathbf{0} & -\mathcal{F}_{2, [2,5]} & \mathbf{0} & \mathbf{0} & \mathcal{F}_{2, [2,5]} \\
\hline
\mathbf{0} & \mathbf{0} & -\mathcal{F}_{3, [3,4]} & \mathcal{F}_{3, [3,4]} & \mathbf{0} \\
\mathbf{0} & \mathbf{0} & -\mathcal{F}_{3, [3,5]} & \mathbf{0} & \mathcal{F}_{3, [3,5]} \\
\hline
\mathbf{0} & \mathbf{0} & \mathbf{0} & -\mathcal{F}_{4, [4,5]} & \mathcal{F}_{4, [4,5]} \\
\end{array}\right]
\end{equation*}
consists rows that correspond to rows $\mathbf{C}_1, \mathbf{C}_2, \mathbf{C}_3, \mathbf{C}_4, \mathbf{C}_6, \mathbf{C}_7, \mathbf{C}_8, \mathbf{C}_{10}, \mathbf{C}_{11}, \mathbf{C}_{13}$ of $\mathbf{C}$. In particular, we have $\dim_\mathbb{R}(L_\mathcal{F}) = 3 \cdot |V| - {\rm rank}(\mathbf{C}) = 3 \cdot 6 - 12 = 6$.
\end{exam.}
\begin{proof}
By assumption, the rank of $\mathcal{F}$ is $3$. Because $G = K_6$ is complete, all row vectors $\mathcal{F}_{i,[i,j]}$ can be spanned by the set $\{ \mathcal{F}_{1,[1,n]} \ | \ n = 2, ..., 6 \}$ of row vectors. Namely, all edges can be uniquely represented by the ordered sequence
\begin{equation*}
\overbrace{[v_1, v_2], [v_1, v_3], ..., [v_1, v_n]}^{n-1 \ \text{edges}}, \overbrace{[v_2, v_3], [v_2, v_4], ..., [v_2, v_n]}^{n-2 \ \text{edges}}, ..., \overbrace{[v_{n-1}, v_n]}^{1 \ \text{edge}}.    
\end{equation*}
Without loss of generality, we may assume that $\mathcal{F}_{1, [1,2]}, \mathcal{F}_{1, [1,3]}, \mathcal{F}_{1, [1,4]}$ are linearly independent. In particular, all row vectors $\mathcal{F}_{i,[i,j]}$ can be spanned by the set $\{ \mathcal{F}_{1, [1,2]}, \mathcal{F}_{1, [1,3]}, \mathcal{F}_{1, [1,4]} \}$.  Similar to the proof in Example \ref{Example: K5 example}, we employ column and row operations on $\mathbf{C}$, leading to the matrix    
\begin{equation*}
\overline{\mathbf{C}} = \left[\begin{array}{cccccc}
\mathbf{0} & \mathcal{F}_{1, [1,2]} & \mathbf{0} & \mathbf{0} & \mathbf{0}  & \mathbf{0}  \\
\mathbf{0} & \mathbf{0} & \mathcal{F}_{1, [1,3]} & \mathbf{0} & \mathbf{0}  & \mathbf{0}  \\
\mathbf{0} & \mathbf{0} & \mathbf{0} & \mathcal{F}_{1, [1,4]} & \mathbf{0}  & \mathbf{0}  \\
\mathbf{0} & \mathbf{0} & \mathbf{0} & \mathbf{0}  & \mathcal{F}_{1, [1,5]} & \mathbf{0}  \\
\mathbf{0} & \mathbf{0} & \mathbf{0} & \mathbf{0}  & \mathbf{0} & \mathcal{F}_{1, [1,6]}  \\
\hline
\mathbf{0} & -\mathcal{F}_{1, [1,3]} & -\mathcal{F}_{1, [1,2]} & \mathbf{0} & \mathbf{0} & \mathbf{0} \\
\mathbf{0} & -\mathcal{F}_{1, [1,4]} & \mathbf{0} & -\mathcal{F}_{1, [1,2]} & \mathbf{0} & \mathbf{0} \\
\mathbf{0} & -\mathcal{F}_{1, [1,5]} & \mathbf{0} & \mathbf{0} & -\mathcal{F}_{1, [1,2]} & \mathbf{0} \\
\mathbf{0} & -\mathcal{F}_{1, [1,6]} & \mathbf{0} & \mathbf{0} & \mathbf{0} & -\mathcal{F}_{1, [1,2]} \\
\hline
\mathbf{0} & \mathbf{0} & -\mathcal{F}_{1, [1,4]} & -\mathcal{F}_{1, [1,3]} & \mathbf{0} & \mathbf{0} \\
\mathbf{0} & \mathbf{0} & -\mathcal{F}_{1, [1,5]} & \mathbf{0} & -\mathcal{F}_{1, [1,3]} & \mathbf{0} \\
\mathbf{0} & \mathbf{0} & -\mathcal{F}_{1, [1,6]} & \mathbf{0} & \mathbf{0} & -\mathcal{F}_{1, [1,3]} \\
\hline
\mathbf{0} & \mathbf{0} & \mathbf{0} & -\mathcal{F}_{1, [1,5]} & -\mathcal{F}_{1, [1,4]} & \mathbf{0} \\
\mathbf{0} & \mathbf{0} & \mathbf{0} & -\mathcal{F}_{1, [1,6]} & \mathbf{0} & -\mathcal{F}_{1, [1,4]} \\
\hline
\mathbf{0} & \mathbf{0} & \mathbf{0} & \mathbf{0} & -\mathcal{F}_{1, [1,6]} & -\mathcal{F}_{1, [1,5]} \\
\end{array} \right].    
\end{equation*}
Moreover, based on the same column and row operations, the submatrix $\mathbf{D}$ becomes
\begin{equation*}
\overline{\mathbf{D}} = \left[\begin{array}{cccccc}
\mathbf{0} & \mathcal{F}_{1, [1,2]} & \mathbf{0} & \mathbf{0} & \mathbf{0} \\
\mathbf{0} & \mathbf{0} & \mathcal{F}_{1, [1,3]} & \mathbf{0} & \mathbf{0} \\
\mathbf{0} & \mathbf{0} & \mathbf{0} & \mathcal{F}_{1, [1,4]} & \mathbf{0} \\
\mathbf{0} & \mathbf{0} & \mathbf{0} & \mathbf{0}  & \mathcal{F}_{1, [1,5]} \\
\hline
\mathbf{0} & -\mathcal{F}_{1, [1,3]} & -\mathcal{F}_{1, [1,2]} & \mathbf{0} & \mathbf{0} \\
\mathbf{0} & -\mathcal{F}_{1, [1,4]} & \mathbf{0} & -\mathcal{F}_{1, [1,2]} & \mathbf{0} \\
\mathbf{0} & -\mathcal{F}_{1, [1,5]} & \mathbf{0} & \mathbf{0} & -\mathcal{F}_{1, [1,2]} \\
\hline
\mathbf{0} & \mathbf{0} & -\mathcal{F}_{1, [1,4]} & -\mathcal{F}_{1, [1,3]} & \mathbf{0} \\
\mathbf{0} & \mathbf{0} & -\mathcal{F}_{1, [1,5]} & \mathbf{0} & -\mathcal{F}_{1, [1,3]} \\
\hline
\mathbf{0} & \mathbf{0} & \mathbf{0} & -\mathcal{F}_{1, [1,5]} & -\mathcal{F}_{1, [1,4]} \\
\end{array} \right]. 
\end{equation*}
Let $\overline{\mathbf{C}}_k$ with $k = 1,2, ..., 15$ be the rows of $\overline{\mathbf{C}}$. By the proof of Example \ref{Example: K5 example}, the row vector $\overline{\mathbf{C}}_5 = \begin{bmatrix}
\label{Eq. submatrix of Cbar in K6 case}
\mathbf{0} & \mathbf{0} & \mathbf{0} & \mathbf{0}  & \mathbf{0} & \mathcal{F}_{1, [1,6]} \\
\end{bmatrix}$ is generated by the row vectors of the submatrix of $\overline{\mathbf{C}}$:
\begin{equation}
\left[\begin{array}{cccccc}
\mathbf{0} & \mathcal{F}_{1, [1,2]} & \mathbf{0} & \mathbf{0} & \mathbf{0}  & \mathbf{0}  \\
\mathbf{0} & \mathbf{0} & \mathcal{F}_{1, [1,3]} & \mathbf{0} & \mathbf{0}  & \mathbf{0}  \\
\mathbf{0} & \mathbf{0} & \mathbf{0} & \mathcal{F}_{1, [1,4]} & \mathbf{0}  & \mathbf{0}  \\
\hline
\mathbf{0} & -\mathcal{F}_{1, [1,3]} & -\mathcal{F}_{1, [1,2]} & \mathbf{0} & \mathbf{0} & \mathbf{0} \\
\mathbf{0} & -\mathcal{F}_{1, [1,4]} & \mathbf{0} & -\mathcal{F}_{1, [1,2]} & \mathbf{0} & \mathbf{0} \\
\mathbf{0} & -\mathcal{F}_{1, [1,6]} & \mathbf{0} & \mathbf{0} & \mathbf{0} & -\mathcal{F}_{1, [1,2]} \\
\hline
\mathbf{0} & \mathbf{0} & -\mathcal{F}_{1, [1,4]} & -\mathcal{F}_{1, [1,3]} & \mathbf{0} & \mathbf{0} \\
\mathbf{0} & \mathbf{0} & -\mathcal{F}_{1, [1,6]} & \mathbf{0} & \mathbf{0} & -\mathcal{F}_{1, [1,3]} \\
\hline
\mathbf{0} & \mathbf{0} & \mathbf{0} & -\mathcal{F}_{1, [1,6]} & \mathbf{0} & -\mathcal{F}_{1, [1,4]} \\
\end{array} \right]. 
\end{equation}
Note that rows in the submatrix only involve linear maps $\mathcal{F}_{i,[i,j]}$ with $i < j$ in $\{ 1, 2, 3, 4, 5, 6 \} \setminus \{ 5 \}$. In particular, the row $\overline{\mathbf{C}}_5$ can be spanned by rows in $\overline{\mathbf{C}}$ that do not involve linear transformation $\mathcal{F}_{1,[1,5]}$. Because the set $\{ \mathcal{F}_{1, [1,2]}, \mathcal{F}_{1, [1,3]}, \mathcal{F}_{1, [1,4]} \}$ spans all the row vectors $\mathcal{F}_{i,[i,j]}$ with $i < j$, we may represent $\mathcal{F}_{1, [1,5]}$ and $\mathcal{F}_{1, [1,6]}$ by
\begin{equation*}
\mathcal{F}_{1,[1,5]} = \sum_{i = 1}^3 \alpha_i \mathcal{F}_{1,[1,i]} \text{ and } \mathcal{F}_{1,[1,6]} = \sum_{i = 1}^3 \beta_i \mathcal{F}_{1,[1,i]}.
\end{equation*}
By row operations, we obtain
\begin{equation*}
\begin{split}
\overline{\mathbf{C}} &\longrightarrow \left[\begin{array}{cccccc}
\mathbf{0} & \mathcal{F}_{1, [1,2]} & \mathbf{0} & \mathbf{0} & \mathbf{0}  & \mathbf{0}  \\
\mathbf{0} & \mathbf{0} & \mathcal{F}_{1, [1,3]} & \mathbf{0} & \mathbf{0}  & \mathbf{0}  \\
\mathbf{0} & \mathbf{0} & \mathbf{0} & \mathcal{F}_{1, [1,4]} & \mathbf{0}  & \mathbf{0}  \\
\mathbf{0} & \mathbf{0} & \mathbf{0} & \mathbf{0}  & \mathcal{F}_{1, [1,5]} & \mathbf{0}  \\
\mathbf{0} & \mathbf{0} & \mathbf{0} & \mathbf{0}  & \mathbf{0} & \mathbf{0}  \\
\hline
\mathbf{0} & -\mathcal{F}_{1, [1,3]} & -\mathcal{F}_{1, [1,2]} & \mathbf{0} & \mathbf{0} & \mathbf{0} \\
\mathbf{0} & -\mathcal{F}_{1, [1,4]} & \mathbf{0} & -\mathcal{F}_{1, [1,2]} & \mathbf{0} & \mathbf{0} \\
\mathbf{0} & -\mathcal{F}_{1, [1,5]} & \mathbf{0} & \mathbf{0} & -\mathcal{F}_{1, [1,2]} & \mathbf{0} \\
\mathbf{0} & -\mathcal{F}_{1, [1,6]} & \mathbf{0} & \mathbf{0} & \mathbf{0} & -\mathcal{F}_{1, [1,2]} \\
\hline
\mathbf{0} & \mathbf{0} & -\mathcal{F}_{1, [1,4]} & -\mathcal{F}_{1, [1,3]} & \mathbf{0} & \mathbf{0} \\
\mathbf{0} & \mathbf{0} & -\mathcal{F}_{1, [1,5]} & \mathbf{0} & -\mathcal{F}_{1, [1,3]} & \mathbf{0} \\
\mathbf{0} & \mathbf{0} & -\mathcal{F}_{1, [1,6]} & \mathbf{0} & \mathbf{0} & -\mathcal{F}_{1, [1,3]} \\
\hline
\mathbf{0} & \mathbf{0} & \mathbf{0} & -\mathcal{F}_{1, [1,5]} & -\mathcal{F}_{1, [1,4]} & \mathbf{0} \\
\mathbf{0} & \mathbf{0} & \mathbf{0} & -\mathcal{F}_{1, [1,6]} & \mathbf{0} & -\mathcal{F}_{1, [1,4]} \\
\hline
\mathbf{0} & \mathbf{0} & \mathbf{0} & \mathbf{0} & -\mathcal{F}_{1, [1,6]} & -\mathcal{F}_{1, [1,5]} \\
\end{array} \right] \\
&\longrightarrow \left[\begin{array}{cccccc}
\mathbf{0} & \mathcal{F}_{1, [1,2]} & \mathbf{0} & \mathbf{0} & \mathbf{0}  & \mathbf{0}  \\
\mathbf{0} & \mathbf{0} & \mathcal{F}_{1, [1,3]} & \mathbf{0} & \mathbf{0}  & \mathbf{0}  \\
\mathbf{0} & \mathbf{0} & \mathbf{0} & \mathcal{F}_{1, [1,4]} & \mathbf{0}  & \mathbf{0}  \\
\mathbf{0} & \mathbf{0} & \mathbf{0} & \mathbf{0}  & \mathcal{F}_{1, [1,5]} & \mathbf{0}  \\
\mathbf{0} & \mathbf{0} & \mathbf{0} & \mathbf{0}  & \mathbf{0} & \mathbf{0}  \\
\hline
\mathbf{0} & -\mathcal{F}_{1, [1,3]} & -\mathcal{F}_{1, [1,2]} & \mathbf{0} & \mathbf{0} & \mathbf{0} \\
\mathbf{0} & -\mathcal{F}_{1, [1,4]} & \mathbf{0} & -\mathcal{F}_{1, [1,2]} & \mathbf{0} & \mathbf{0} \\
\mathbf{0} & -\mathcal{F}_{1, [1,5]} & \mathbf{0} & \mathbf{0} & -\mathcal{F}_{1, [1,2]} & \mathbf{0} \\
\mathbf{0} & -\mathcal{F}_{1, [1,6]} & \mathbf{0} & \mathbf{0} & \mathbf{0} & -\mathcal{F}_{1, [1,2]} \\
\hline
\mathbf{0} & \mathbf{0} & -\mathcal{F}_{1, [1,4]} & -\mathcal{F}_{1, [1,3]} & \mathbf{0} & \mathbf{0} \\
\mathbf{0} & \mathbf{0} & -\mathcal{F}_{1, [1,5]} & \mathbf{0} & -\mathcal{F}_{1, [1,3]} & \mathbf{0} \\
\mathbf{0} & \mathbf{0} & -\mathcal{F}_{1, [1,6]} & \mathbf{0} & \mathbf{0} & -\mathcal{F}_{1, [1,3]} \\
\hline
\mathbf{0} & \mathbf{0} & \mathbf{0} & -\mathcal{F}_{1, [1,5]} & -\mathcal{F}_{1, [1,4]} & \mathbf{0} \\
\mathbf{0} & \mathbf{0} & \mathbf{0} & -\mathcal{F}_{1, [1,6]} & \mathbf{0} & -\mathcal{F}_{1, [1,4]} \\
\hline
\mathbf{0} & \bfu & \bfv & \bfw & \mathbf{0} & \mathbf{0} \\
\end{array} \right] = \overline{\overline{\mathbf{C}}},
\end{split}
\end{equation*}
where $\bfu = \beta_2\mathcal{F}_{1,[1,5]} + \alpha_2\mathcal{F}_{1,[1,6]}$, $\bfv = \beta_3\mathcal{F}_{1,[1,5]} + \alpha_3\mathcal{F}_{1,[1,6]}$, and $\bfw = \beta_4\mathcal{F}_{1,[1,5]} + \alpha_4\mathcal{F}_{1,[1,6]}$. By expanding the terms $\mathcal{F}_{1,[1,5]}$ and $\mathcal{F}_{1,[1,6]}$ as linear combinations of $\mathcal{F}_{1, [1,2]}$, $\mathcal{F}_{1, [1,3]}$, and $\mathcal{F}_{1, [1,4]}$, 
\begin{equation*}
\begin{split}
\bfu &= (\alpha_2\beta_2 + \alpha_2\beta_2) \cdot \mathcal{F}_{1,[1,2]} + (\alpha_3\beta_2 + \alpha_2\beta_3) \cdot \mathcal{F}_{1,[1,3]} + (\alpha_4\beta_2 + \alpha_2\beta_4) \cdot \mathcal{F}_{1,[1,4]}, \\
\bfv &= (\alpha_2\beta_3 + \alpha_3\beta_2) \cdot \mathcal{F}_{1,[1,2]} + (\alpha_3\beta_3 + \alpha_3\beta_3) \cdot \mathcal{F}_{1,[1,3]} + (\alpha_4\beta_3 + \alpha_3\beta_4) \cdot \mathcal{F}_{1,[1,4]}, \\
\bfw &= (\alpha_2\beta_4 + \alpha_4\beta_2) \cdot \mathcal{F}_{1,[1,2]} + (\alpha_3\beta_4 + \alpha_4\beta_3) \cdot \mathcal{F}_{1,[1,3]} + (\alpha_4\beta_4 + \alpha_4\beta_4) \cdot \mathcal{F}_{1,[1,4]}. \\
\end{split}    
\end{equation*}
By adding the row vector combination 
\begin{equation}
\label{Eq. K6 linear combination}
\begin{split}
&-2\alpha_2\beta_2 \overline{\mathbf{C}}_1 -2\alpha_3\beta_3 \overline{\mathbf{C}}_2
-2\alpha_4\beta_4 \overline{\mathbf{C}}_3 + (\alpha_3\beta_2 + \alpha_2\beta_3) \cdot \overline{\mathbf{C}}_6 + (\alpha_4\beta_2 + \alpha_2\beta_4) \cdot \overline{\mathbf{C}}_7 + (\alpha_4\beta_3 + \alpha_3\beta_4) \cdot \overline{\mathbf{C}}_{10}     
\end{split}
\end{equation}
to the row vector $\overline{\overline{\mathbf{C}}}_{15} = \begin{bmatrix}
\mathbf{0} & \bfu & \bfv & \bfw & \mathbf{0} & \mathbf{0} \\
\end{bmatrix}$. More precisely, Let $\bfs_1, \bfs_2, \bfs_3, \bfs_4, \bfs_5, \bfs_6$ be the corresponding $1 \times 3$ vectors of the row vector in \eqref{Eq. K6 linear combination}, then we have
\begin{equation*}
\begin{split}
\bfs_1 &= \bfs_5 = \bfs_6 = \mathbf{0}, \\
\bfs_2 &= -2\alpha_2\beta_2 \cdot \mathcal{F}_{1, [1,2]} -(\alpha_3\beta_2 + \alpha_2\beta_3) \cdot \mathcal{F}_{1, [1,3]} - (\alpha_4\beta_2 + \alpha_2\beta_4) \cdot \mathcal{F}_{1, [1,4]} = -\bfu, \\
\bfs_3 &= -2\alpha_3\beta_3 \cdot \mathcal{F}_{1, [1,3]} -(\alpha_3\beta_2 + \alpha_2\beta_3) \cdot \mathcal{F}_{1, [1,2]} - (\alpha_4\beta_3 + \alpha_3\beta_4) \cdot \mathcal{F}_{1, [1,4]} = -\mathbf{v}, \\
\bfs_4 &= -2\alpha_4\beta_4 \cdot \mathcal{F}_{1, [1,4]} - (\alpha_4\beta_2 + \alpha_2\beta_4) \cdot \mathcal{F}_{1, [1,2]} - (\alpha_4\beta_3 + \alpha_3\beta_4) \cdot \mathcal{F}_{1, [1,3]} = -\mathbf{w}. 
\end{split}    
\end{equation*}
In particular, these equations imply that the matrix $\overline{\overline{\mathbf{C}}}$ becomes
\begin{equation*}
\widehat{\mathbf{C}} = 
\left[\begin{array}{cccccc}
\mathbf{0} & \mathcal{F}_{1, [1,2]} & \mathbf{0} & \mathbf{0} & \mathbf{0}  & \mathbf{0}  \\
\mathbf{0} & \mathbf{0} & \mathcal{F}_{1, [1,3]} & \mathbf{0} & \mathbf{0}  & \mathbf{0}  \\
\mathbf{0} & \mathbf{0} & \mathbf{0} & \mathcal{F}_{1, [1,4]} & \mathbf{0}  & \mathbf{0}  \\
\mathbf{0} & \mathbf{0} & \mathbf{0} & \mathbf{0}  & \mathcal{F}_{1, [1,5]} & \mathbf{0}  \\
\mathbf{0} & \mathbf{0} & \mathbf{0} & \mathbf{0}  & \mathbf{0} & \mathbf{0}  \\
\hline
\mathbf{0} & -\mathcal{F}_{1, [1,3]} & -\mathcal{F}_{1, [1,2]} & \mathbf{0} & \mathbf{0} & \mathbf{0} \\
\mathbf{0} & -\mathcal{F}_{1, [1,4]} & \mathbf{0} & -\mathcal{F}_{1, [1,2]} & \mathbf{0} & \mathbf{0} \\
\mathbf{0} & -\mathcal{F}_{1, [1,5]} & \mathbf{0} & \mathbf{0} & -\mathcal{F}_{1, [1,2]} & \mathbf{0} \\
\mathbf{0} & -\mathcal{F}_{1, [1,6]} & \mathbf{0} & \mathbf{0} & \mathbf{0} & -\mathcal{F}_{1, [1,2]} \\
\hline
\mathbf{0} & \mathbf{0} & -\mathcal{F}_{1, [1,4]} & -\mathcal{F}_{1, [1,3]} & \mathbf{0} & \mathbf{0} \\
\mathbf{0} & \mathbf{0} & -\mathcal{F}_{1, [1,5]} & \mathbf{0} & -\mathcal{F}_{1, [1,3]} & \mathbf{0} \\
\mathbf{0} & \mathbf{0} & -\mathcal{F}_{1, [1,6]} & \mathbf{0} & \mathbf{0} & -\mathcal{F}_{1, [1,3]} \\
\hline
\mathbf{0} & \mathbf{0} & \mathbf{0} & -\mathcal{F}_{1, [1,5]} & -\mathcal{F}_{1, [1,4]} & \mathbf{0} \\
\mathbf{0} & \mathbf{0} & \mathbf{0} & -\mathcal{F}_{1, [1,6]} & \mathbf{0} & -\mathcal{F}_{1, [1,4]} \\
\hline
\mathbf{0} & \mathbf{0} & \mathbf{0} & \mathbf{0} & \mathbf{0} & \mathbf{0} \\
\end{array} \right].    
\end{equation*}
Because $\mathcal{F}_{1,[1,2]}, \mathcal{F}_{1,[1,3]}, \mathcal{F}_{1,[1,4]}$ are linearly independent, ${\rm rank}(\widehat{\mathbf{C}}) = {\rm rank}(\mathbf{D}) + 3$. By Example \ref{Example: K5 example}, ${\rm rank}(\mathbf{D}) = 9$. Therefore, ${\rm rank}(\mathbf{C}) = {\rm rank}(\widehat{\mathbf{C}}) = 9 + 3 = 12$. 
\end{proof}
Building on the findings from Examples \ref{Example: K5 example} and \ref{Example: K6 example}, we are now prepared to demonstrate that any rank $3$ sheaves $\mathcal{F}: (K_n,\leq) \rightarrow \textup{\textsf{Vect}}_\mathbb{R}$, as described in Equation~\eqref{Eq. Anisotropic sheaf definition-form 2} on any $K_n$, necessarily possess a $0$-th coboundary matrix $\mathcal{F}: C^0(K,\mathcal{F}) \rightarrow C^1(K,\mathcal{F})$ with rank $3 \cdot (n-2)$.
\begin{theorem}\label{Theorem: Main result 3-}
\label{Theorem: Kn example}
Let $n \geq 4$, $G = K_n$ with $V = \{ 1, 2, ...,n \}$, and let $\mathcal{F}: (G,\leq) \rightarrow \textup{\textsf{Vect}}_\mathbb{R}$ be defined as in Equation~\eqref{Eq. Anisotropic sheaf definition-form 2}. If ${\rm rank}(\mathcal{F}) = 3$, then the $0$-th coboundary matrix $\mathbf{C}: C^0(G,\mathcal{F}) \rightarrow C^1(G,\mathcal{F})$ has rank $3 \cdot (n-2)$. In particular, we have $\dim_\mathbb{R}(L_\mathcal{F}) = 3 \cdot |V| - {\rm rank}(\mathbf{C}) = 3 \cdot n - (3 \cdot n - 6) = 6$.    
\end{theorem}
\begin{proof}
For $n = 4$, the theorem follows by Example \ref{Example: K4 example}. We use mathematical induction to prove the theorem.  Suppose the theorem holds for some $n \geq 4$. By column and row operations, we focus on the matrix
\begin{equation*}
\overline{\mathbf{C}} = \left[ \begin{array}{cccccc|c}
\mathbf{0} & \mathcal{F}_{1, [1,2]} & \mathbf{0} & \mathbf{0} & \cdots & \mathbf{0}  & \mathbf{0}  \\
\mathbf{0} & \mathbf{0} & \mathcal{F}_{1, [1,3]} & \mathbf{0} & \cdots & \mathbf{0}  & \mathbf{0}  \\
\vdots & \vdots & \vdots & \vdots & \vdots & \vdots  & \vdots  \\
\mathbf{0} & \mathbf{0} & \mathbf{0} & \mathbf{0} & \cdots  & \mathcal{F}_{1, [1,n]} & \mathbf{0}  \\
\mathbf{0} & \mathbf{0} & \mathbf{0} & \mathbf{0} & \cdots & \mathbf{0}  & \mathcal{F}_{1, [1,n+1]}  \\
\hline
\mathbf{0} & -\mathcal{F}_{1, [1,3]} & -\mathcal{F}_{1, [1,2]} & \mathbf{0} & \cdots & \mathbf{0} & \mathbf{0} \\
\mathbf{0} & -\mathcal{F}_{1, [1,4]} & \mathbf{0} & -\mathcal{F}_{1, [1,2]} & \cdots & \mathbf{0} & \mathbf{0} \\
\vdots & \vdots & \vdots & \vdots & \vdots & \vdots  & \vdots  \\
\mathbf{0} & -\mathcal{F}_{1, [1,n]} & \mathbf{0} & \mathbf{0} & \cdots & -\mathcal{F}_{1, [1,2]} & \mathbf{0} \\
\mathbf{0} & -\mathcal{F}_{1, [1,n+1]} & \mathbf{0} & \mathbf{0} & \cdots & \mathbf{0} & -\mathcal{F}_{1, [1,2]} \\
\hline
\mathbf{0} & \mathbf{0} & -\mathcal{F}_{1, [1,4]} & -\mathcal{F}_{1, [1,3]} & \cdots & \mathbf{0} & \mathbf{0} \\
\vdots & \vdots & \vdots & \vdots & \vdots & \vdots  & \vdots  \\
\mathbf{0} & \mathbf{0} & -\mathcal{F}_{1, [1,n]} & \mathbf{0} & \cdots & -\mathcal{F}_{1, [1,3]} & \mathbf{0} \\
\mathbf{0} & \mathbf{0} & -\mathcal{F}_{1, [1,n+1]} & \mathbf{0} & \cdots & \mathbf{0} & -\mathcal{F}_{1, [1,3]} \\
\hline
\vdots & \vdots & \vdots & \vdots & \vdots & \vdots  & \vdots  \\
\hline
\mathbf{0} & \mathbf{0} & \mathbf{0} & \mathbf{0} & \cdots & -\mathcal{F}_{1, [1,n+1]} & -\mathcal{F}_{1, [1,n]} \\
\end{array}\right].   
\end{equation*}
By assumption, the rank of $\mathcal{F}$ is $3$. Because $G = K_{n+1}$ is complete, all row vectors $\mathcal{F}_{i,[i,j]}$ can be spanned by the set $\{ \mathcal{F}_{1,[1,j]} \ | \ j = 2, ..., n+1 \}$ of row vectors.  Without loss of generality, we may assume that $\mathcal{F}_{1, [1,2]}, \mathcal{F}_{1, [1,3]}, \mathcal{F}_{1, [1,4]}$ are linearly independent.  By the proof of Example \ref{Example: K5 example}, the vector
\begin{equation*}
\left[ \begin{array}{cccccc|c}
\mathbf{0} & \mathbf{0} & \mathbf{0} & \mathbf{0} & \cdots & \mathbf{0}  & \mathcal{F}_{1, [1,n+1]}  \\
\end{array}\right]
\end{equation*}
is generated by the row vectors of the submatrix with size $9 \times 3 \cdot n$:
\begin{equation}
\label{Eq. Submatrix A1}
\mathbf{A}_1 = 
\left[ \begin{array}{ccccccc|c}
\mathbf{0} & \mathcal{F}_{1, [1,2]} & \mathbf{0} & \mathbf{0} & \mathbf{0} & \cdots & \mathbf{0} & \mathbf{0} \\
\mathbf{0} & \mathbf{0} & \mathcal{F}_{1, [1,3]} & \mathbf{0} & \mathbf{0} & \cdots & \mathbf{0} & \mathbf{0} \\
\mathbf{0} & \mathbf{0} & \mathbf{0} & \mathcal{F}_{1, [1,4]} & \mathbf{0} & \cdots & \mathbf{0} & \mathbf{0} \\
\hline
\mathbf{0} & -\mathcal{F}_{1, [1,3]} & - \mathcal{F}_{1, [1,2]} & \mathbf{0} & \mathbf{0} & \cdots & \mathbf{0} & \mathbf{0} \\
\mathbf{0} & -\mathcal{F}_{1, [1,4]} & \mathbf{0} & -\mathcal{F}_{1, [1,2]} & \mathbf{0} & \cdots & \mathbf{0} & \mathbf{0} \\
\mathbf{0} & -\mathcal{F}_{1, [1,n+1]} & \mathbf{0} & \mathbf{0} & \mathbf{0} & \cdots & \mathbf{0} & -\mathcal{F}_{1, [1,2]} \\
\hline
\mathbf{0} & \mathbf{0} & -\mathcal{F}_{1, [1,4]} & -\mathcal{F}_{1, [1,3]} & \mathbf{0} & \cdots & \mathbf{0} & \mathbf{0} \\
\mathbf{0} & \mathbf{0} & -\mathcal{F}_{1, [1,n+1]} & \mathbf{0} &  \mathbf{0} & \cdots & \mathbf{0} & -\mathcal{F}_{1, [1,3]} \\
\hline
\mathbf{0} & \mathbf{0} & \mathbf{0} & -\mathcal{F}_{1, [1,n+1]} & \mathbf{0} & \cdots & \mathbf{0} & -\mathcal{F}_{1, [1,4]} \\
\end{array}\right].    
\end{equation}
On the other hand, by the proof of Example \ref{Example: K6 example}, for every $j \in \{ 5, ..., n \}$, the row vector 
\begin{equation*}
\left[ \begin{array}{ccccccc|c}
\mathbf{0} & \mathbf{0} & \cdots & -\mathcal{F}_{1, [1,n+1]} & \mathbf{0} & \cdots & \mathbf{0}  & -\mathcal{F}_{1, [1,j]}  \\
\end{array}\right]
\end{equation*}
with the vector at the $j$th location of the row is generated by the row vectors of the submatrix with size $13 \times 3 \cdot  n$:
\begin{equation*}
\label{Eq. Submatrix A2}
\mathbf{A}_2 = 
\left[\begin{array}{cccccccc|c}
\mathbf{0} & \mathcal{F}_{1, [1,2]} & \mathbf{0} & \mathbf{0} & \mathbf{0} & \cdots & \cdots & \mathbf{0}  & \mathbf{0}  \\
\mathbf{0} & \mathbf{0} & \mathcal{F}_{1, [1,3]} & \mathbf{0} & \mathbf{0}  & \cdots & \cdots & \mathbf{0} & \mathbf{0}  \\
\mathbf{0} & \mathbf{0} & \mathbf{0} & \mathcal{F}_{1, [1,4]} & \mathbf{0}  & \cdots & \cdots & \mathbf{0} & \mathbf{0}  \\
\mathbf{0} & \mathbf{0} & \mathbf{0} & \mathbf{0}  & \cdots & \mathcal{F}_{1, [1,j]} & \cdots & \mathbf{0} & \mathbf{0} \\
\hline
\mathbf{0} & -\mathcal{F}_{1, [1,3]} & -\mathcal{F}_{1, [1,2]} & \mathbf{0} & \mathbf{0} & \mathbf{0} & \cdots & \mathbf{0} & \mathbf{0} \\
\mathbf{0} & -\mathcal{F}_{1, [1,4]} & \mathbf{0} & -\mathcal{F}_{1, [1,2]} & \mathbf{0} & \cdots & \cdots & \mathbf{0} & \mathbf{0} \\
\mathbf{0} & -\mathcal{F}_{1, [1,j]} & \mathbf{0} & \mathbf{0} & \cdots & -\mathcal{F}_{1, [1,2]} & \cdots & \mathbf{0} & \mathbf{0} \\
\mathbf{0} & -\mathcal{F}_{1, [1,n+1]} & \mathbf{0} & \mathbf{0} & \mathbf{0} & \mathbf{0} & \cdots & \mathbf{0} & -\mathcal{F}_{1, [1,2]} \\
\hline
\mathbf{0} & \mathbf{0} & -\mathcal{F}_{1, [1,4]} & -\mathcal{F}_{1, [1,3]} & \mathbf{0} & \mathbf{0} & \cdots & \mathbf{0} & \mathbf{0} \\
\mathbf{0} & \mathbf{0} & -\mathcal{F}_{1, [1,j]} & \mathbf{0} & \cdots & -\mathcal{F}_{1, [1,3]} & \cdots & \mathbf{0} & \mathbf{0} \\
\mathbf{0} & \mathbf{0} & -\mathcal{F}_{1, [1,n+1]} & \mathbf{0} & \mathbf{0} & \cdots & \cdots & \mathbf{0} & -\mathcal{F}_{1, [1,3]} \\
\hline
\mathbf{0} & \mathbf{0} & \mathbf{0} & -\mathcal{F}_{1, [1,j]} & \cdots & -\mathcal{F}_{1, [1,4]} & \cdots  & \mathbf{0} & \mathbf{0} \\
\mathbf{0} & \mathbf{0} & \mathbf{0} & -\mathcal{F}_{1, [1,n+1]} & \mathbf{0} & \mathbf{0} & \cdots & \mathbf{0} & -\mathcal{F}_{1, [1,4]} \\
\end{array}\right].    
\end{equation*}
Note that the matrix $\overline{\mathbf{C}}$ is formed by augmenting $\overline{\mathbf{D}}$---the matrix corresponding to the complete graph $K_n$ with vertex set $\{1, 2, \ldots, n \}$---with $n$ new rows, each a $1 \times 3n$ vector. Given that $\mathcal{F}_{1,[1,2]}, \mathcal{F}_{1,[1,3]}, \mathcal{F}_{1,[1,4]}$ are linearly independent, and by the induction hypothesis, the rank of $\overline{\mathbf{D}}$ is $3n - 6$. Observing the rows in matrices $\mathbf{A}_1$ and $\mathbf{A}_2$, we note that, apart from the rows in matrix $\overline{\mathbf{D}}$, only three additional linearly independent vectors are required to generate all the rows in $\overline{\mathbf{C}}$. This shows that ${\rm rank}(\overline{\mathbf{C}}) = {\rm rank}(\overline{\mathbf{D}}) + 3 = 3n - 3 = 3(n+1) - 6$. By mathematical induction, the theorem follows.
\end{proof}
\section{Simplicial Complexes and Their Geometric Realizations}
This paper uses two typical representations of a finite simplicial complex in $\mathbb{R}^3$. First, we represent a simplicial complex in $\mathbb{R}^3$ as a collection $K$ of geometric simplices embedded in $\mathbb{R}^3$, equipped with a partial order $\leq$ that based on the face relations of the simplices; that is, as a partially ordered set $(K, \leq)$ of geometric simplices in $\mathbb{R}^3$. Second, we consider the simplicial complex as a subspace of $\mathbb{R}^3$, known as its geometric realization in $\mathbb{R}^3$ and denoted by $|K|$. The former representation focuses primarily on the combinatorial face relations of simplices in $K$, while the latter embeds the entire finite simplicial complex as a compact subset of $\mathbb{R}^3$, formed by the union of geometric simplices in $\mathbb{R}^3$. We briefly recap these formal definitions as follows.
\begin{def.}
A \textbf{simplicial complex} in \( \mathbb{R}^3 \) is a collection \( K \) of simplices, where each simplex is the convex hull \( \sigma = \mathrm{conv}(S) \) for some affinely independent set \( S \subseteq \mathbb{R}^3 \). Two simplices $\sigma$ and $\tau$ satisfy the relation $\sigma \leq \tau$ if $\sigma$ is a face of $\tau$. The collection \( K \) satisfies the following properties: {\rm (a)} if \( \sigma \in K \) and \( \tau \leq \sigma \), then \( \tau \in K \); and {\rm (b)} if \( \sigma, \tau \in K \), then \( \sigma \cap \tau \in K \).
\end{def.}
A simplicial complex in $\mathbb{R}^3$, as defined above, is a collection of subsets of $\mathbb{R}^3$, with an emphasis on the combinatorial properties of a collection of simplices within $\mathbb{R}^3$. On the other hand, when analyzing the global structure of a simplicial complex embedded in $\mathbb{R}^3$, its geometric realization is considered, which is defined as follows.
\begin{def.}
Let $K$ be a finite simplicial complex in $\mathbb{R}^3$. Then, the \textbf{geometric realization} of $K$, denoted as $|K|$, is the union of all simplices in $\mathbb{R}^3$, i.e., $|K| = \bigcup_{\sigma \in K} \sigma \subseteq \mathbb{R}^3$.
\end{def.}
To investigate the geometric and topological properties of $|K|$, we briefly recap the following well-known properties of simplices within $\mathbb{R}^3$, focusing particularly on their boundary, interior, and supporting hyperplane properties.
\begin{prop.}
\label{Proposition: Elementary Munkres properties}
Let $S = \{ \bfx_0, ..., \bfx_q \}$ be an affinely independent set in $\mathbb{R}^3$ and let $\sigma = \conv(S)$ be the simplex generated by $S$. Then the following assertions hold.
\begin{itemize}
\item[\rm (a)] The relative interior of $\sigma$, denoted by  $\reInt(\sigma)$, is $\{ \sum_{i = 0}^q t_i \bfx_i \ | \ t_i \in (0, 1), \sum_{i = 0}^q t_i = 1 \}$.
\item[\rm (b)] If $\sigma$ is an $3$-simplex, then the interior and relative interior of $\sigma$ coincide, i.e., $\reInt(\sigma) = \Int(\sigma)$.
\item[\rm (c)] The all $(q-1)$-faces of $\sigma$ are $\conv(S_i)$, where $S_i = S \setminus \{ \bfx_i \}$ for $i \in \{ 0, ..., q \}$.
\item[\rm (d)] The relative boundary of $\sigma$, denoted by $\reBd(\sigma)$, is the union of all $(q-1)$-faces of $\sigma$.
\item[\rm (e)] If $\sigma$ is a $3$-simplex, then the boundary and relative boundary of $\sigma$ coincide, i.e., $\reBd(\sigma) = \bd(\sigma)$.
\end{itemize}
\end{prop.}
\begin{proof}
These properties are not limited to the case of $N = 3$; they hold when $S = \{ \bfx_0, \dots, \bfx_q \}$ is a set of $q+1$ affinely independent vectors in $\mathbb{R}^N$ for any $N, q \in \mathbb{N}$ with $q \leq N$ \cite{munkres2018elements}.
\end{proof}
Furthermore, due to the convexity of simplices, every $3$-simplex in $\mathbb{R}^3$ can be separated by a hyperplane spanned by its faces. More precisely, for a $3$-simplex \( \sigma = \mathrm{conv}(S) \) generated by an affinely independent set \( S = \{ \mathbf{x}_0, \mathbf{x}_1, \mathbf{x}_2, \mathbf{x}_3 \} \subseteq \mathbb{R}^3 \), any $2$-face spans a hyperplane in \( \mathbb{R}^3 \). For instance, for the $2$-face \( \tau = \mathrm{conv}(\mathbf{x}_0, \mathbf{x}_1, \mathbf{x}_2) \) of \( \sigma \), the set \( \{ \mathbf{x}_1 - \mathbf{x}_0, \mathbf{x}_2 - \mathbf{x}_0 \} \) is linearly independent, and the hyperplane spanned by \( \tau \) is defined as the affine hull of \( \mathbf{x}_0, \mathbf{x}_1, \mathbf{x}_2 \), i.e., $\mathrm{aff}(\mathbf{x}_0, \mathbf{x}_1, \mathbf{x}_2) = \mathbf{x}_0 + \mathrm{span}_{\mathbb{R}} \{ \mathbf{x}_1 - \mathbf{x}_0, \mathbf{x}_2 - \mathbf{x}_0 \}$. In addition, if $\bfv$ is a vector that is perpendicular to vectors $\bfx_1 - \bfx_0, \bfx_2 - \bfx_0$, then
\begin{equation*}
\bfx_0 + {\rm span}_{\mathbb{R}} \{ \bfx_1 - \bfx_0, \bfx_2 - \bfx_0 \} = H(\bfx_0, \bfv) := \{ \bfx \in \mathbb{R}^3 \ | \ \langle \bfv, \bfx - \bfx_0 \rangle = 0 \}.    
\end{equation*}
Furthermore, if one chooses $\bfv$ with the property that $\langle \bfv, \bfx_3 - \bfx_0 \rangle > 0$, then for scalars $t_0, t_1, t_2, t_3 \in [0,1]$ with $t_0 + t_1 + t_2 + t_3 = 1$, one obtains
\begin{equation*}
\begin{split}
\langle \bfv, (t_0\bfx_0 + t_1\bfx_1 + t_2\bfx_2 + t_3\bfx_3) -\bfx_0 \rangle &= \sum_{i = 0}^3 t_i \cdot \langle \bfv, \bfx_i - \bfx_0 \rangle = t_3 \cdot \langle \bfv, \bfx_3 - \bfx_0 \rangle \geq 0.
\end{split}
\end{equation*}
In other words, the entire simplex $\sigma$ is contained in the \textit{closed half-space} $H(\mathbf{x}_0, \mathbf{v})_{\geq 0} := \{ \mathbf{x} \in \mathbb{R}^3 \mid \langle \mathbf{v}, \mathbf{x} - \mathbf{x}_0 \rangle \geq 0 \}$. Furthermore, the interior of \(\sigma\) is contained in the \textit{open half-space} \( H(\mathbf{x}_0, \mathbf{v})_{> 0} := \{ \mathbf{x} \in \mathbb{R}^3 \mid \langle \mathbf{v}, \mathbf{x} - \mathbf{x}_0 \rangle > 0 \} \). Similarly, the opposite closed and open half-spaces \( H(\mathbf{x}_0, \mathbf{v})_{\leq 0} := \{ \mathbf{x} \in \mathbb{R}^3 \mid \langle \mathbf{v}, \mathbf{x} - \mathbf{x}_0 \rangle \leq 0 \} \) and \( H(\mathbf{x}_0, \mathbf{v})_{< 0} := \{ \mathbf{x} \in \mathbb{R}^3 \mid \langle \mathbf{v}, \mathbf{x} - \mathbf{x}_0 \rangle < 0 \} \) are defined. With this geometric intuition, the intersection \( \sigma \cap H(\mathbf{x}_0, \mathbf{v}) \) is often referred to as the \textit{face} of \( \sigma \) opposite to \( \mathbf{x}_3 \)~\cite{munkres2018elements}. We summarize this observation in the following proposition, which will be beneficial in investigating simplicial complexes in \( \mathbb{R}^3 \).
\begin{prop.}
\label{Proposition: Supporting hyperplane of simplex on face}
Let \( S = \{ \mathbf{x}_0, \mathbf{x}_1, \mathbf{x}_2, \mathbf{x}_3 \} \) be an affinely independent set in \(\mathbb{R}^3\), and let \(\sigma = \operatorname{conv}(S)\) be the $3$-simplex generated by \( S \). Let \(\mathbf{v}\) be a vector in \(\mathbb{R}^3\) that is perpendicular to the vectors \(\mathbf{x}_1 - \mathbf{x}_0\) and \(\mathbf{x}_2 - \mathbf{x}_0\) and satisfies the property \(\langle \mathbf{v}, \mathbf{x}_3 - \mathbf{x}_0 \rangle > 0 \). Then, 
\begin{itemize}
\item[\rm (a)] $\sigma \subseteq H(\mathbf{x}_0, \mathbf{v})_{\geq 0} := \{ \mathbf{x} \in \mathbb{R}^3 \mid \langle \mathbf{v}, \mathbf{x} - \mathbf{x}_0 \rangle \geq 0 \}$;
\item[\rm (b)] $\Int(\sigma) \subseteq H(\mathbf{x}_0, \mathbf{v})_{> 0} := \{ \mathbf{x} \in \mathbb{R}^3 \mid \langle \mathbf{v}, \mathbf{x} - \mathbf{x}_0 \rangle > 0 \}$;
\item[\rm (c)] each $\bfw \in \reInt(\conv(\mathbf{x}_0, \mathbf{x}_1, \mathbf{x}_2))$ admits an open ball $B_{\epsilon}(\bfw) := \{ \bfx \in \mathbb{R}^3 \ | \ \Vert \bfx - \bfw \Vert < \epsilon \}$ centered at $\bfw$ with radius $\epsilon$ such that $B_{\epsilon}(\bfw) \cap H(\mathbf{x}_0, \mathbf{v})_{> 0} \subseteq \Int(\sigma)$;
\item[\rm (d)] if $\bfy$ is another point in $H(\mathbf{x}_0, \mathbf{v})_{> 0}$ such that $T = \{ \mathbf{x}_0, \mathbf{x}_1, \mathbf{x}_2, \bfy \}$ is an affinely independent set and $\tau = \conv(T)$, then $\Int(\tau) \cap \Int(\sigma) \neq \emptyset$; in particular, $\tau$ and $\sigma$ share the $2$-face $\conv(\mathbf{x}_0, \mathbf{x}_1, \mathbf{x}_2)$.
\end{itemize}
\end{prop.}
\begin{proof}
Generalized statements for properties (a), (b), and (c) in the case of simplices of arbitrary dimension can be found in \cite{munkres2018elements}. Leveraging (a), (b), and (c), we prove property (d), which helps explore homogeneous simplicial $3$-complexes in $\mathbb{R}^3$.  Let $\bfy$ be another point in $H(\mathbf{x}_0, \mathbf{v})_{> 0}$ such that $T = \{ \mathbf{x}_0, \mathbf{x}_1, \mathbf{x}_2, \bfy \}$ is an affinely independent set and $\tau = \conv(T)$. Then, both $\sigma$ and $\tau$ satisfy properties (a), (b), and (c). Choose $\bfw \in \reInt(\conv(\mathbf{x}_0, \mathbf{x}_1, \mathbf{x}_2))$, then there are $\epsilon_1 > 0$ and $\epsilon_2 > 0$ such that $B_{\epsilon_1}(\bfw) \cap H(\mathbf{x}_0, \mathbf{v})_{> 0} \subseteq \Int(\sigma)$ and $B_{\epsilon_2}(\bfw) \cap H(\mathbf{x}_0, \mathbf{v})_{> 0} \subseteq \Int(\tau)$. By choosing $\epsilon = \min\{ \epsilon_1, \epsilon_2 \}$, then $\Int(\tau) \cap \Int(\sigma) \neq \emptyset$ since $\emptyset \neq B_{\epsilon}(\bfw) \cap H(\mathbf{x}_0, \mathbf{v})_{> 0} \subseteq \Int(\sigma)$ and $\emptyset \neq B_{\epsilon}(\bfw) \cap H(\mathbf{x}_0, \mathbf{v})_{> 0} \subseteq \Int(\tau)$.
\end{proof}
In addition to representing a $3$-simplex \( \sigma \subseteq \mathbb{R}^3 \) as the convex hull of affinely independent vectors \( \mathbf{x}_0, \mathbf{x}_1, \mathbf{x}_2, \mathbf{x}_3 \), one can also represent a $3$-simplex as the intersection of closed half-spaces defined by its faces. Actually, any convex polytope is the intersection of closed half-spaces defined by its faces (cf. \cite{ewald1996combinatorial}). We formalize this specific fact for the case of $3$-simplices in the following proposition.
\begin{prop.}
\label{Proposition: sigma as the intersection of all its face spaces}
Let $\sigma \subseteq \mathbb{R}^3$ be the $3$-simplex defined as the convex hull of an affinely independent set $S = \{ \mathbf{x}_0, \mathbf{x}_1, \mathbf{x}_2, \mathbf{x}_3 \}$. For each $i \in \{ 0, 1, 2, 3 \}$, there is a normal vector $\mathbf{v}_i \in \mathbb{R}^3$ as in Proposition \ref{Proposition: Supporting hyperplane of simplex on face} such that $\sigma \subseteq H(\mathbf{x}_i, \mathbf{v}_i)_{\geq 0}$ and $\sigma = \bigcap_{i=0}^3 H(\mathbf{x}_i, \mathbf{v}_i)_{\geq 0}$.
\end{prop.}
The intersection formula in Proposition \ref{Proposition: sigma as the intersection of all its face spaces} has an alternative geometric explanation. More precisely, by choosing the vectors $\mathbf{v}_i$ properly, the supporting hyperplanes $H(\mathbf{x}_i, \mathbf{v}_i)$ in Proposition \ref{Proposition: sigma as the intersection of all its face spaces} are exact the hyperplanes spanned by the $2$-faces of $\sigma$. For instance, if $\mathbf{x}_0, \mathbf{x}_1, \mathbf{x}_2, \mathbf{x}_3$ and $\mathbf{v} \in \mathbb{R}^3$ are defined as in Proposition \ref{Proposition: Supporting hyperplane of simplex on face}, then $H(\mathbf{x}_0, \mathbf{v}) = {\rm aff}(\mathbf{x}_0, \mathbf{x}_1, \mathbf{x}_2)$, where ${\rm aff}(S)$ denotes the affine hull spanned by a set $S \subseteq \mathbb{R}^3$.
\begin{coro}
\label{Corollary: A point outside the sigma and supporting hyperplane thm}
Let $\sigma \subseteq \mathbb{R}^3$ be the $3$-simplex defined as the convex hull of an affinely independent set $S = \{ \mathbf{x}_0, \mathbf{x}_1, \mathbf{x}_2, \mathbf{x}_3 \}$. For every $\mathbf{x} \in \mathbb{R}^3 \setminus \sigma$, there is an $i \in \{ 0, 1, 2, 3 \}$ such that $H := {\rm aff}(S \setminus \{ \mathbf{x}_i \})$ is a supporting hyperplane that separates $\bfx$ and $\sigma$ into the disjoint open and closed half-spaces. 
\end{coro}
\begin{proof}
By Proposition \ref{Proposition: sigma as the intersection of all its face spaces}, by choosing the vectors $\mathbf{v}_0, \mathbf{v}_1, \mathbf{v}_2, \mathbf{v}_3$ as required in the proposition, the $3$-simplex $\sigma$ can be represented as the intersection $\sigma = \bigcap_{i=0}^3 H(\mathbf{x}_i, \mathbf{v}_i)_{\geq 0}$. Then,
\begin{equation*}
\mathbf{x} \in \mathbb{R}^3 \setminus \left( \bigcap_{i=0}^3 H(\mathbf{x}_i, \mathbf{v}_i)_{\geq 0} \right) = \bigcup_{i=0}^3 \mathbb{R}^3 \setminus H(\mathbf{x}_i, \mathbf{v}_i)_{\geq 0} = \bigcup _{i=0}^3 H(\mathbf{x}_i, \mathbf{v}_i)_{< 0}.
\end{equation*}
Say $\mathbf{x} \in H(\mathbf{x}_i, \mathbf{v}_i)_{< 0}$ for some $i \in \{ 0, 1, 2, 3 \}$ and set $H = H(\mathbf{x}_i, \mathbf{v}_i)$. Because $\sigma \subseteq H(\mathbf{x}_i, \mathbf{v}_i)_{\geq 0}$, the supporting hyperplane separates $\sigma$ and $\mathbf{x}$ into two different half-spaces.
\end{proof}

\end{document}